# AUDIT

# DU SYSTÈME D'INFORMATION ET DU MODÈLE DE GOUVERNANCE DE LA BIBLIOTHÈQUE NUMÉRIQUE DE L'ESPACE UNIVERSITAIRE FRANCOPHONE (BNEUF)

## Projet IDNEUF

**Initiative pour le Développement du Numérique dans l'Espace Universitaire Francophone**

*Établi pour l'Agence universitaire de la Francophonie*

**Mokhtar BEN HENDA**

**Janvier 2023**

# Table des matières

# PARTIE I : ANALYSE DES FONCTIONS ET SERVICES

## 1 PRÉAMBULE

Ce rapport s'inscrit dans la continuité du dernier état des lieux de la Bibliothèque Numérique de l'Espace Universitaire Francophone (BNEUF), réalisé le 17 novembre 2021 par Cheikh Ahmed Tidiane Niang, Responsable conception développement et déploiement d'applications à la Direction du numérique (DN) de l'AUF, Département Solutions Applicatives et Innovation (DSAI)[1]. Dans ce dernier état des lieux, après une description du cycle de vie et un diagnostic des fonctionnalités et usages de la BNEUF, il a été surtout question d'identifier les forces et les faiblesses qui ont fait que la BNEUF, malgré son énorme potentiel, n'a pas connu le succès escompté en termes d'usage, de contributions collectives et d'impact scientifique.

Dans l'optique d'appuyer la nouvelle stratégie de l'AUF pour 2021-2025, avec ses nouveaux fondements structurels et de gouvernance pour la réalisation du projet de la Francophonie scientifique, il a été décidé d'aller vers un réaménagement des ressources et services numériques existants et à venir en vue de les incorporer à la future plateforme collaborative mondiale de services intégrés. L'objectif de ce réaménagement est d'introduire une nouvelle approche permettant à la fois une meilleure organisation collaborative des contenus (mise en valeur de collections réparties), une meilleure efficacité dans les usages des contenus (de l'information à la connaissance), une meilleure ergonomie dans les interfaces homme-machine (Ux-design[2]) et une meilleure articulation avec les services connexes (Atlas des experts et réseaux sociaux).

Dans le présent document, il sera question d'un diagnostic de la structure globale du dispositif BNEUF et de son mode opératoire. Dans un premier temps il sera question de rappeler les conclusions du rapport de diagnostic interne fait par la Direction du numérique de l'AUF. Dans un deuxième temps, il sera question de proposer un diagnostic externe (complémentaire) qui proposera de nouvelles formes d'organisation et d'usage du dispositif BNEUF. L'objectif ultime est de fournir à l'équipe projet de l'AUF de nouvelles pistes d'une gestion optimisée des ressources numériques de la BNEUF en lien avec l'Atlas de l'expertise et le Réseau social francophone.

Il faut souligner aussi que le présent rapport ne couvrira pas les aspects techniques relatifs aux développements informatiques et réseautiques ni aux aspects de gouvernance comme les partenariats avec les éditeurs ou le modèle économique pour le financement du dispositif. La question de la gouvernance sera toutefois traitée dans un deuxième document complémentaire une fois le contenu du présent document est revu et mis à jour.

## 2 DIAGNOSTIC INTERNE PAR LA DIRECTION DU NUMÉRIQUE DE L'AUF

Le rapport de diagnostic de la Direction du numérique de l'AUF constitue l'une des ressources bibliographiques essentielles du présent document. Le rapport de la DN a permis d'établir avec précision un état des lieux critique de la BNEUF. Il sera donc inutile de reproduire ici les mêmes détails concernant un dispositif qui n'a pas subi de transformations majeures dans sa structure ni dans sa

[1] Cheikh Ahmed Tidiane Niang. « Etat des lieux de la BNEUF ».

[2] Ux-design ou « conception de l'expérience utilisateur » est le processus de définition de l'expérience qu'un utilisateur vivrait lorsqu'il interagit avec un produit numérique ou un site Web

gouvernance. Il est néanmoins utile d'en rappeler les éléments les plus significatifs, notamment les points forts et les points faibles, pour construire dessus des solutions de renforcement et des alternatives de remédiation. Car, il s'agit avant tout de capitaliser sur l'existant et proposer des améliorations dans une optique qui s'inscrit dans les orientations de la nouvelle stratégie de l'AUF.

## 2.1 Les points forts à retenir

La BNEUF a sans doute un potentiel important qu'il faudrait entretenir selon les nouvelles tendances technologiques et les nouvelles procédures de gestion des données numériques à l'échelle mondiale. Notons principalement :

- **Son architecture ORI/OAI** qui répond aux caractéristiques d'un système numérique libre et ouvert permettant de gérer tous les documents numériques produits par les établissements universitaires membres, de les partager avec d'autres établissements, de les valoriser par une indexation professionnelle, et de les rendre accessibles à distance selon des droits définis à travers des interfaces ergonomiques. Ce modèle nécessite toutefois de s'ouvrir davantage sur les spécificités nouvelles du Web des données et des donnée liées ;
- **Sa migration vers un système nuagique** qui combine des services multiples (i.e. IBM Cloudant & Redis, ElasticSearch, Azure AD, ec.) sous forme de base de données distribuée entièrement gérée et optimisée pour les charges de travail lourdes (ressources & utilisateurs) et les applications Web et mobiles à croissance rapide. La nature entièrement gérée de ces services permet de se concentrer sur le développement des applications (APIs) au lieu d'avoir à se soucier de l'infrastructure serveur ;
- **Son potentiel de construire des réseaux de portails inter-universitaires** pour échanger et partager les patrimoines numériques de chaque établissement ;
- **Sa capacité de moissonnage** par un protocole (OAI-PMH) capable d'indexer des entrepôts contenant des millions de ressources à des fréquences régulières, de permettre l'échange des métadonnées sur Internet entre plusieurs institutions et de multiplier les accès aux documents numériques ;
- **Son caractère multidisciplinaire** qui couvre un large éventail de domaines pouvant satisfaire un grand nombre d'utilisateurs francophones ;
- **Ses fonctionnalités de recherche et de filtrage** des résultats, d'évaluation et de partage de ressources (bibliothèque personnalisée, classeurs thématiques et partageable, etc.) ;
- **Une structuration autour d'un écosystème** fournissant des services connexes complémentaires dont un Atlas d'experts et un réseau social francophone qui favorisent un environnement numérique intégré pour le partage de ressources et l'échange collaboratifs, notamment dans des communautés privées ;
- **Une configuration qui favorise la production de marques blanches** en vue de rallier des produits ou services connexes créés par des partenaires externes dans des domaines spécifiques ;

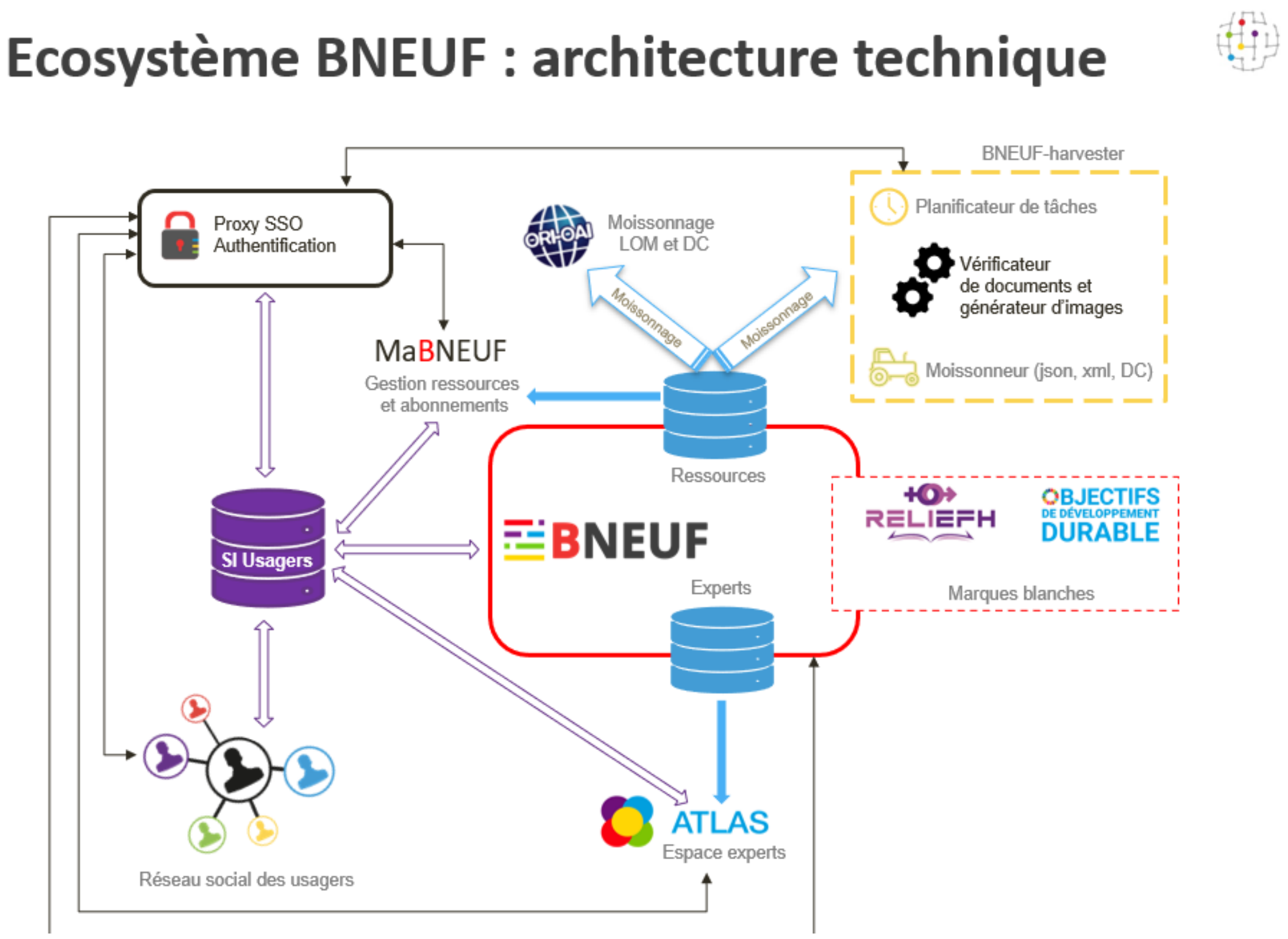

Source : « Cartographie de l'écosystème BNEUF ». Cheikh Ahmed Tidiane Niang

Ces acquis traduisent une maturité avancée dans le processus de conception et mise en œuvre du dispositif BNEUF qu'il faudrait néanmoins retravailler par une analyse qualitative interne et mettre au diapason des nouvelles tendances technologiques des systèmes d'information numériques.

## 2.2 Les points faibles à traiter

Le rapport de diagnostic de la DN a mis le doigt sur des formes de dysfonctionnement issues principalement des changements d'orientation et de choix stratégiques des équipes dirigeantes de l'AUF depuis le lancement du projet IDNEUF en 2015. Ces formes de dysfonctionnement sont synthétisées essentiellement dans les points suivants :

- **Écart entre le volume de l'offre documentaire et les taux des usages** : l'accroissement accéléré du volume de la BNEUF a fini par ne plus correspondre aux besoins réels des principaux utilisateurs visés qui sont les étudiants et les enseignants-chercheurs. Cet état souligne l'existence d'un problème dans la stratégie de sélection et d'alimentation du portail en ressources numériques. À force de vouloir renflouer les BD du dispositif pour lui donner de la consistance, des ressources non adaptées à la catégorie des utilisateurs ont fini par s'infiltrer créant ce que l'on appelle des « déchets documentaires ». Une démarche de filtrage à la source, et de nettoyage de l'existant (selon le principe de désherbage dans les bibliothèques classiques) est à prévoir au sein d'une politique générale de ravitaillement rentable ;

- **Absence de politique éditoriale** : Le rapport souligne l'absence d'une politique éditoriale claire et documentée pour BNEUF. Une politique éditoriale définit le cadre général de l'information collectée et diffusée ainsi que les normes et critères d'édition, de gestion et de présentation des contenus. Il est dès lors fortement recommandé d'en produire une qui couvrirait l'ensemble de la chaine de production/collecte, traitement, conservation, récupération et diffusion des ressources dans BNEUF ;

- **Incohérence dans la gestion des données de la propriété intellectuelle des ressources** : puisque la BNEUF contient des ressources libres de droits et d'autres protégées par des droits d'auteurs, ces spécifications sensibles de protection de la propriété intellectuelle ne suivent pas toujours les règles en vigueur. Or, dans un cadre d'usage public et ouvert, ces paramètres sont à prendre avec beaucoup de précaution ;

- **Manque de clartés sur l'accès libre et gratuit aux ressources** : ce dysfonctionnement ramène au problème de modèle économique hybride mal géré entre ressources libres et payantes de la BNEUF. Initialement fondée sur le principe des ressources éducatives libres (REL), BNEUF a été renforcé en 2018 par un partenariat avec des éditeurs privées pour alimenter ses fonds documentaires. Or, les ressources dans BNEUF ne sont pas systématiquement tagguées avec les indicateurs qui permettent aux utilisateurs d'en faire la distinction durant les opérations de recherche ;

- **Incohérence dans le processus de référencement par métadonnées :** ce dysfonctionnement est attribué à un problème de référencement irrégulier des ressources par des métadonnées aléatoires qui, avec la massification des contenus et l'accroissement des usages, rend la recherche très contaminée par les deux phénomènes de « bruit » et de « silence », c.à.d. restitution de ressources non-pertinentes et inversement, non-restitution de ressources pertinentes. Or, des métadonnées structurées, normalisées et représentatives au cœur d'un système d'information sont la condition principale de sa réussite ;

- **Absence de partenariats avec les fournisseurs de ressources gratuites**. « L'intégralité des ressources gratuites exposées sur la BNEUF a été moissonné à partir d'entrepôts publics. Cependant, aucun accord de partenariat n'a été noué avec les différents fournisseurs de ces ressources. Par conséquent, aucune visibilité sur les modalités de constitution des catalogues n'est donnée. Le portail ne mentionne aucun partenaire ou modalité de recueil ». La négligence du caractère public des ressources BNEUF, victimes de leur gratuité, est pami les causes qui ont mal impacté la qualité de BNEUF ;

- **L'absence d'une politique incitative avec les contributeurs universitaires francophones à la BNEUF** : l'absence d'un cadre réglementaire de collaboration avec les universités partenaires pour alimenter la BNEUF avec des ressources locales est de nature à fausser le rôle de la BNEUF sur deux principaux aspects : d'abord un défaut dans la pertinence des ressources de la BNEUF par rapport aux attentes et besoins de tous les acteurs. Les fonds provenant majoritairement d'un moissonnage systématique d'entrepôts et de portails publics essentiellement du Nord, manqueraient de contextualité pour un large public des pays du Sud ; ensuite la collecte aléatoire de ressources provenant d'établissements universitaires membres de l'AUF reste souvent muette sur la contribution des enseignants et formateurs dans ces établissements qui pourraient par ricochet alimenter l'annuaire des experts ;

- **Absence d'un cadre stable et permanent de gouvernance de la BNEUF, autrement un « comité éthique et organisationnel de gestion de la BNEUF » :** c'est l'une des principales recommandations du dernier rapport de la Direction du numérique. BNEUF nécessite une structure de pilotage sous le contrôle de la Direction du numérique avec la contribution de services connexes agissant sur les questions financières et ressources humaines. Cette structure devrait disposer d'un cadre de travail reconnu dans l'organigramme de l'AUF.

Le rapport de la DN s'est terminé par une série de recommandations « non exhaustives » qu'il serait utile de synthétiser ci-après.

## 2.3 Recommandations du rapport de la DN

Au vu du constat critique de la BNEUF, le rapport interne prévoit que l'équipe projet de l'AUF agisse comme suit :

- **Miser sur une stratégie d'étapes** à trois temps avec des objectifs à court, moyen, et long terme. Une proposition détaillée sur cette planification prospective sera décrite ultérieurement ;
- **Bien identifier les publics cibles** en étudiant tous les profils utilisateurs auxquels la BNEUF doit s'adresser. Il est entendu ici sans doute un meilleur contrôle sur les usages via une politique d'accès contrôlée aux services et ressources de la BNEUF. Ce point pose la controverse des architectures des portails avec ENT intégré : espace Internet public Vs espace Intranet réservé. Quel schéma la BNEUF devrait suivre pour un meilleur contrôle des usages et des usagers ?
- **Améliorer les services actuels et identifier de nouveaux services** de sorte à offrir des fonctionnalités adaptées qui répondraient au mieux aux besoins des différents publics cibles. Ce point est fondamental pour que BNEUF évolue vers un seuil d'innovation conforme aux nouvelles techniques des systèmes d'information documentaires actuels ;
- **Définir un périmètre fonctionnel et une feuille de route** de déclinaison opérationnelle des services à fournir à court, moyen et long terme. Cette proposition est liée à celle relative à une stratégie d'étape pour BNEUF sur le court, moyen et long terme ;
- **Définir une architecture organisationnelle de gestion et de suivi** du projet avec un mode de gouvernance standard/adapté qui impliquerait toutes les parties prenantes, y compris les représentants des usagers. Cette suggestion englobe une série de mesures liées à la création d'une communauté de pratique organisée autour d'un organe de contrôle, de pilotage et de promotion du portail, ci-haut identifié comme un « comité éthique et organisationnel de gestion de la BNEUF ». A cette communauté seront associés dans l'ordre (dixit rapport DN) :
  - Les établissements pour s'assurer de la bonne adoption de la BNEUF et être, de ce fait, des ambassadeurs de la plateforme ;
  - Les fournisseurs de ressources [publiques et payantes] en tant que partenaires industriels privilégiés ;
  - Les enseignants-chercheurs en tant que contributeurs et prescripteurs ;
  - Les étudiants dans la conception et les retours d'expériences sur les services ;
- **Trouver un modèle économique souple et viable** qui permettrait de financer d'abord l'accès aux ressources payantes, mais aussi compenser les efforts de contribution des établissements partenaires et des enseignants-chercheurs producteurs de ressources. Une modalité de « tiers-payeurs » à l'image des processus de la publication dorée des archives ouvertes pourrait être envisagée. La voie dorée des publications en libre accès consiste à publier chez un éditeur commercial une ressource numérique (livre ou article de revue) en Open Access après une période négociée d'embargo et moyennant des frais de publications (APC ou BPC) qui peuvent être pris en charge par l'AUF, l'établissement de tutelle ou par un fonds d'aide à la publication ;

Au vu des résultats de ce diagnostic interne, la section suivante tentera de proposer une lecture approfondie des formes de dysfonctionnement et proposer des pistes complémentaires de remédiation autour d'une nouvelle stratégie pour une BNEUF 2.0.

# 3 DIAGNOSTIC EXTERNE DE L'ÉCOSYSTÈME BNEUF

Ce diagnostic externe (un audit ou une analyse systémique) couvrira les trois modules du dispositif BNEUF : « Ressources numériques », « Atlas de l'expertise » et « Réseau social » et les articulations qui les lient. Il permettra notamment de faire l'analyse de l'ensemble des fonctions, modes, paramètres et options à maintenir, à ajouter ou à retravailler dans la perspective d'une version améliorée du système d'information BNEUF. Ces paramètres seront centrés sur les usages, l'efficience et la valeur ajoutée des services plus que les couches techniques et informatiques du système. Ils concerneront des aspects opérationnels et ergonomiques comme la découvrabilité et la livraison des ressources ainsi que la gestion des ressources humaines à travers un Atlas d'experts et l'animation communautaire via un réseau social francophone.

## 3.1 Interface d'accueil du dispositif

Le premier contact avec la BNEUF est sa page d'accueil développée de manière sobre et ergonomique dans un agencement harmonieux de couleurs entre noir et blanc et une répartition spatiale allégée et équilibrée.

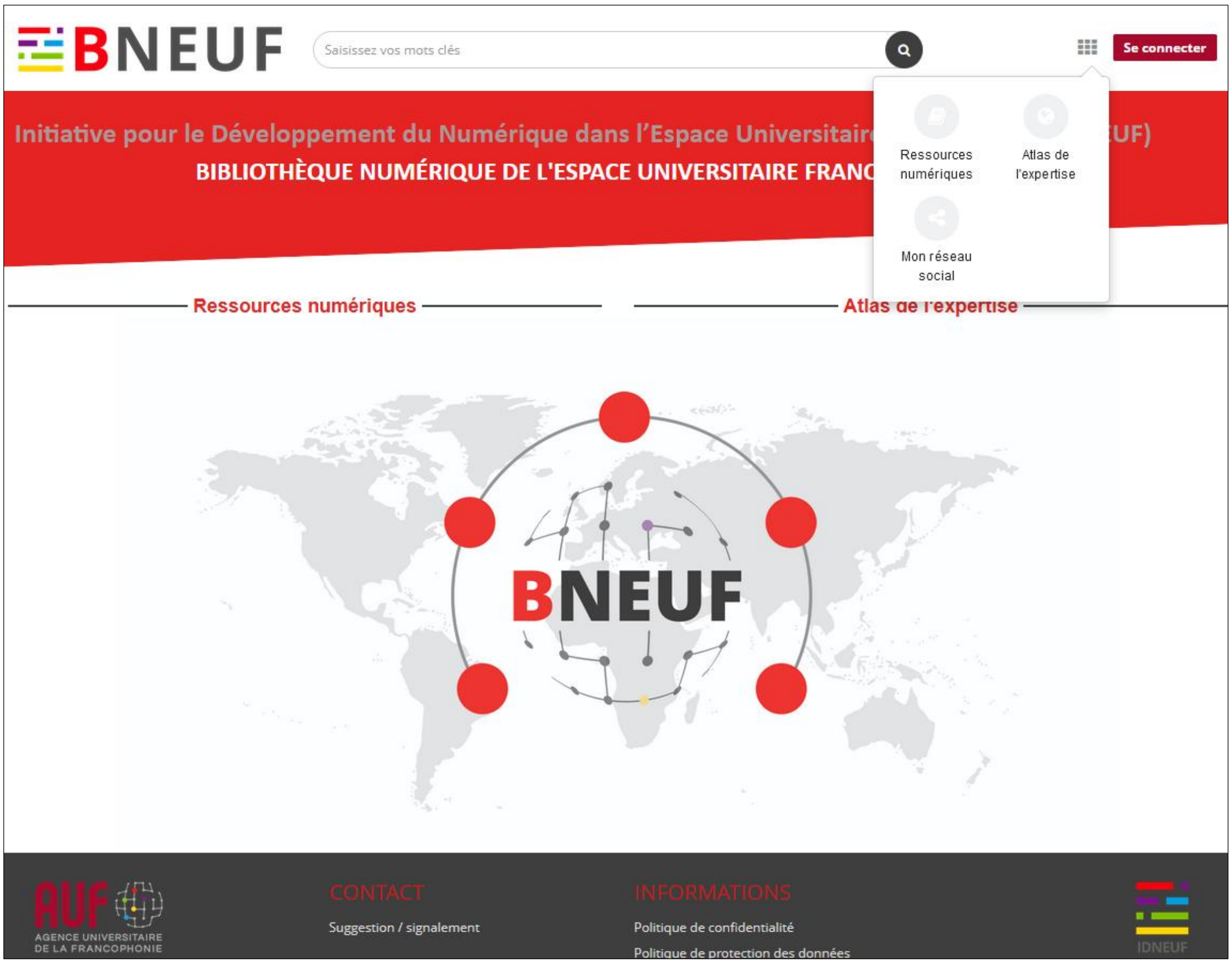

L'interface d'accueil pourrait cependant avoir quelques réadaptations d'un point de vue de son design et de ses fonctionnalités. Sur le plan ergonomique les points suivants sont à considérer :

- Un logo BNEUF donne clairement l'identité du produit alors que le logo de l'AUF qui informe sur le commanditaire ou la « tutelle » institutionnelle du produit est très peu visible, mis à l'écart en marge droite et en pied de page sur un fond noir (contraste) ;

- A propos des logos, il faudrait aussi revoir si l'articulation AUF / IDNEUF / BNEUF est toujours d'actualité d'autant que le logo IDNEUF en bas de page ne pointe plus sur un site « IDNEUF » quelconque. D'ailleurs une recherche par « IDNEUF » sur Google pointe directement sur BNEUF, ce qui accentue l'ambiguïté d'une part entre les trois entités AUF/IDNEUF/BNEUF et d'autre part entre un logo IDNEUF qui ne pointe vers rien et un intitulé BNEUF en haut de page qui n'a plus une raison d'être ;

- Sur les trois principaux services, à savoir les Ressources numériques, l'Atlas de l'expertise et le Réseau social, seuls les deux premiers sont proposés en clair sur la page d'accueil alors que le réseau social n'est détecté qu'en cliquant sur le bouton damier en haut à droite de l'écran. Cette approche relègue le réseau social à un deuxième niveau d'importance dans le dispositif ;

- Au centre de l'écran, la sphère BNEUF avec les 5 points rouges et la lettre B en rouge n'est plus d'actualité. Elles symbolisent l'identité du module de la bibliothèque numérique au sein de l'écosystème IDNEUF mis en place par l'AUF dans sa stratégie précédente (2017-2021) autour des 4 modules « Ateliers du Numérique », « Bibliothèque Numérique », « Campus Numériques » et « Développements Numériques ». Cet écosystème n'existe plus. La symbolique du nombre et de la couleur de l'initiale « B » ne tient donc plus. L'acronyme BNEUF est-il d'ailleurs encore utile sachant que la bibliothèque numérique constitue l'un des 3 modules qui composent un système intégré incluant aussi un Atlas et un réseau social ?

Une réadaptation peut prendre forme d'une réduction des bulles rouges de cinq à trois et d'en faire des points d'accès vers les 3 modules de ce système intégré. Les deux rubriques « Ressources numériques » et « Atlas de l'expertise » et du bouton damier à gauche du bouton « se connecter » peuvent être gardés comme points d'accès supplémentaires.

Du point de vue des fonctionnalités, les deux rubriques « Ressources numériques » et « Atlas de l'expertise » ne dévoilent leur importance qu'après l'accomplissement d'une recherche. Elles ont un double rôle d'accès aux modules concernés et d'affichage des statistiques de recherche pour les ressources et les experts.

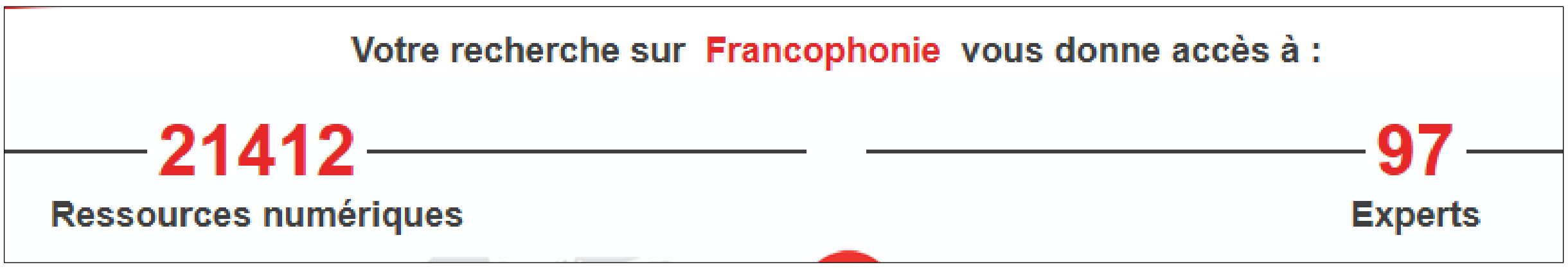

La page d'accueil propose aussi un module transversal « Moteur de recherche » standard qui sera analysé plus loin.

La page d'accueil propose aussi un bouton de connexion pour des abonnées au dispositif. La différence entre un statut d'abonné et un accès anonyme aux trois modules sera aussi pris en compte durant l'analyse du système.

La BNEUF nécessite un tutoriel d'utilisation qui devrait être disponible dès l'arrivée sur la page d'accueil. Ce tutoriel peut prendre forme d'un lien vers une page Web, un prospectus PDF ou des vidéos de vulgarisation. Une rubrique adaptée (tutoriel ; mode d'utilisation, etc.) peut prendre place en bas de page de l'accueil sous la rubrique « Informations ».

## 3.2 Le module « Ressources numériques »

L’analyse du module documentaire « ressources numériques » peut être faite sur deux niveaux : celui de la collection globale et de ses modes d’organisation, puis celui de l’unité documentaire (fiche ressource) dans ses aspects structurels, ergonomiques et accessibilité.

### *3.2.1 Analyse globale de la collection*

À la fin de l’année 2022, la collection recense plus de dix-sept millions de références, ce qui constitue un réservoir de données colossal qui nécessite des études d’impacts par une remontée statistique d’usage et de satisfaction. La collection nécessite aussi une page d’introduction pour expliquer la nature de ses contenus, ses objectifs, son mode opérationnel, ses traits d’originalité, sa distinction des dispositifs, etc. Bref, une page marketing et de promotion de ses services dans un but de fidélisation des usagers.

Il faut rappeler que la BNEUF est actuellement un « fournisseur de services » plutôt qu’un « fournisseur de données ». Elle n’est pas dépositaire de documents primaires et fonctionne avec des métadonnées issues quasi exclusivement d’un moissonnage OAI-PMH dans 34 catalogues de ressources. La BNEUF ne peut ainsi prétendre à un contrôle sur la qualité des données moissonnées et leur harmonisation qu’à condition d’appliquer au niveau de ses propres mécanismes (i.e. BNEUF *Harvester*) un traitement supplémentaire pour reformatter, reconvertir, réordonner et reclasser les données avant de les intégrer dans ses propres bases pour les livrer de façon harmonieuse et adaptée aux besoins de ses utilisateurs.

La collection BNEUF aurait besoin d’un traitement en back-office avant d’être proposée à la consultation. Deux parmi les premiers problèmes identifiés au niveau de la collection sont les doublons et les déchets documentaires :

❖ ***Les doublons***

Plusieurs ressources dans BNEUF sont répertoriées avec des métadonnées en double (voire plus) mais qui pointent vers la même source :

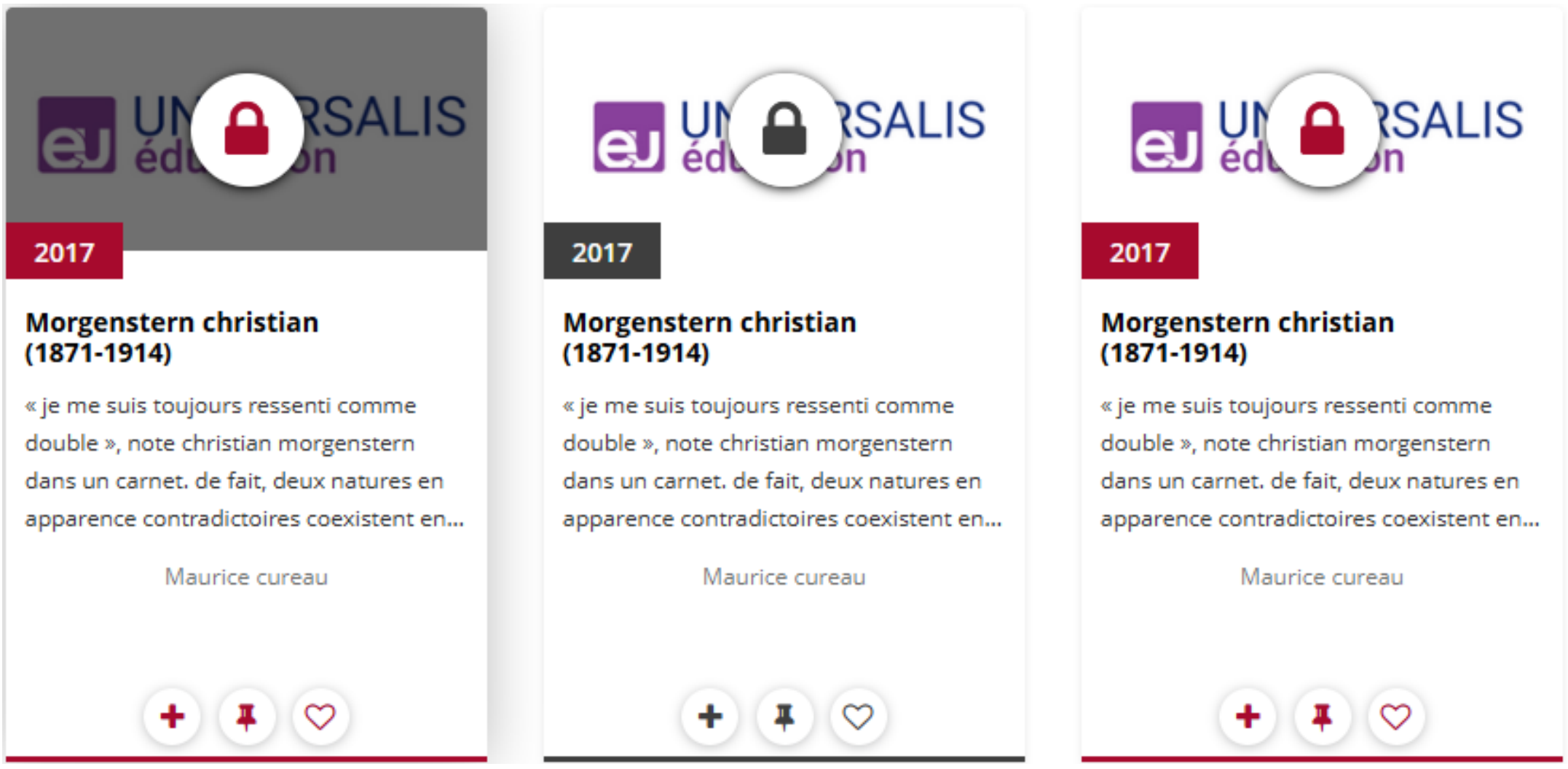

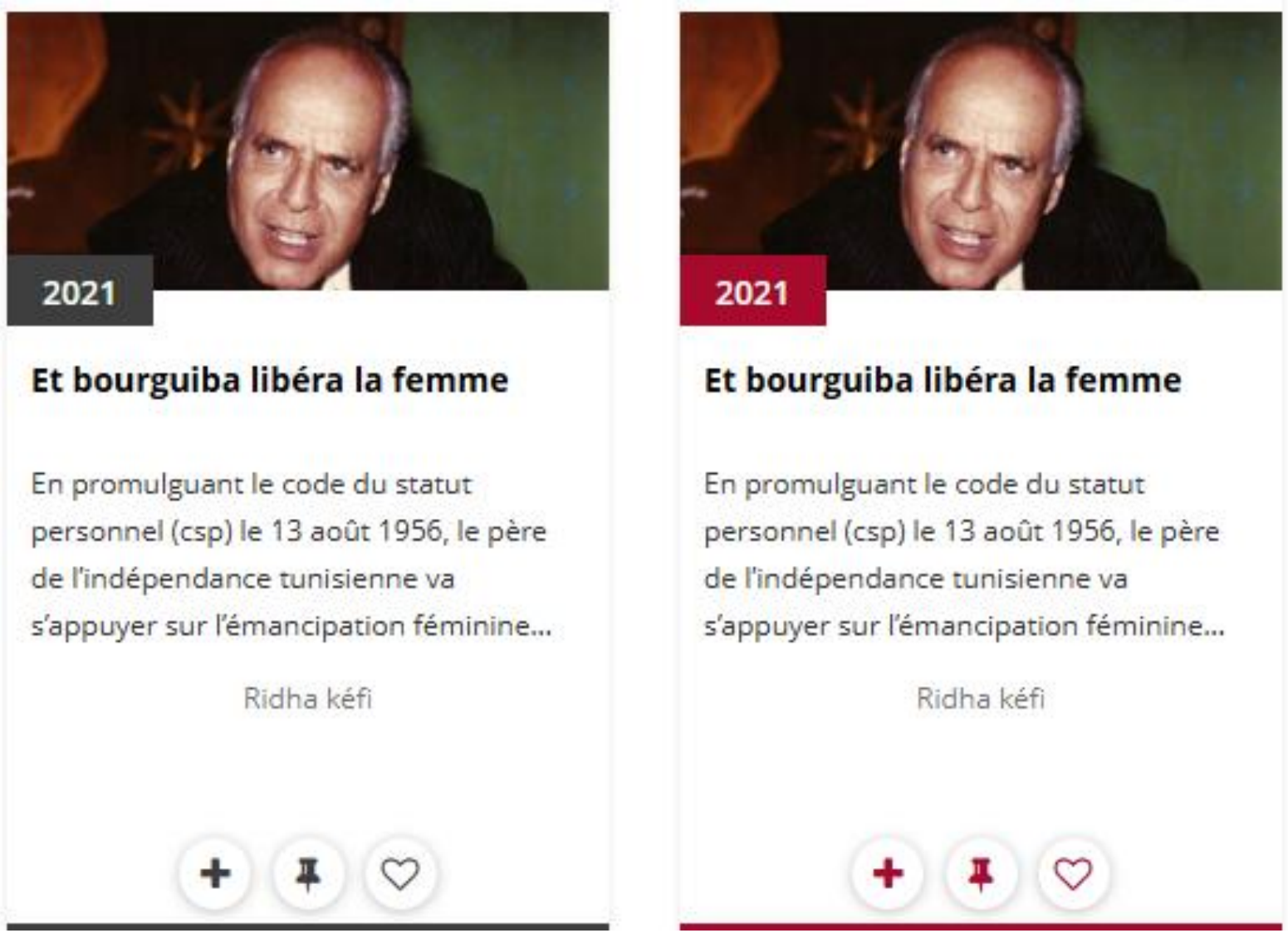

❖ ***Les déchets documentaires***

Plusieurs ressources sont mal indexées, voire vides, de contenu. Une opération de nettoyage au niveau des métadonnées moissonnées est indispensable pour « épurer » la base BNEUF de ces déchets documentaires à l'image de l'exemple suivant :

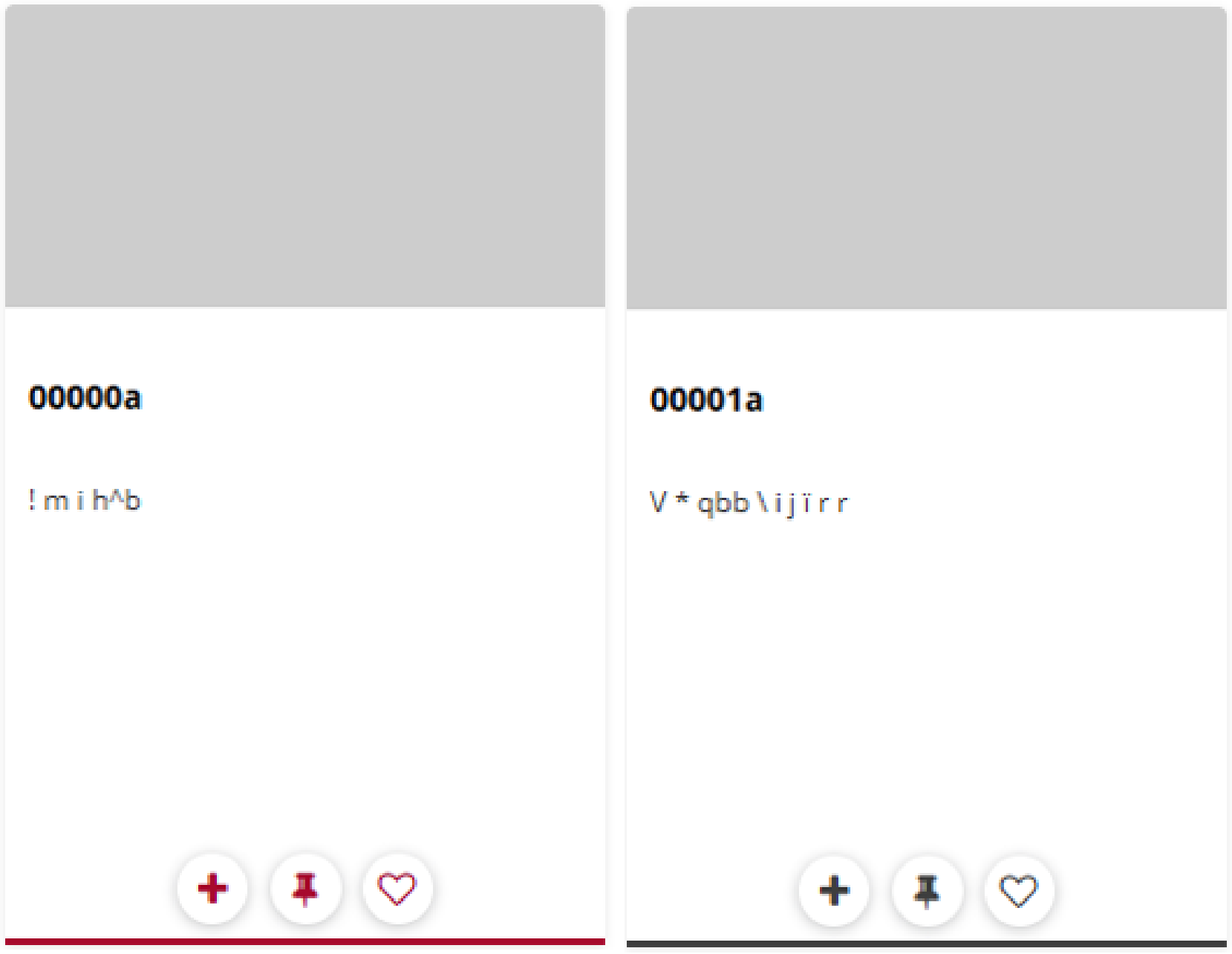

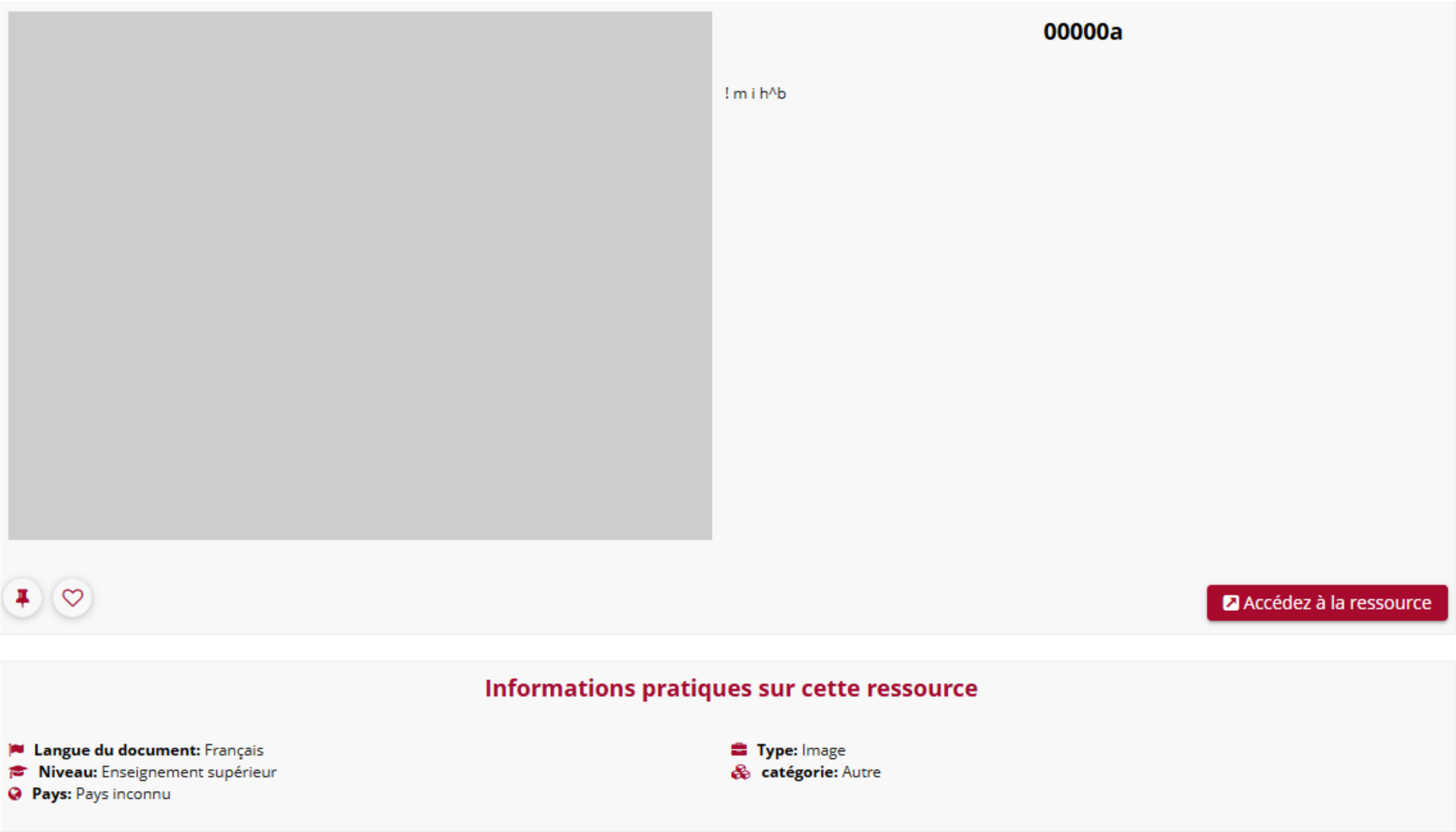

D'autres types de ressources sont incomplètes malgré l'existence de contenus. Un tri alphabétique permet d'avoir des ressources sans titres ni résumés bien que ces informations existent dans les documents sources.

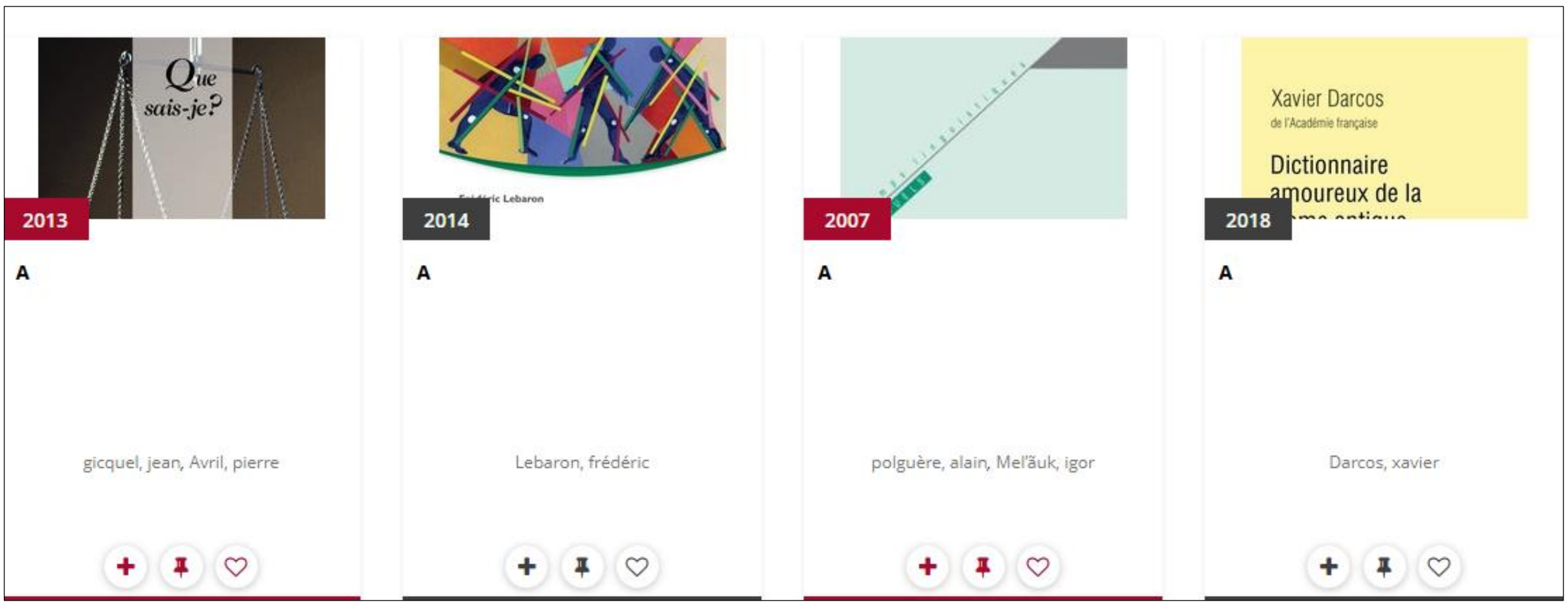

Ces problèmes techniques sont tributaires en grande partie de la variété des catalogues sources qui utilisent des modalités différentes de traitement et dont l'impact se traduit dans la BNEUF sous ces formes d'incohérence. Il est nécessaire, en termes de démarche qualité, d'agir au niveau du module BNEUF *Harvester* pour une meilleur contrôle/filtrage des métadonnées au moment et après le moissonnage JSON, XML et DC.

### *3.2.2 Analyse des fonctions de l'interface BNEUF*

Le point fort de l'interface BNEUF est dans la disposition des ressources en mode mosaïque et le menus de filtrage multicritères des ressources. L'interface de la BNEUF présente toutefois des anomalies et des redondances au niveau des modes d'accès (recherche et navigation) qu'il serait important de souligner.

La BNEUF dispose de deux modes de filtrage et d'un moteur de recherche comme moyen d'accès aux ressources :

— Un menu latéral gauche pour filtrer les ressources (recherche multicritères) ;
— Une liste déroulante à droite pour choisir un ordre d'affichage des ressources (critères de tri) ;
— Un module de recherche transversal hérité du système ORI-OAI ;

Ces différents modes de filtrage et de recherche nécessitent quelques améliorations.

❖ ***Le menu des filtres de recherche (zone de gauche)***

Plusieurs remarques d'ordre général sont à faire avant d'entrer dans les détails des options des filtres :

1) *Première remarque – Redondance des filtres* : Le choix des filtres est normalement une projection de la nature et du format des métadonnées moissonnées. Un filtre existe parce qu'il y a une métadonnée qui lui correspond. La question est donc : Pourquoi ces filtres et pourquoi cet ordre ? Une étude de pertinence via les usages a-t-elle eu lieu ? Une telle étude permettrait de connaître les meilleurs filtres à employer, selon les tendances d'usage, et l'ordre dans lequel ils sont disposés. Des redondances sont observées entre les filtres proposés ;

2) *Deuxième remarque – recherche à variable unique* : La méthode de recherche proposée par filtrages néglige un principe de base dans les stratégies de recherche d'information : la recherche combinatoire. Si le mode de filtrage en cours permet une recherche multicritère c.à.d. la combinaison de plusieurs critères pour construire une équation de recherche (i.e. chercher par thème, par date et par langue), ce mode de filtrage ne permet pas de combiner deux variables d'un même filtre. Un utilisateur ne peut pas chercher, par exemple, des ressources provenant de deux catalogues différents, ou publiées en deux années différentes, ou en deux formats différents, etc. Pourtant, la solution est techniquement possible : afficher les options des filtres sous forme d'une liste à choix multiples (case à cocher). L'utilisateur peut alors choisir une ou plusieurs variables d'un même filtre qu'il pourra combiner à une ou plusieurs variables d'un autre filtre ;

3) *Troisième remarque – filtre auteur* : Tous les systèmes documentaires proposent des recherches par auteur. Or, les filtres proposés dans BNEUF ne proposent pas ce champ stratégique. Or, toute ressource, de n'importe quel type, est logiquement créée par une entité physique ou morale qui en détient les droits de propriété intellectuelle. Un utilisateur peut toujours chercher les publications d'un auteur ou d'un organisme référent. En l'absence d'une entité « auteur », les systèmes d'information utilisent souvent l'expression « Anonyme » ;

4) *Quatrième remarque – tri alphabétique* : Si les options sous chacun des filtres sont listées avec le nombre de leurs occurrences dans la collection (les plus nombreux sont placés en premier), il est plus judicieux de procéder par un tri alphabétique pour faciliter la localisation d'un filtre particulier dans une longue liste de filtres. Une permutation entre un ordre alphabétique et un ordre statistique peut aussi être envisagée, mais le tri alphabétique reste généralement le mieux adapté pour les utilisateurs dans les opérations de recherches ;

Chaque filtre du menu latéral gauche présente aussi des anomalies qu'il faudra traiter :

- « **Discipline** » : l'ordre proposé ne correspond pas à un plan de classification standard des sciences comme la CDD (Classification Décimale de Dewey) ou la CDU (Classification Décimale Universelle). L'importance de ce point sera démontrée dans d'autres lieux de ce rapport ;

- « **Niveau** » : la liste des options comporte beaucoup d'ambigüité et de redondance : exemples « enseignement en ligne » n'est pas un niveau, « CEGEP » est exprimé en plusieurs formats. Ceci confirme la divergence des modes d'indexation utilisés par les différents catalogues de ressources moissonnées, mais il indique surtout l'absence d'un traitement post-moissonnage au sein de la BNEUF pour le nettoyage et l'épuration ;

- « **Format** » : ce filtre présente les mêmes signes de redondance et d'imprécision (deux exemples de variables inappropriées : « publication », « objet »). Un filtrage par une technique de mappage (mise en correspondance avec des termes contrôlés) est indispensable pour maîtriser la diversité des nomenclatures employées dans les catalogues des sources avec une liste BNEUF homogénéisée ;

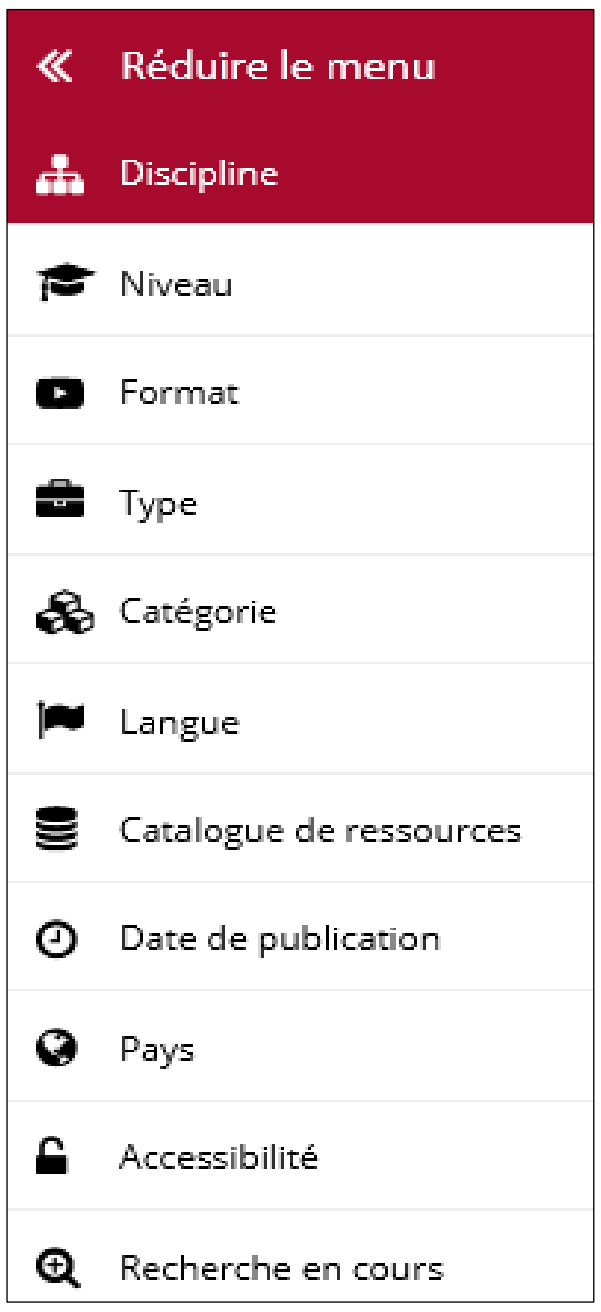

- « **Type** » : ce filtre présente un double problème, d'abord dans la redondance entre les options du même filtre comme « musique » et « partition », « livre » et « ouvrage », etc., puis dans les double-emplois avec le filtre « Format ». Par exemple, les options « texte », « image » et « son » existent sous les deux filtres ;

- « **Catégorie** » : ce filtre présente également un double problème, d'abord pour la redondance entre des options similaires exprimées en deux langues (français et anglais), puis pour la redondance avec le filtre « Discipline ». Un filtrage via un plan de classification détaillé (hiérarchique) serait beaucoup plus pertinent qu'une liste redondante et non alphabétique ;

- « **Langue** » : sous ce filtre, l'option « autres langues » affiche un nombre plus important que l'anglais. Il serait utile de faire une analyse de fréquence plus poussée des langues les plus présentes dans ce lot et de les ajouter comme filtres linguistiques supplémentaires puisque ces langues indiquent l'existence d'un corpus multilingue important dans la BNEUF comme l'espagnol qui occupe une place non négligeable. Ce passage sou silence de certaines langues comme l'espagnol peut fort probablement résulter d'une négligence dans l'indexation du champ « langue » dans les catalogues des ressources moissonnées ;

- « **Catalogue de ressources** » : vue la longueur de la liste, autant ajouter à ce filtre une fonction de tri alphabétique pour faciliter le repérage des catalogues sources. Ce filtre nécessite un traitement en profondeur lié au choix des Catalogues retenus pour le moissonnage. Plusieurs peuvent contenir les mêmes ressources, ce qui génère les doublons en cas où un contrôle de dédoublement n'est pas mis en place pour supprimer les métadonnées en double ou les fusionner ;

- « **Date de publication** » : ce filtre présente quelques ambiguïtés dans les énoncés. L'expression « antérieur à 2019 » suppose que le filtrage chronologique s'arrête à 2018, ce qui est le cas. En revanche, le filtre « entre 2019 et 2022 » prête à confusion car il peut signifier que les deux années sont inclusives, c.à.d. que le filtre couvre les ressources publiées de janvier 2019 jusqu'à décembre 2022. Or, la limite de ce filtre s'arrête à décembre 2021. L'expression « De 2019 à 2021 » est plus adaptée, d'autant que l'ordre annuel suivant reprend avec l'année 2022 puis 2023. Il est d'ailleurs recommandé de réordonner la période entre 2019 et 2021 par année pour permettre des filtrages chronologiques plus fins ;

- « **Pays** » : le filtre pays pose lui-aussi un double problème. D'abord il nécessite une option de tri alphabétique au cas où un utilisateur voudrait chercher par pays qu'il aura du mal à trouver dans une longue liste non triée. Ensuite il est mal exploité si l'on juge par le volume de l'option « Autre » qui dépasse de loin celui des autres pays listés. Ceci explique qu'il y a beaucoup de pays non identifiés par l'indexation du champ « Pays » dans les catalogues des ressources ;

- « **Accessibilité** » : les options de ce filtre sont très arbitraires et redondantes : par exemple, l'option « notice bibliographique » ne peut pas être considérée comme un critère d'accessibilité ; l'option « Accès ouvert » est affichée en double sous ce filtre et chacune donne des résultats différents ; l'option « Accès retardé » n'est pas un concept clair ; l'option « Via abonnement » restitue uniquement des ressources d'accès restreint de l'encyclopédie *Universalis* alors que cette exclusivité pourrait être intégrée à l'option « accès fermé ». Entre « Accès fermé », « Via abonnement » et « Restrict », il y a probablement des nuances qui ne sont pas évidents pour les utilisateurs. Ces options de filtrage nécessitent un nouveau paramétrage, voire une fusion, en utilisant des concepts standards de l'accès aux ressources ;
- **« Recherche en cours » :** C'est le dernier item du menu de filtrage qui affiche une équation de recherche combinée comme construite par l'utilisateur. Cette fonction présente à son tour des anomalies importantes :
  - L'équation de recherche est remise à zéro dès que l'on quitte la page d'affichage des ressources, par exemple en cliquant par inadvertance sur le logo BNEUF ou en faisant un retour arrière dans la barre de menu du navigateur. Il n'y pas moyen de garder en mémoire une combinaison de filtrage en construction ;
  - Il n'y pas non plus de fonction d'enregistrement d'une équation de recherche dans l'historique de recherche d'un utilisateur inscrit ou dans la session d'un utilisateur anonyme ;
  - Le module Recherche ne dispose pas d'une fonction MAJEURE d'export des résultats :

> La fonction « Export des résultats » dans des formats bibliographiques standardisés est extrêmement importante pour constituer des bibliographies personnalisées. Les formats d'exports les plus courants sont BibTex, TEI, Dublin Core, DCTerms, EndNote, Datacite, APA, MLA, etc. Ces formats bibliographiques ont le potentiel d'être interopérables avec la majorité des systèmes de gestion bibliographique et seraient d'un apport considérable pour les enseignants-chercheurs afin de constituer leurs dossiers pédagogiques ou de recherche. Cette fonction n'existe nulle part ailleurs dans BNEUF.

***Le Menu des critères d'affichage (liste déroulante à droite)***

Pour la liste déroulante de droite (ordre d'affichage) les options « Plus récents », « Plus pertinents », « Ordre alphabétique », « Mieux notées », « Plus consultées », l'utilisateur devrait pouvoir les combiner pour mieux lister les ressources filtrées. Par exemple, un utilisateur peut vouloir lister les ressources « plus récentes », les « plus pertinentes » et les « mieux notées » dans un « ordre alphabétique » (NB : penser à harmoniser les accords féminin ou masculin des adjectifs de cette liste d'options). Sans tri multicritères, il ne pourra pas le faire. C'est pareil pour les options des filtres du menu de gauche, il serait plus simple de cocher en une seule fois plusieurs options d'un même filtre que de le refaire plusieurs fois.

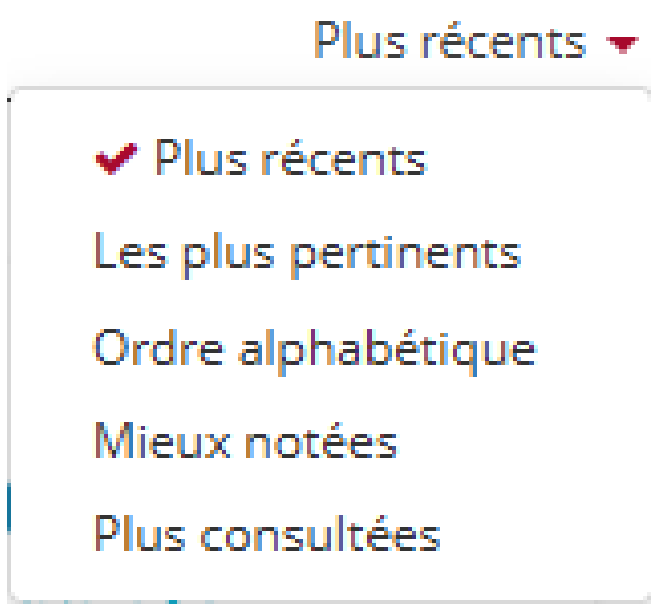

### ❖ *Le moteur de recherche*

À l'instar des CMS du marché (Drupal, Wordpress ou Joomla), BNEUF propose un moteur basique, n'offrant aucune possibilité de recherche avancée (combinatoire, sous-ensembles, etc.) ni de syntaxe de filtrage évolué (troncature, expression exacte en langage naturel, etc.).

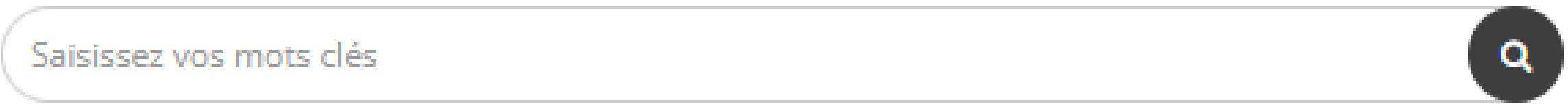

Il est pourtant possible d'améliorer cette fonction basique en lui ajoutant des fonctionnalités de recherches avancées.

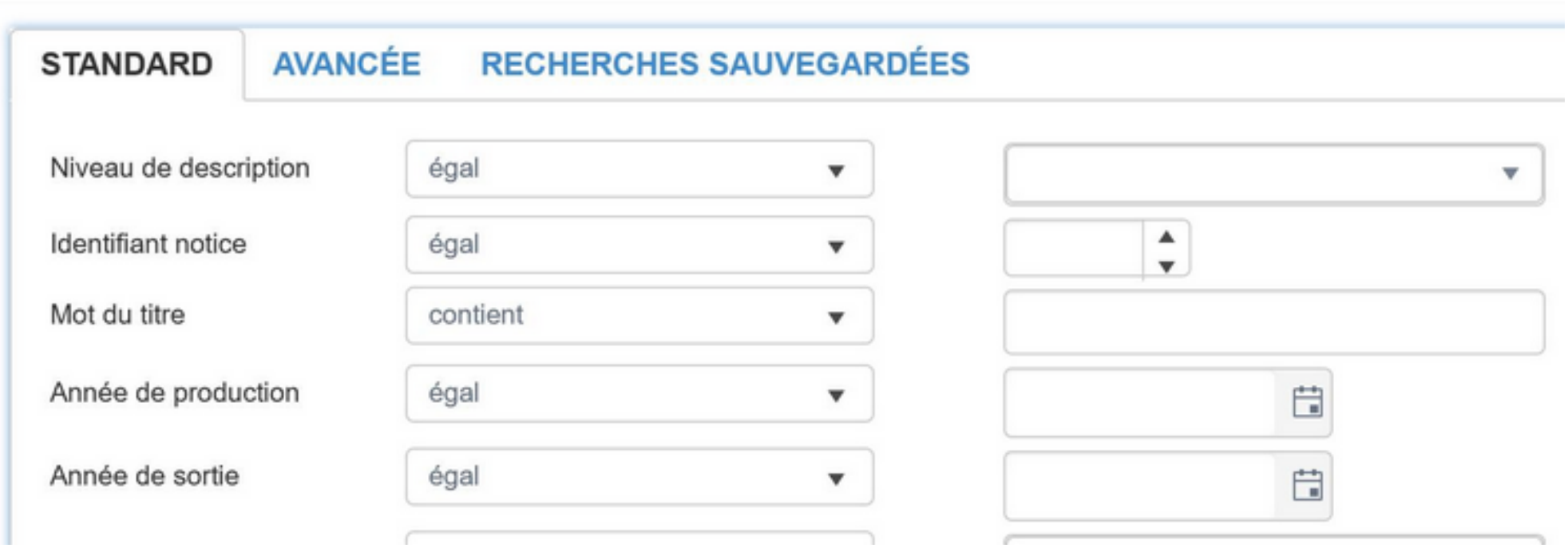

L'ajout d'extensions/modules ou de solutions logicielles spécifiques permettrait de proposer une fonction de recherche plus riche. Trois types d'alternatives peuvent être étudiées : les solutions d'éditeurs comme Exalead ou Oracle Endeca, les solutions SaaS comme Algolia ou Amazon CloudSearch, les solutions Open Source comme ElasticSearch ou Solr.

## *3.2.3 Analyse de l'entité « ressource »*

Avant d'atteindre le contenu intégral des ressources, celles-ci sont présentées en mode mosaïque dans un format de vignettes interactives générées par DBNEUF *Harvester*.

### ❖ *La vignette ressource*

Dans son format optimal, une vignette est composée d'une imagette d'entête, l'année de publication, le titre de la ressource, un bref résumé, un nom d'auteur, un score étoilé, trois boutons pour activer les trois fonctions « plus de détails », « ajouter à un classeur » et « ajouter aux favoris ». Une fonction « Partager » est ajoutée pour les inscrits.

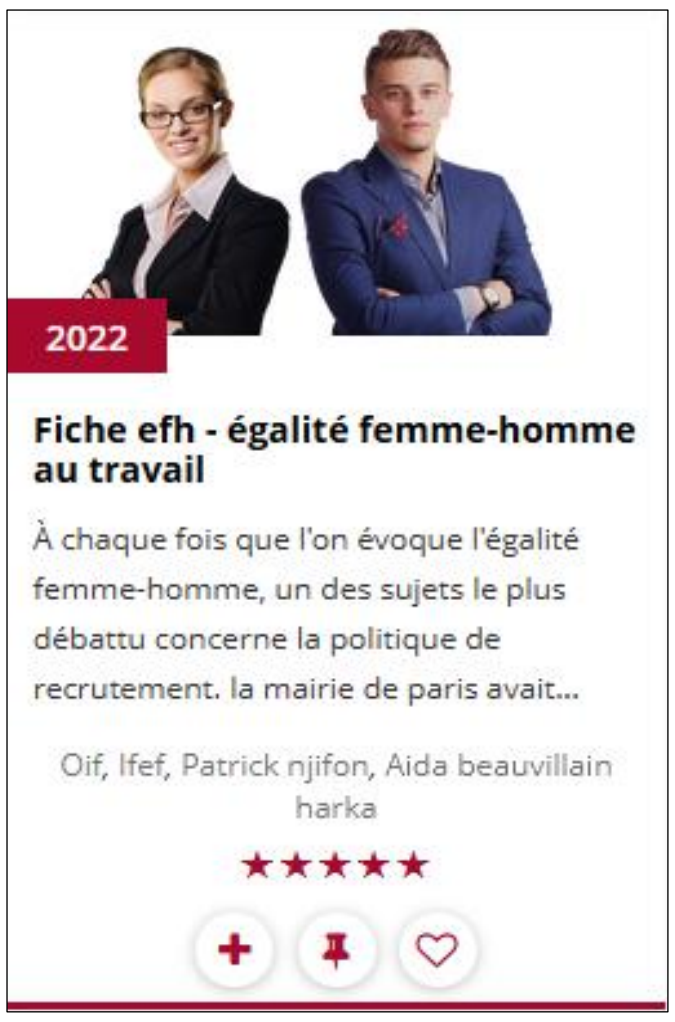

Générées par un module « vérificateur de données et générateur d'images » dans le BNEUF *Harvester* en back office, les vignettes offrent un accès convivial aux ressources comparé à d'autres modes d'affichage, comme le mode listing ou le mode annuaire. Une perspective d'affichage cartographique (avec des liens sémantiques) peut aussi être envisagée si une dimension sémantique (gestion des connaissances par l'usage d'ontologies/thésaurus) serait ajoutée au dispositif. Ce point sera plus détaillé sous Partie II-2.3.

Cependant, la structure en vignettes des ressources, bien qu'elle soit retenue comme standard d'affichage, présentent quelques anomalies. Certains éléments d'information dans les vignettes sont redondants et irréguliers :

- L'identité du catalogue de ressources est souvent utilisée en guise d'auteur de contenu alors qu'un auteur personne physique est indiqué dans la ressource elle-même ;
- L'information auteur disparait complètement des vignettes des ressources de 2023 exclusivement puisées dans le portail FUN alors que le nom du Catalogue source prend la place de l'auteur dans les ressources des autres années. En outre, dans le portail FUN, les ressources ont souvent des auteurs présentés comme « Équipe pédagogique » ;

Une autre incohérence concerne les trois formes de restrictions d'accès (« Accès fermé », « Via abonnement » et « Restrict »). Les trois options du filtre Accessibilité produisent le même effet : elles ne permettent l'accès à la ressource intégrale que sur inscription ou abonnement. Or, contrairement aux restrictions « Via abonnement » et « Restrict » qui sont signalées par un cadenas, « accès fermé » ne l'est pas.

Toutes les anomalies peuvent être justifiées par la différence dans les stratégies d'indexation dans les catalogues de ressources. Elles doivent néanmoins être traitées dans le dispositif de la BNEUF au moment ou après leur moissonnage.

❖ ***Au-delà de la vignette***

Les vignettes des ressources donnent accès à 3 types d'extensions :

1. Une fiche publique de métadonnées plus complète, accessible via le bouton « Plus de détails ». Elle constitue un passage obligé pour arriver jusqu'au contenu intégral de la ressource ;
2. Un classeur individuel dans l'espace abonné pour classer les ressources sélectionnées (bouton « Ajouter à un classeur ») ;
3. Une liste de favoris dans l'espace abonné via le bouton « Ajouter aux favoris » ;

❖ ***La fiche publique de métadonnées***

Elle est censée fournir des « informations pratiques » sur la ressource. Or cette fiche présente à son tour des incohérences dans le choix des éléments de métadonnées et leurs contenus :

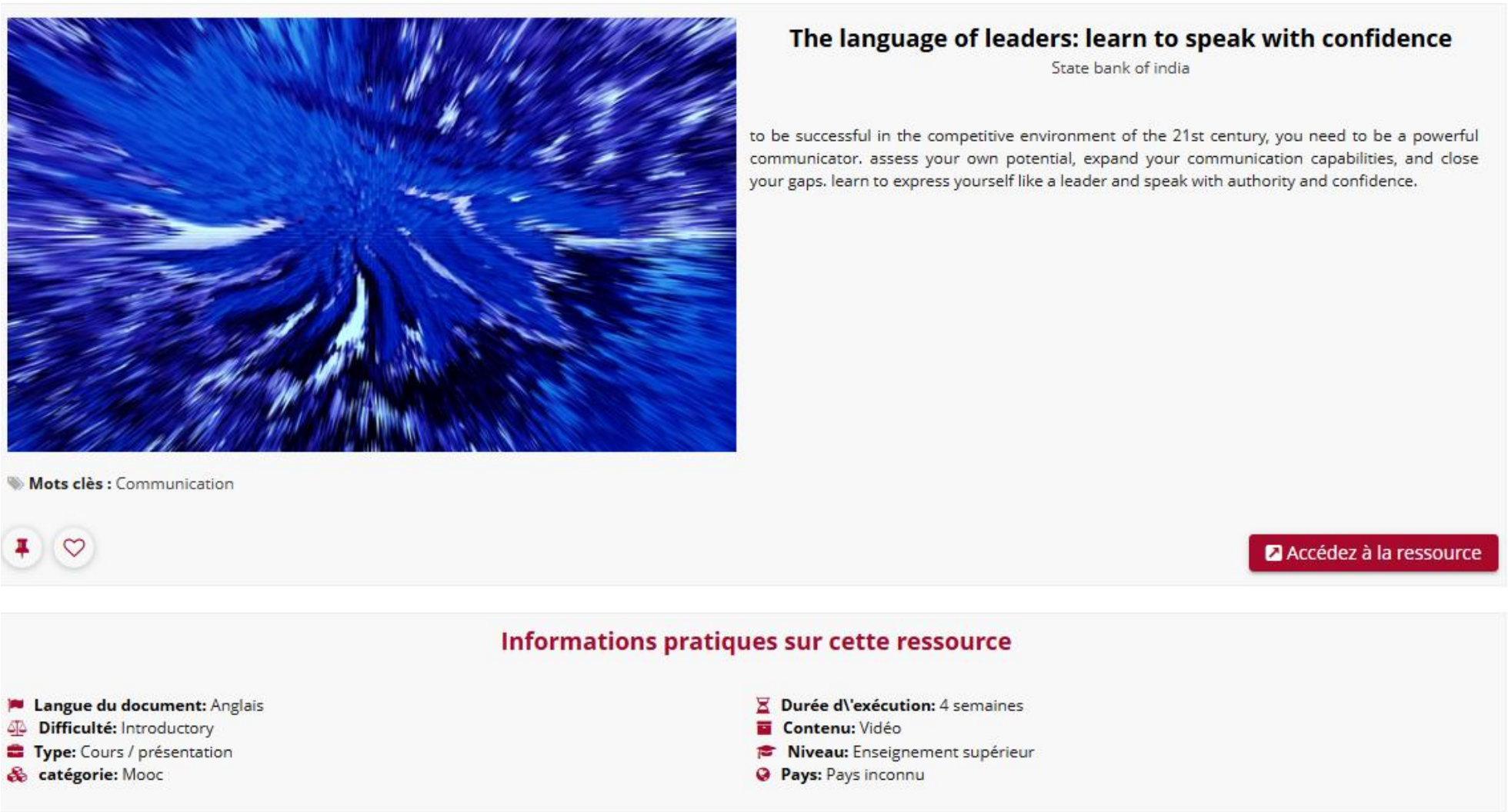

- **Les éléments de métadonnées**

L'observation de différentes fiches de métadonnées laisse envisager l'emploi d'un profil d'application composé des 13 champs suivants : « Titre », « Auteur », « Résumé », « Mots-clés », « Droits d'auteurs », « Langue du document », « Difficultés », « Type », « Catégorie », « Durée d'exécution », « Contenu », « Niveau », « Pays ». Ces 13 éléments de métadonnées sont optionnels et ne renvoient du contenu dans une fiche ou dans une vignette que s'ils contiennent une valeur dans le catalogue source.

La question est de savoir si ces 13 éléments de métadonnées sont communs aux 34 catalogues des ressources moissonnées, ce qui est peu probable. C'est là où un schéma pivot ou un profil d'application de métadonnées propre à la BNEUF s'avère nécessaire pour résoudre certaines formes d'incohérence dans le module documentaire dont voici quelques-unes :

- Les métadonnées en dessous de l'image vignette sont irrégulières et varient selon la présence ou non d'une valeur de métadonnée dans le catalogue source. L'affiche suit le principe qu'un élément de métadonnées vide ne s'affiche pas ;

- L'élément de métadonnée « auteur » sous le titre est parfois absent. Il est souvent remplacé par le nom du catalogue de ressources bien qu'une entité « auteur » existe parfois dans la ressource complète et qu'on peut vérifier en accédant au contenu entier de la ressource par le bouton « Accéder à la ressources ». Il sera probablement compliqué et onéreux pour la BNEUF d'extraire cette information d'un document source si celle-ci n'est pas moissonnée en tant que métadonnée dans un champ spécifique ;

- La dimension de l'image vignette dans la fiche complète des métadonnées n'est pas standardisée. Elle varie d'un format paysage à un format portrait dans des dimensions irrégulières. Puisqu'il s'agit d'une extraction BNEUF à partir du document source, cette zone graphique pourrait avoir des dimensions fixes contrôlées par le module « vérificateur de document et générateur d'image » du BNEUF Harvester ;

- **Les contenus (valeurs) des métadonnées**

Les valeurs des métadonnées (contenus) obéissent à plusieurs règles de forme auxquelles dérogent plusieurs ressources dans la BNEUF. À titre d'exemple :

- **« Mots-clés »** : un mots clés sert essentiellement de pointeur d'accès à une ressource pendant la recherche. Il doit donc avoir une charge sémantique forte pour qu'un utilisateur puisse l'employer comme un terme de recherche. Il ne doit pas être réduit à un terme « valise » ou un terme banalisé qu'aucun utilisateur n'emploiera dans une équation de recherche. C'est le cas de mots-clés du genre « Communication », « Travail », « enseignement », « physique et chimie ». Quoique provenant de l'indexation source, ces mots génériques produisent ce qui est appelé un effet de « Bruit » dans les résultats de recherche, c.à.d. restituer des ressources inappropriées pour la recherche en cours. Par soucis de pertinence, l'épuration du produit de moissonnage dans la BNEUF doit filtrer ces « mots-vides » et les évacuer ;

- **« Difficulté »** : assez rarement indiqué comme élément de métadonnée, ce paramètre est parfois exprimé dans une forme atypique comme « Introductory / Intermediate / Advanced ». Or, la difficulté est très souvent exprimée par des termes plus explicites comme « Facile », « difficile »,

« très difficile » ou « Complexe », etc. Malgré la grande divergence des nomenclatures utilisées dans les catalogues des ressources, un mappage (correspondance) BNEUF doit gérer ces variations pour les faire converger vers une taxonomie unifiée comme celle utilisée dans le formulaire Ma Bneuf : « Facile / Moyen / Difficile » (cf. 3.5.4) ;

- **« Droits d'auteur »** : cet élément est sensible et fondamental. Pourtant, il présente des formes redondantes, contradictoires et non significatives comme dans les exemples suivants :

> © **Droits d'auteur :** . gratuit

> © **Droits d'auteur :** . cairn

> © **Droits d'auteur :** libre de droits , gratuit . droits réservés à l'éditeur et aux auteurs. © craham - centre michel de boüard - umr 6273 unicaen/cnrs

L'élément « droit d'auteur » doit faire l'objet d'une plus grande rigueur dans le choix des licences et la nature des droits. Des propositions sont données sous Partie II-2.5.

❖ ***Le classeur individuel et la liste des favoris***

Deux des boutons interactifs de la fiche de métadonnées ajoutent la ressource respectivement au classeur individuel et à la liste des favoris de l'utilisateur. Les deux font partie du module « Ma bibliothèque » et seront étudiés avec plus de précision sous Partie I-3.6.

### *3.2.4 Recommandations générales pour l'optimisation des services et ressources BNEUF*

D'un point de vue des méthodes de moissonnage et d'organisation des ressources, des paramètres de filtrage et des stratégies de recherche, l'ensemble des services du système BNEUF nécessite une révision pour une meilleure découvrabilité des contenus. Il s'agit essentiellement de prévoir d'autres fonctions productrices de valeurs ajoutées comme le téléchargement des notices, l'export et la conversion des notices, la sauvegarde de l'historique de recherche, l'analyse statistiques, etc. ;

- *Redynamiser le module de crowdsourcing* : la BNEUF devrait éviter d'être juste un fournisseur de services ou un relayeur de métadonnées d'autres catalogues de ressources. Sa spécificité propre est de servir la communauté francophone, notamment celle du Sud, par des ressources contextualisées, originales et adaptées aux besoins réels des usagers. La fonction actuelle de proposer des « ressources éducatives » à partir du module « Ma Bneuf » est quasi invisible et très peu valorisée. Une stratégie de sensibilisation, de motivation et d'accompagnement est nécessaire (cf. Partie II-3.2) ;

- *Agréger de nouvelles sources de métadonnées* : outre la fonction d'autoarchivage proposée dans « Ma Bneuf » pour des initiatives individuelles, la BNEUF gagnerait à moissonner d'autres types de métadonnées institutionnelles provenant de catalogues de bibliothèques des partenaires francophones du Sud. Une politique de partenariat et d'assistance technique est nécessaire pour créer dans ces universités les connecteurs techniques avec la BNEUF ;

-

- *Favoriser autant que possible la production de marques blanches* : avec des réseaux et communautés partenaires en vue de drainer davantage de ressources spécialisées. L'exemple de la marque blanche avec le portail RELIEFH pour les ressources éducatives sur l'égalité femme-homme est à reproduire dans d'autres domaines avec d'autres structures et établissements. Plusieurs sujets d'actualité peuvent constituer un créneau d'intérêt comme les Objectifs de développement durable, le réchauffement climatique, les droits de l'Homme, la migration clandestine, l'énergie renouvelable, la paix ans le monde (cf. Parti II-2.6) ;

- *Établir une connexion avec les systèmes mondiaux de référencement* : « Ma Bneuf » devrait avoir aussi des ouvertures sur les systèmes de publication et de diffusion scientifique et les réseaux sociaux de recherche (i.e. Google Scholar, BNF, WordlCat, Research Gate, Academia, etc.). Elle doit s'ouvrir aussi aux services du libre accès et des archives ouvertes comme OpenEdition, arXiv, HAL, DOAJ, PubMedcentral, OpenAire, REPEC, SSOAR, etc.) ;

- *Créer un système DOI* : La BNEUF gagnerait à proposer un système d'identifiant numérique (DOI) pour les ressources déposées en mode autoarchivage dans Ma Bneuf. Ce système permettra l'échange avec les circuits du référencement scientifique et de la e-Réputation comme SCOPUS, Hal-ID, etc. Ce sera aussi un facteur d'incitation pour motiver plus de chercheurs à signaler leurs ressources dans BNEUF, sachant que ces ressources seraient relayées dans les réseaux de référencement internationaux ;

**Identifiants**

HAL Id : **hal-03739832 , version 1**

DOI : **10.1115/1.4054509**

- *Utiliser un plan de classification disciplinaire* : la BNEUF peut réorganiser ses collections de métadonnées dans un ordre conforme à un plan de classification généraliste ou disciplinaire (e.g. Mesh, Rameau, LCC, CDD, CDU, etc.) ;

- *Ajouter des fonctions génératrices de valeurs ajoutées :* la BNEUF peut proposer une série de fonctions de ce genre comme :

    - La traduction en français de résumés et mots clés en d'autres langues que le français ;

    - La génération de métriques d'impact pour juger de la pertinence des ressources via les statistiques de consultations, de téléchargements, etc. ;

    - L'exportation des notices dans des formats normalisées comme EndNote, BibText, DC, TEI, XML, JSON, etc. pour des usage personnalisées (bibliographies, citations, collections personnalisées, etc.) ;

    - La conversion et le reformatage des métadonnées (formats d'échange), l'analyse sémantique des corpus (production de cartographies pour la DataViz), etc.

## 3.3 Le module « Atlas de l'expertise »

La première remarque concernant la page d'accueil de l'Atlas de l'expertise est l'absence d'une orientation de l'utilisateur pour lui expliquer la manipulation de l'interface. Un lien vers un court tutoriel ou des consignes d'orientation pourraient bien figurer en pied de page sous la rubrique « Information ». Sans cela, l'utilisateur comptera sur sa propre intuition pour comprendre la fonction de chaque élément graphique et trouver ensuite les clés du mécanisme de filtrage/recherche des experts. Il y a donc un souci d'ergonomie de la page d'accueil de l'Atlas.

Une autre remarque concerne l'image d'arrière-plan de la page d'accueil de l'Atlas. Elle signale par une simulation géothermique les points de concentration des experts présents dans l'Atlas. Elle est « zoomable » mais non cliquable pour afficher des détails concernant les zones « chaudes ».

La recherche dans l'Atlas de l'expertise est répartie sur deux rangés de filtres : filtres génériques et filtres spécifiques à caractère géographique.

### 3.3.1 *Filtres génériques*

Quatre cadres à filtres sont rangés horizontalement dont les deux premiers sont sous forme de nuages de mots organisés successivement par entrées « Spécialité » et « Mots-clés » et les deux suivants sont en mode listing par « Établissements » et « Experts » (Note éditoriale : choisir un mode singulier ou pluriel pour les intitulés des cadres).

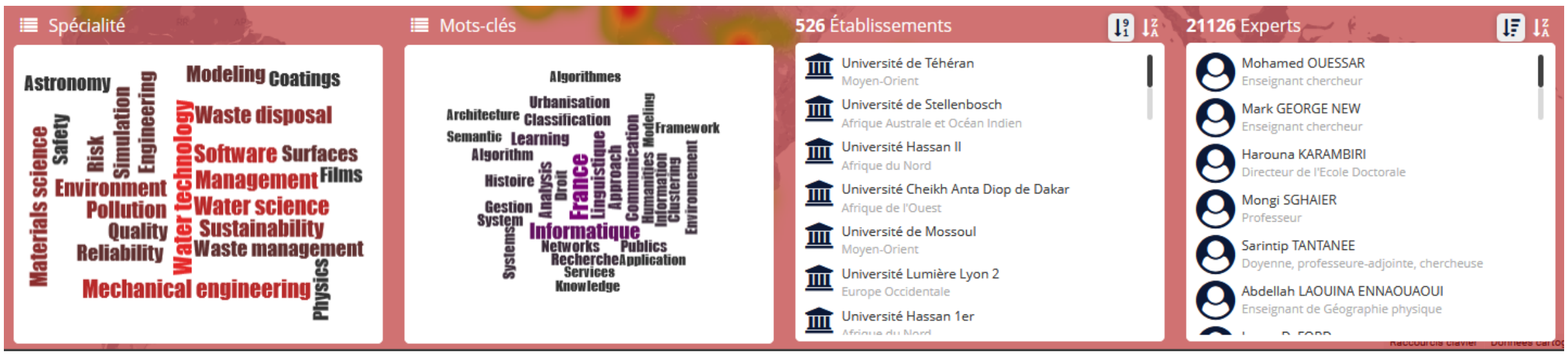

Chacune des deux paires de cadres dispose de deux modes d'affichage différents. Les cadres à nuages peuvent permuter entre mode cloud et mode listing et les deux autres peuvent permuter entre deux modes de tri « par pertinence » et « alphabétique ».

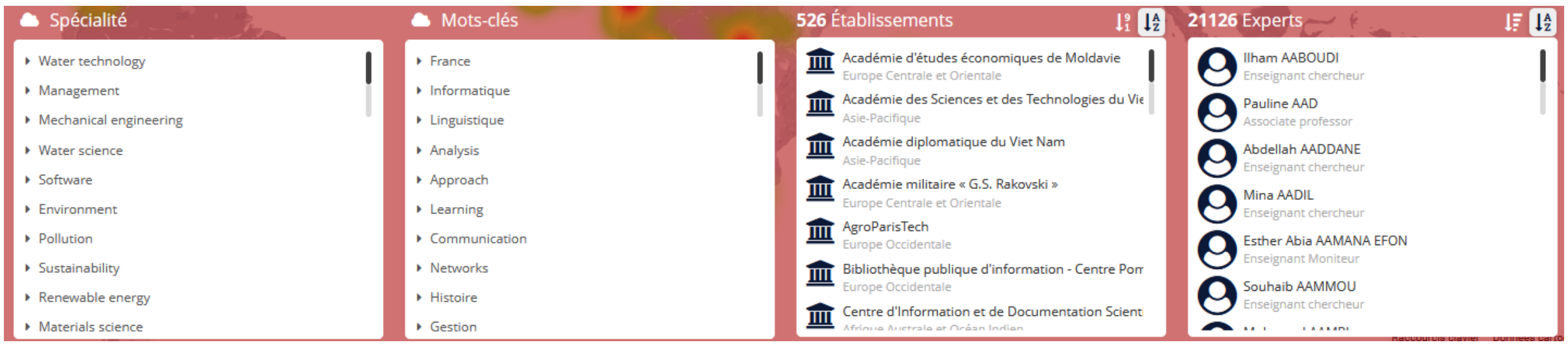

❖ ***Disposition et articulations***

- La pertinence des mots clés dans les nuages de mots des deux premiers cadres n'est pas très intuitive. Elle dépendra du libre arbitre de l'utilisateur comment il interprètera la direction de l'écriture horizontale ou verticale des termes, le sens de la couleur, de la grosseur et l'emplacement de chaque mot dans le nuage. Aucune explication n'aide à comprendre la logique de ce mode d'affichage. Les dispositions verticale et horizontale des mots dans les nuages ne semblent pas non plus avoir une signification en rapport avec la pondération puisqu'elles changent d'orientation à chaque clic ;

- Les quatre cadres fonctionnent en mode cascade : un filtre « spécialité » a une incidence sur les filtres « mots-clés » qui ont une incidence sur la liste des « établissements » qui à son tour a une incidence sur liste des « experts » ;

- Seuls les filtres « Spécialité » et « Mots-clés » ont des incidences réciproques. En revanche, un filtre d'une liste « établissements » ou « experts » n'a pas d'incidence ascendante sur les deux premiers ;

- Seuls les filtres des deux premiers cadres participent de la construction de l'équation de recherche dans la zone « Recherche globale » et agissent sur la carte géothermique. Le choix d'un établissement ou d'un expert ouvre directement la fiche correspondante sans changer l'équation de recherche ;

- Le mode listing des deux premiers cadres ne suit aucun ordre de tri. Il est primordial que cet affichage se fasse en ordre alphabétique pour faciliter le repérage rapide des filtres à utiliser. L'ordre alphabétique est primordial dans l'organisation des données au niveau des interfaces homme-machine. Si l'ordre par défaut est basé sur un calcul de pertinence, autant lui ajouter une permutation avec un tri alphabétique comme c'est le cas avec les filtres « Établissements » et « Experts » ;

- Les deux cadres en mode listing pour « Établissements » et « Experts » proposent deux ordres de tri complémentaires : « pertinence » et « alphabétique ». Ceci démontre que la programmation d'un ordre alphabétique est techniquement faisable dans d'autres emplacements du système BNEUF notamment le Module « Ressources numériques ». Il est fortement recommandé que ce soit ainsi chaque fois qu'il s'agit de donner la possibilité à l'utilisateur de faire une ou plusieurs sélections. Le mode alphabétique est le seul moyen (avec la numérotation séquentielle) que l'humain peut facilement et rapidement assimiler et utiliser ;

- Les quatre cadres manquent d'une fonction importante pour la recherche : la combinaison entre plusieurs filtres d'une même catégorie : par exemple chercher plusieurs experts ou plusieurs établissements dans un même domaine ou spécialité, ou encore identifier des experts d'un même établissement qui travaillent sur un même sujet. Ces combinaisons sont importantes pour constituer des groupes d'experts avec des spécialités et des affinités identiques ou rapprochées ;

❖ ***Fonctionnement et ergonomie***

- L'accès aux données de l'Atlas de l'expertise commence avec la fiche « Établissement » qui fournit un maximum d'information dans une organisation riche et variée. On y trouve les coordonnées de l'Université : logo à dimension standardisée (300 X 250), adresse postale, courriel du président de l'université, URL du site Web, numéro de téléphone et un synopsis de présentation ;

- On y trouve aussi des onglets consacrés aux structures de l'établissement, ses spécialités, les mots-clés qui lui sont associés et la liste de ses experts. Ces onglets sont dotés de deux fonctions importantes : un tri alphabétique du contenu et un mini-moteur de recherche locale pour trouver une occurrence dans une longue liste d'items comme la liste des structures, des spécialités ou des experts affiliés à l'établissement. Cette technique est très utile et il est conseillé de la reproduire autant que possible dans les formulaires du dispositif BNEUF ;

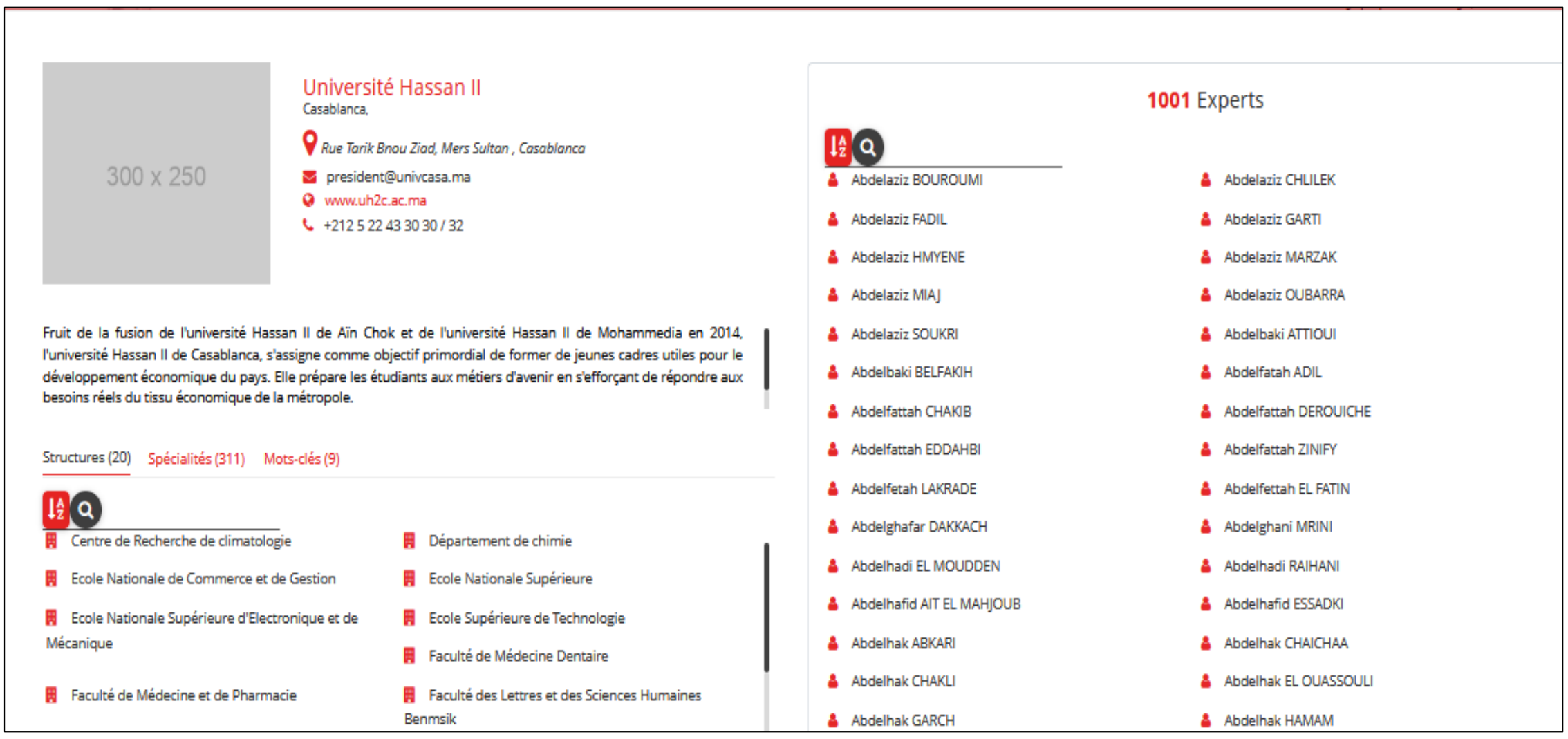

La richesse des fonctionnalités de la fiche « Établissement » est exemplaire et à reproduire ailleurs, particulièrement dans le module documentaire « Ressources numériques » où elle sera très utile pour la qualité de la recherche et du filtrage d'information.

- La fiche « Experts » fournit un ensemble d'informations convenables avec des liens dynamiques. On y trouve un cadre photo à dimensions standards, le nom de l'expert, son statut et son affiliation, puis une liste dynamique de ses spécialités. On y trouve également trois onglets qui listent ses projets, ses publications et ses structures d'attache ;

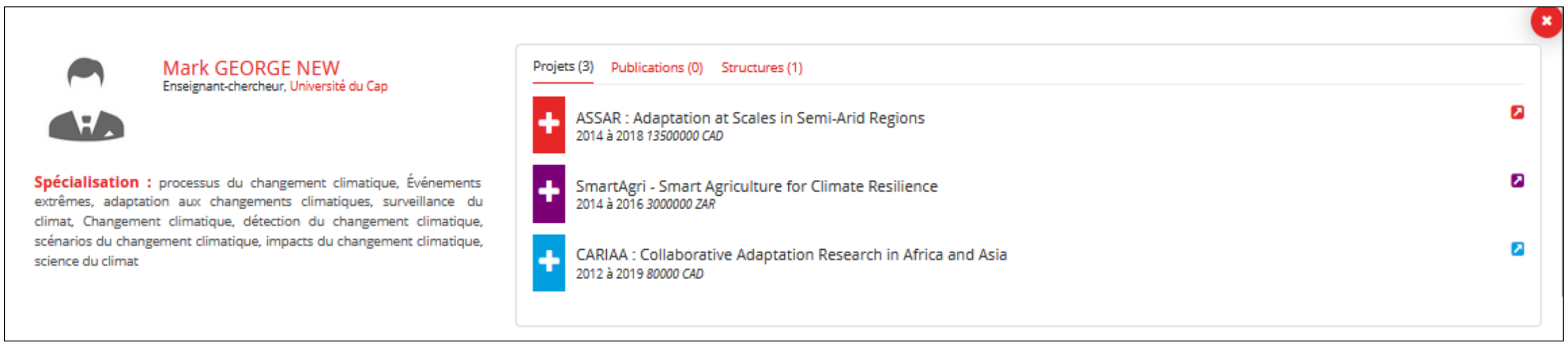

- La fiche manque d'une adresse mail de l'expert qui devrait figurer sous son statut et son affiliation ;

- L'onglet « projets » est bien garni en métadonnée avec des mots-clés dynamiques qui génèrent une équation de recherche peu utile à ce stade de consultation ;

- L'onglet « publications » affiche une liste non triée de publications sans style bibliographique normalisé. L'accès aux publications ne se fait pas par un lien sur le titre de la publication, mais par un bouton en face qui génère un bug de style d'affichage ;

- L'onglet « Structure » contient la liste dynamique des structures d'affiliation de l'expert, sauf que les liens des structures ne sont pas fonctionnels. Ils renvoient vers un cadre vide dans lequel les ongles « projets » et « Publications » ne sont plus réactifs.

### *3.3.2 Filtres spécifiques à caractère géographique*

Les quatre cadres de filtres de deuxième niveau (à caractère géographique : Région/Pays/Ville/Établissement) sont de moindre performance, mais présentent une meilleure articulation et une exclusivité importante à reproduire : la recherche multicritère.

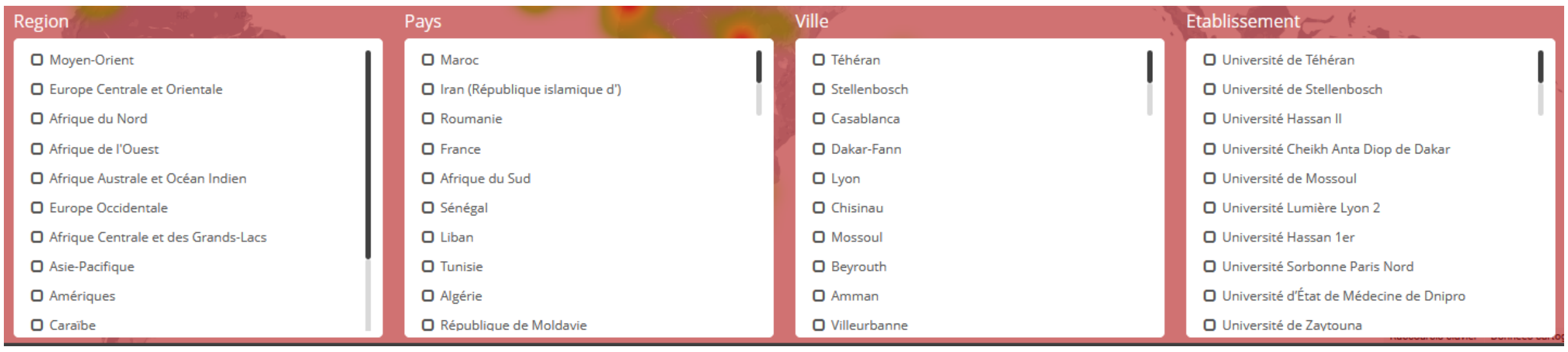

❖ ***Disposition et articulations***

- Les quatre groupes de filtres utilisent un seul style d'affichage : le mode listing ;

- Les quatre groupes de filtres sont bien liés hiérarchiquement : « Région », « Pays », « Ville » et « Établissement » ;

- Les quatre groupes de filtres n'ont pas de mode de tri alphabétique, ce qui compliquent leur utilisation quand il s'agit de trouver une entrée particulière dans une longue liste non triée. Un tri alphabétique serait très utile notamment pour le groupe « Région » qui a une incidence de haut niveau sur les autres filtres ;

❖ ***Fonctionnement et pertinence***

- L'articulation entre les niveaux hiérarchiques des quatre blocs de filtres fonctionne de manière rigoureuse : la sélection d'une région produit automatiquement une incidence simultanée sur les trois autres blocs. Le choix d'une région produit un filtrage des pays de cette région, puis les villes de ces pays et enfin les établissements dans ces pays. Cet ordre hiérarchique ne fonctionne pas dans le sens inverse ;

> Les quatre blocs de filtres proposent une exception intéressante : le **tri multicritère** à partir d'une **liste à choix multiples**. Cet exemple montre que l'option de liste à choix multiples est techniquement possible et qu'il est donc très recommandé d'en faire usage dans les autres zones de la BNEUF où une recherche avancée est nécessaire.

### *3.3.3 Commentaires sur l'Atlas de l'expertise*

Les deux entités principales de l'Atlas de l'expertise sont l'Expert et l'Établissement. Les deux sont le résultat d'une procédure d'inscription d'Experts via le module « Ma Bneuf ». Un lien intrinsèque de filiation existe normalement entre les deux entités, mais rien dans la procédure d'inscription ne démontre que l'affiliation est une donnée conditionnelle pour valider une inscription d'expert. Elle est laissée optionnelle après la validation de l'inscription par une simple adresse de messagerie ou via un compte Facebook ou Gmail. Une telle démarche risque de créer des « Experts orphelins », sans affiliation connue à un établissement déterminé ;

Plusieurs incohérences dans la procédure d'inscription seront soulevées sous « Ma Bneuf » dans la section suivante 3.4.

## 3.4 Le module « Ma Bneuf »

Accessible uniquement aux inscrits, « Ma Bneuf » est un module bilingue externe à l'environnement de la BNEUF avec laquelle il interagit via des API. On y accède après un processus d'inscription et de validation par messagerie.

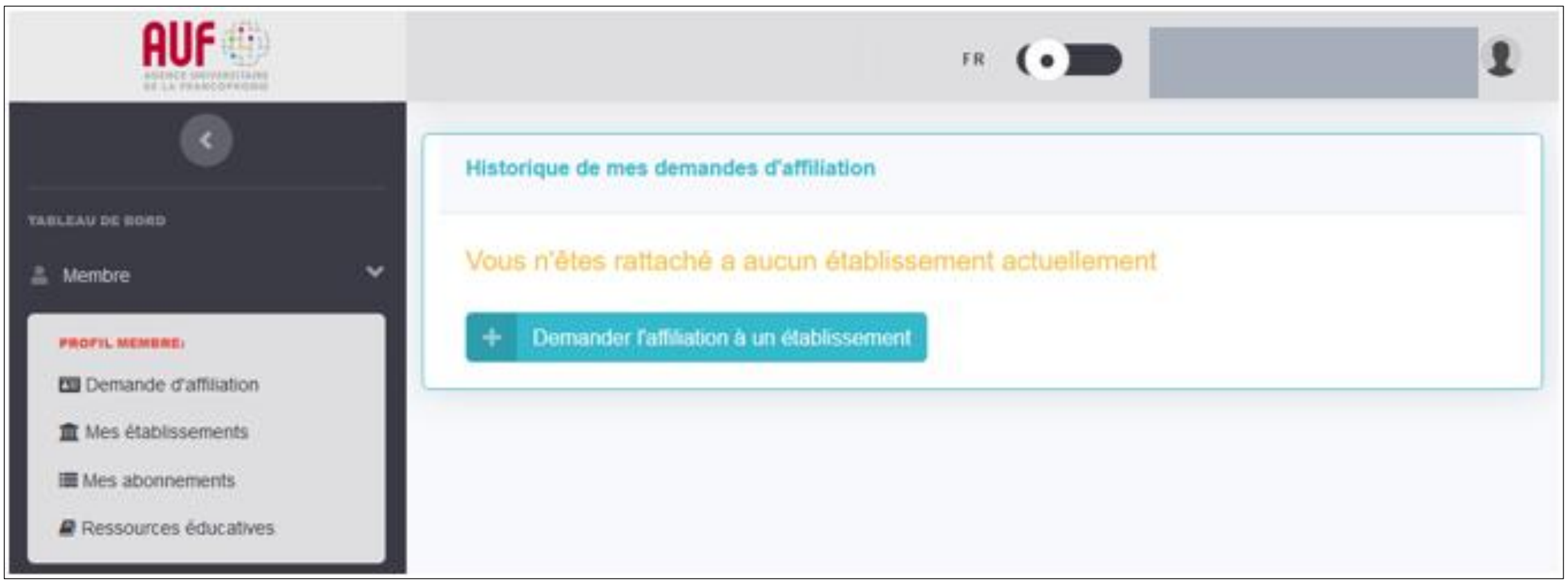

L'interface de « Ma Bneuf » dispose d'un ruban d'entête et d'un Tableau de bord latéral avec 04 rubriques. Ces zones sont analysées ci-après :

### *3.4.1 Le ruban d'entête*

- « Ma Bneuf » porte un logo différent de la BNEUF. Le logo de l'AUF est également obsolète et doit être remplacé par le nouveau. Un logo BNEUF pourrait aussi être conçu pour le dispositif entier ;

- Le bouton de changement de langue n'affecte pas la totalité des informations du module ;

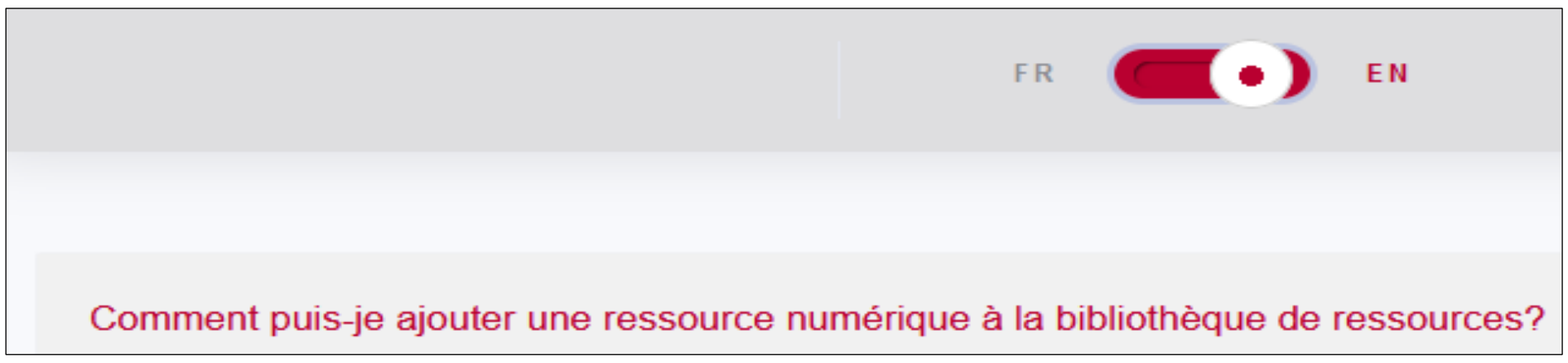

- Le menu sous la photo de l'expert est pauvre en informations. Il propose une seule rubrique d'aide qui se limite à un court passage sur l'ajout de ressources numérique.

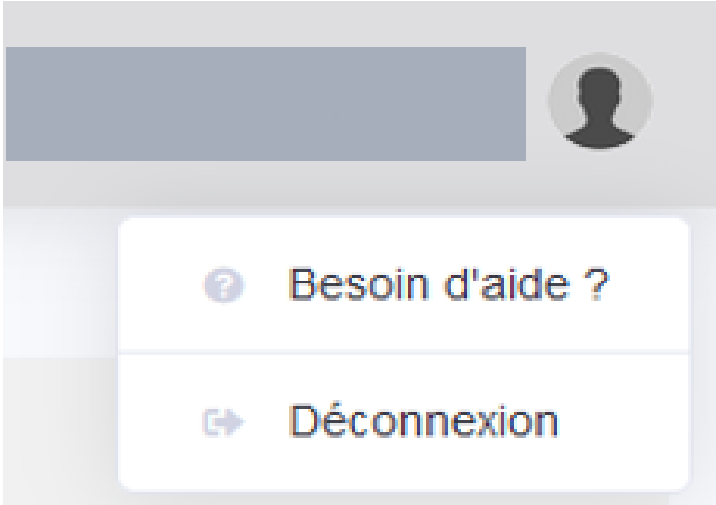

- Le menu de la photo doit normalement proposer beaucoup plus de renseignements et de services comme une rubrique d'affichage et de modification des détails du profil : photos, adresse mail, changement de mot de passe, date d'inscription, etc. ;

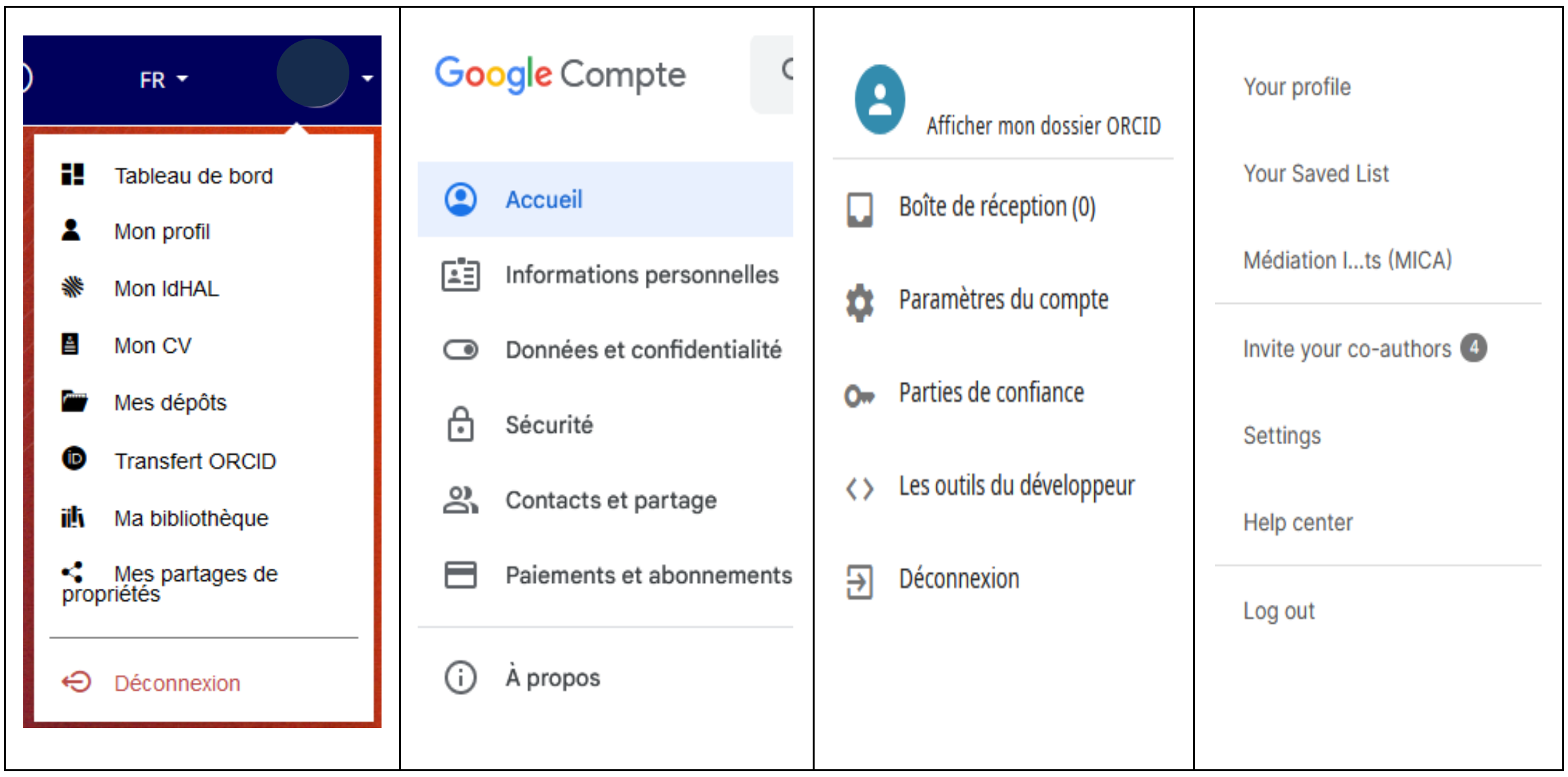

### 3.4.2 *Le tableau de bord*

Le tableau de bord est constitué de quatre rubriques :

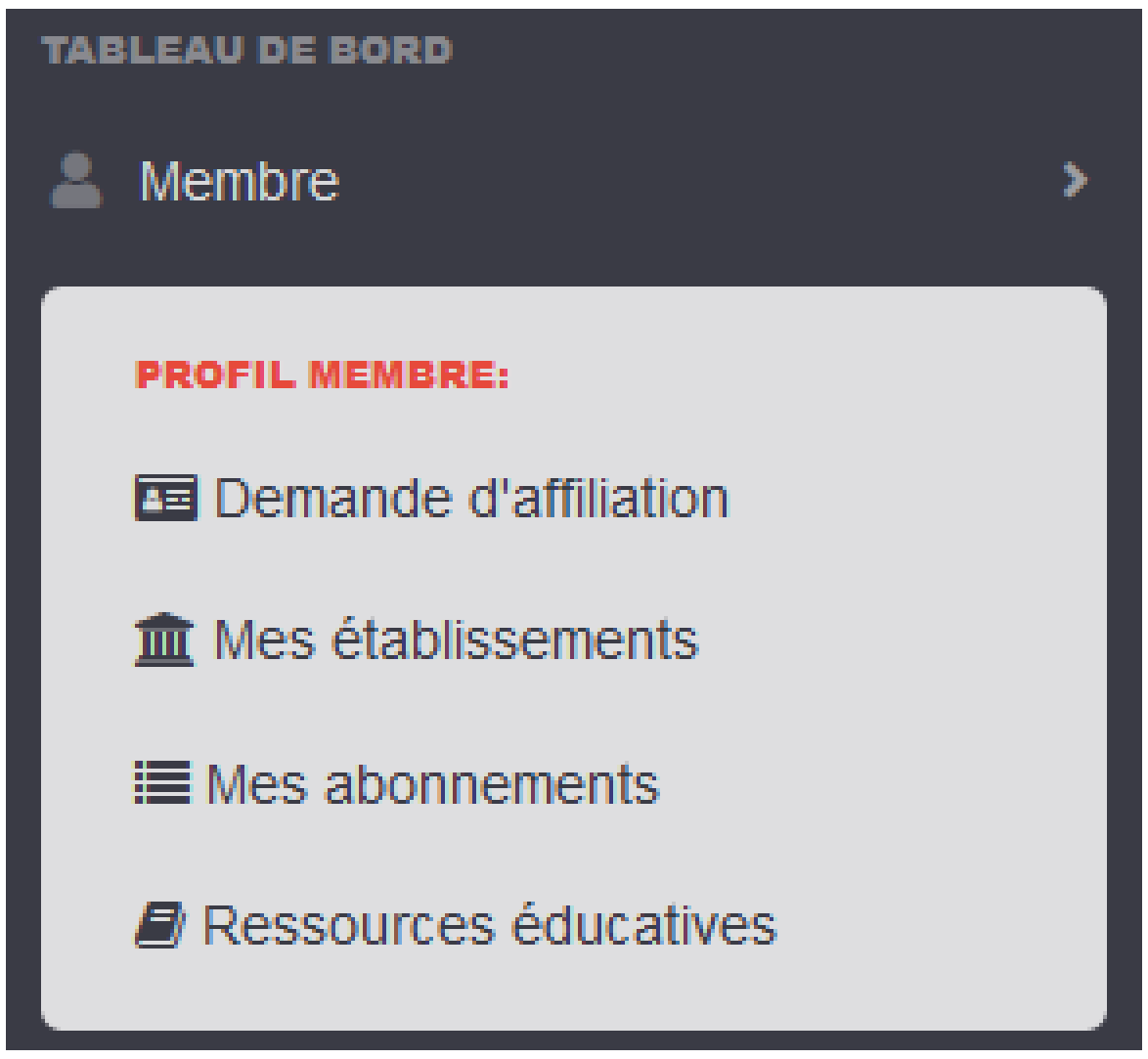

❖ ***Demande d'affiliation***

- La demande d'affiliation à un établissement se fait par formulaire à l'intérieur de Ma Bneuf. Il faut au préalable avoir créé un compte sur Bneuf par voie de messagerie ou par un compte Facebook.

> L'affiliation auprès d'un établissement doit normalement être exigée comme condition de validation d'une inscription dans l'Atlas de l'expertise.

- Le formulaire actuel propose des champs obligatoires et des champs optionnels. Certains sont libres et d'autres contrôlés ;

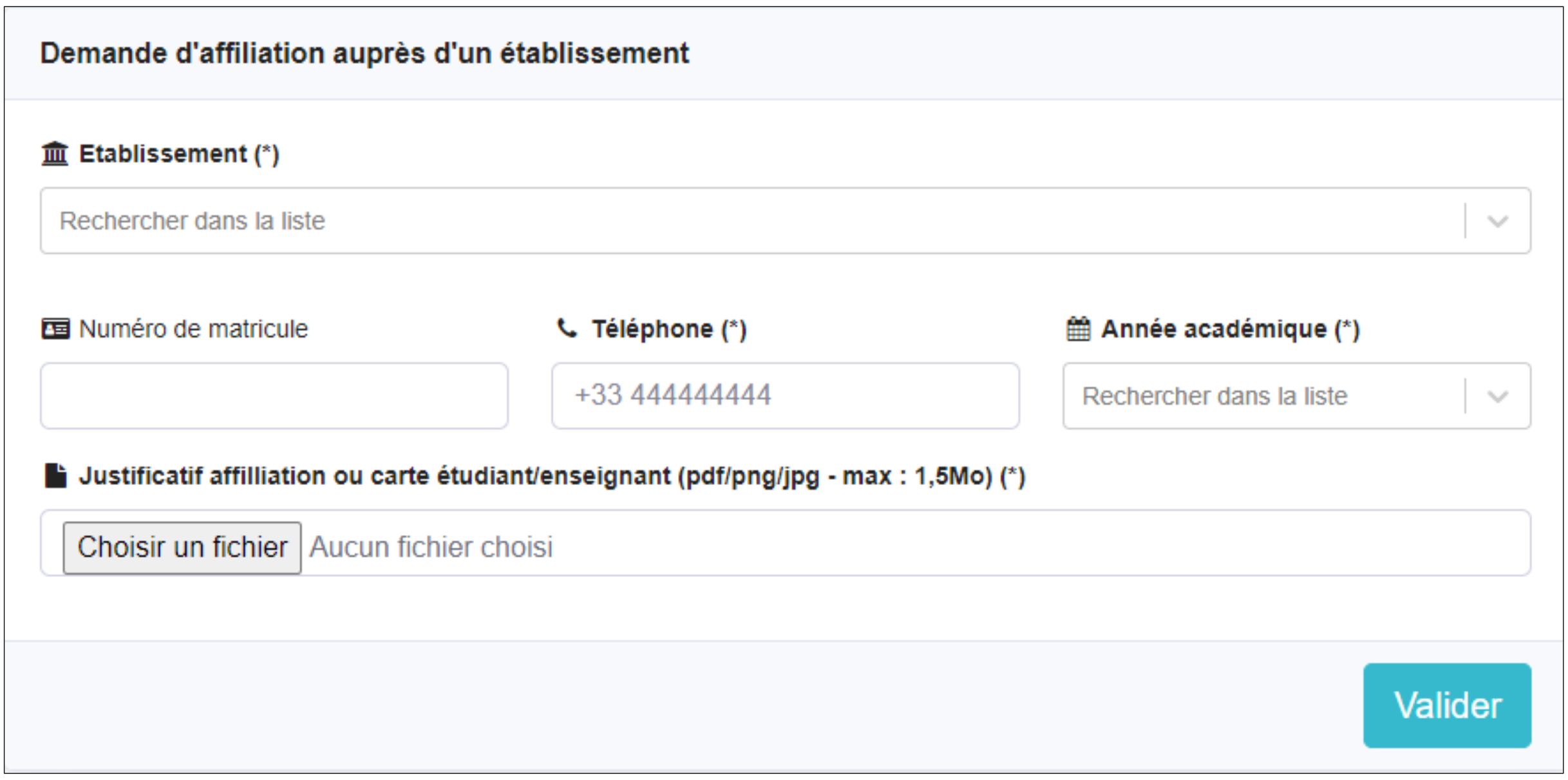

- La liste proposée des établissements n'est ni commode ni complète ni alphabétique. Un tri alphabétique par pays (code/nom de pays mis en préfixe de chaque établissement) aurait facilité le repérage d'un établissement dans une longue liste ;

- La source de la liste proposée des établissements n'est pas identifiée comme source statique ou dynamique, c.à.d. actualisée manuellement ou à partir d'une source liée. Par souci d'exhaustivité et de pertinence, elle doit provenir d'une source dynamique qui permet sa mise à jour systématique à chaque modification de la source ;

- La sélection d'un établissement dans la liste pourrait proposer une liste des structures internes appartenant à cet établissement (ces données sont disponibles dans la fiche de chaque établissement), d'autant que les experts sont plutôt membres d'une structure (département, UFR, Laboratoire, etc.) dans un établissement ;

- « Le numéro de matricule » n'est pas explicite. Aucune indication sur sa nature ou comment on l'obtient. Un bouton d'aide aurait pu donner une explication sur la valeur à saisir dans ce champ ;

- Le masque de saisie du numéro de téléphone propose un format français avec indicatif de pays. Un établissement peut toutefois être dans un autre pays que la France et son système téléphonique peut avoir une structure différente (plus longue ou plus courte). Même la saisie d'un numéro français selon le format proposé génère une erreur (pas d'espace après l'indicatif) !

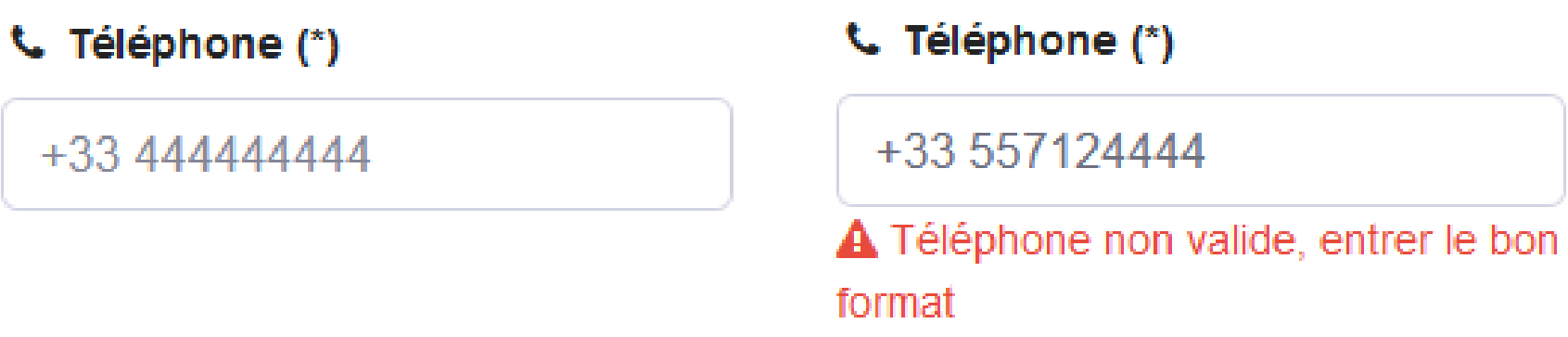

❖ ***Mes établissements***

- L'historique des demandes d'affiliation pressente une anomalie de date : une demande envoyée à une date indiquée, est enregistrée avec des dates fixes de commencement et de fin. Ces dates sont figées sur la période d'une année entre octobre et septembre. La logique de cette périodicité n'est pas évidente d'un point de vue durée (pourquoi une année ?) ni dates (pourquoi 30 septembre/01 octobre ?) ;

**Historique de mes demandes d'affiliation**

| Etablissement | Commence le | Fin | Statut |
|---|---|---|---|
| | | | |
| Université Bordeaux Montaigne | 30/09/2022 | 01/10/2023 | En attente |

❖ ***Mes abonnements***

- Sans explication de ce qui est entendu comme « abonnement », cette rubrique reste imprécise. L'ambiguïté entre la rubrique principale « MES abonnements » puis l'intitulé « Liste de d'abonnements de VOS établissements » accentue le flou. La colonne « collection dans le tableau laisse comprendre qu'il s'agit d'abonnements à des revues, mais ce point n'est pas mis en évidence ailleurs dans le système BNEUF. Cette rubrique a besoin d'une bulle d'aide ;

**Liste des abonnements de vos établissements**

| Etablissement | Commence le | Fin | Collections |
|---|---|---|---|
| | | | |

❖ ***Ressources éducatives***

- C'est une rubrique très importante puisqu'elle constitue la voie par laquelle la BNEUF peut développer du crowdsourcing ou de l'autoarchivage ;

- La rubrique affiche la liste des ressources déposées en autoarchivage avec un bouton pour en proposer de nouvelles ;

**Liste des ressources éducatives**

+ Proposer

Afficher 10 éléments                Rechercher:

| Titre | Date | Statut | Actions |
|---|---|---|---|
| Indexation des ressources pédagogiques | 04/02/2023 | En Attente de validation | |

Précédent 1 Suivant

- Toutes les colonnes disposent d'une fonction de tri à double sens (ascendant/descendant). C'est une autre preuve que techniquement, cette fonction de tri alphabétique (ou chronologique) est possible, mais qu'elle n'a pas été utilisée à bon escient là où il faut ;

- La proposition d'une nouvelle ressource passe par un formulaire à 15 champs de métadonnées décrits et commentés dans le tableau suivant :

| Champ | Commentaires et suggestions |
|---|---|
| **Titre *** | - Champ obligatoire par défaut<br>- Un titre parallèle (équivalent linguistique) pourrait être proposé pour les ressources bilingues |
| **Url *** | - Champ obligatoire du moment que la BNEUF gère uniquement les métadonnées des ressources gardées à leurs emplacements d'origine |
| **Collection *** | - Champ obligatoire à base d'une liste fermée ;<br>- La liste d'options est très limitée et la variable « Autres » n'apporte pas d'utilité au référencement ;<br>- Mieux paramétrer ce champ pour un meilleur contrôle qualité (des options à récupérer dans une source dynamique interne ou externe à BNEUF) ; |
| **Description *** | - Champ obligatoire mais le terme « description » est ambigu. En langage documentaire il implique une description matérielle (catalogage) de la ressource. Le terme « résumé » serait moins redondant. Une longueur maximale en caractère pourrait aider à harmoniser la longueur de ce champ pour toutes les ressources ; |
| **Mots clés** | - Champ optionnel démuni de tout contrôle de consistance sémantique ;<br>- Les mots-clés peuvent être proposés à partir d'une taxonomie contrôlée, d'une liste d'autorités vedettes matières comme Rameau ou d'un plan de classement comme CDD ou CDU ;<br>- Le problème des mots-clés laissés en langage libre est la prolifération des mots-vides (Europe, Travail, Transport, pluies, climat, etc.) qui ne constituent pas des pointeurs de recherche ;<br>- Limiter les mots-clés à un nombre maximum ;<br>- Proposer une saisie linéaire avec des séparateurs logiques (ponctuation) plutôt que d'avoir des champs séparés pour chaque mot-clé. La ponctuation aidera à les séparer et les trier par ordre alphabétique dans le processus d'indexation ; |
| **Auteur(s)** | - Champ optionnel sur la base qu'une ressource peut être anonyme ;<br>- Champ pouvant aussi être obligatoire et lié à une liste d'auteurs prédéfinie. En cas d'absence d'auteur, l'indication « anonyme » peut jouer le rôle de vedette auteur pour un besoin de tri. La saisie de nouveaux auteurs se fait au moment du catalogage selon des règles et une syntaxe précise (Nom, prénom). Cette liste est alimentée au fur et à mesure des besoins documentaires. Les nouveaux auteurs sont créés en fonction des documents à indexer ; |
| **Pays** | - Champ optionnel rempli à parti d'une liste fermée prédéfinie de noms ou codes (ISO 3166-1 alpha-3) de pays ;<br>- La recherche se fait à partir des premiers caractères saisis ; |

<table>
<tr><td></td><td>

- La liste des pays est triée par ordre alphabétique sauf exception de pays prioritaires mis au début (France) !?

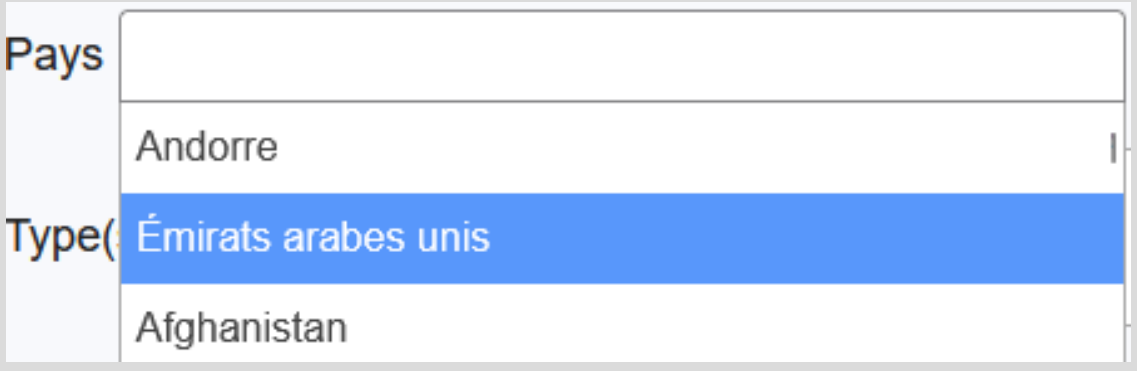

</td></tr>
<tr><td>Type(s)</td><td>

- La typologie des documents est actuellement prédéfinie dans une liste sans tri alphabétique et d'une grande redondance dans la définition des types de documents (cf. exemples suivants) ;
- Une typologie fermée à base d'un langage contrôlé éviterait cette redondance ;

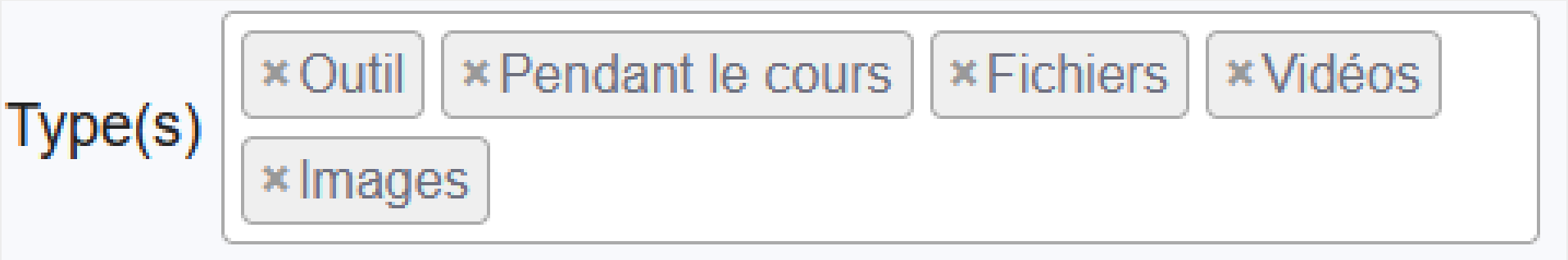

</td></tr>
<tr><td>Format(s)</td><td>

- Le contenu du champ format est prédéfini dans une liste sans tri alphabétique et dans une grande redondance des choix comme le montre l'exemple suivant : « Images » et « Vidéos » sont à la fois des types et des formats. « Pendant le cours » est un non-sens comme format !

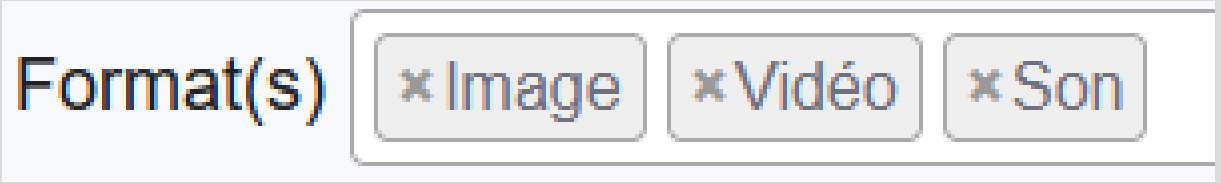

</td></tr>
<tr><td>Niveau(x)</td><td>

- Le contenu du champ « niveau » est aussi prédéfini dans une liste sans tri alphabétique et dans une grande redondance des choix comme le montre l'exemple suivant :

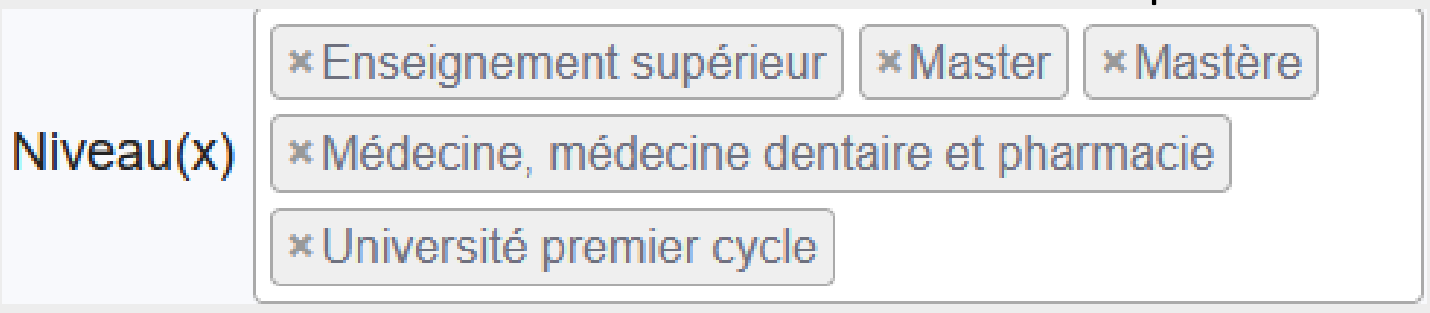

</td></tr>
<tr><td>Discipline(s)</td><td>

- Le contenu du champ « discipline » est également prédéfini dans une liste sans tri alphabétique et dans une grande redondance des choix comme le montre l'exemple suivant où « label », « Internet », « Le savoir » etc. ne peuvent constituer des disciplines ;
- Pourtant le module « Ressources numériques » de la BNEUF propose une liste mieux organisée de « Disciplines » qui pourraient servir à remplir ce champs !

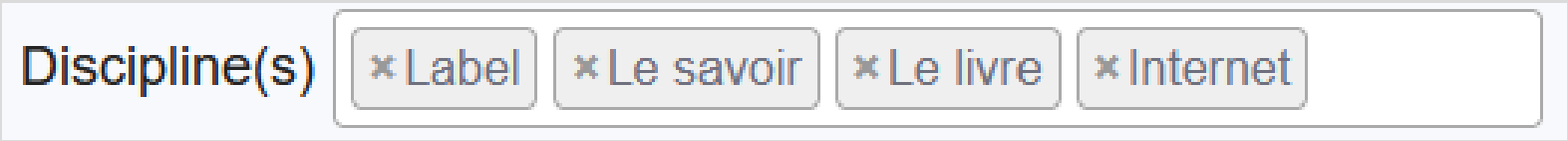

</td></tr>
<tr><td>Licence</td><td>

- Le champ licence est très mal renseigné avec une liste fermée à trois variables non représentatives. « Libre de droit », « payant » et « gratuit » ne représentent pas les vraies alternatives utilisées pour renseigner une licence. Les options doivent être « Libre », « ouverte » ou « propriétaire » sinon des options plus détaillées comme « Domaine public », « *Creative commons* », « Droit d'auteurs » ;

</td></tr>
</table>

| | |
|---|---|
| **Difficulté** | - Les trois options « facile », « Moyen », « Difficile » sont acceptables ;<br>- Les options du LOM (Learning Object Metadata) pour ce genre de métadonnées sont au nombre de 05 : « très facile », « facile », « moyen », « difficile », « très difficile » ; |
| **Langue** | - Le champ « langues » est géré de façon aléatoire et sans tri alors qu'il pourrait se ressourcer dans une liste normée (ISO 639) de codes langues ; |
| **Url de l'image d'illustration de la ressource** | - Ce champ pourrait aussi proposer le chargement d'une image locale |

### *3.4.3 Remarques générales sur « Ma Bneuf »*

Ma Bneuf est un module stratégique dans le dispositif global BNEUF. Il nécessite des soins particuliers. Il produit et gère les métadonnées des acteurs du dispositif à savoir les Experts et les Établissements. Ces deux acteurs nécessitent des traitements spécifiques comme :

- Développer un système d'attribution d'identifiants numériques aux chercheurs francophones (ID-BNEUF), équivalent aux autres systèmes d'ID numériques comme ORCID, SCOPUS, ID-HAL, ResearcherID, ISNI, IDRef, etc.) ;

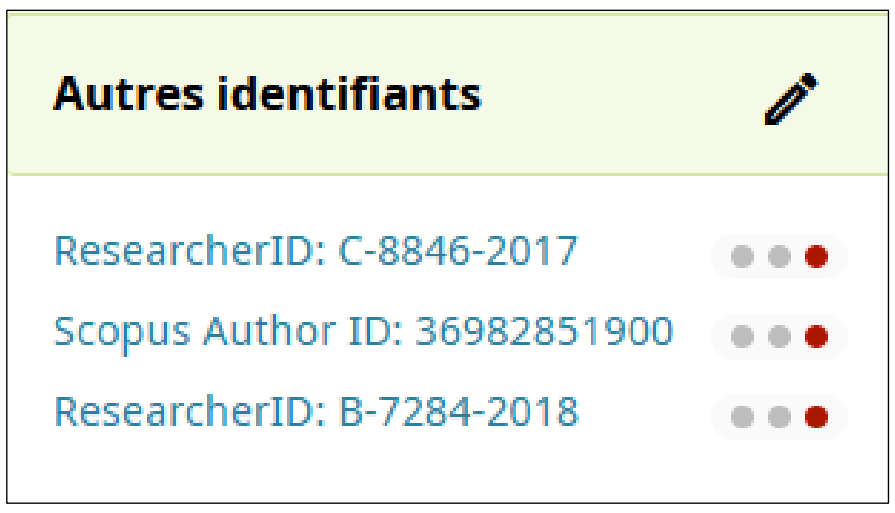

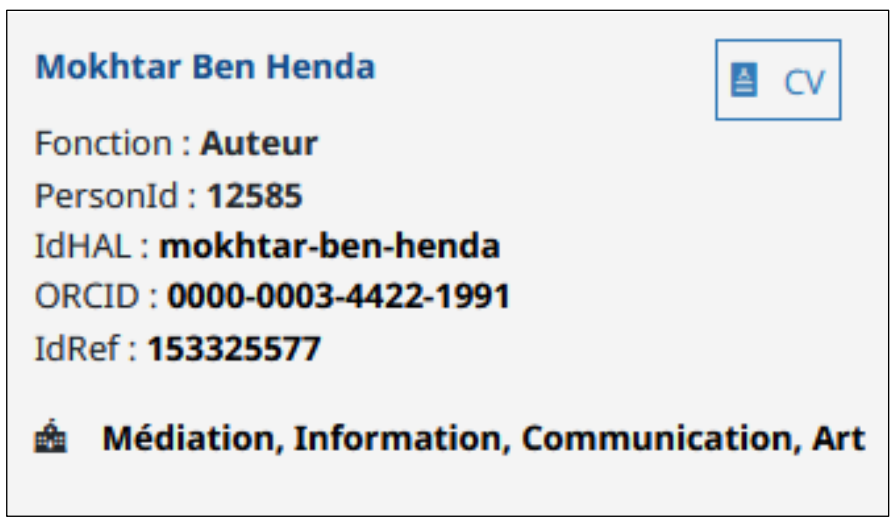

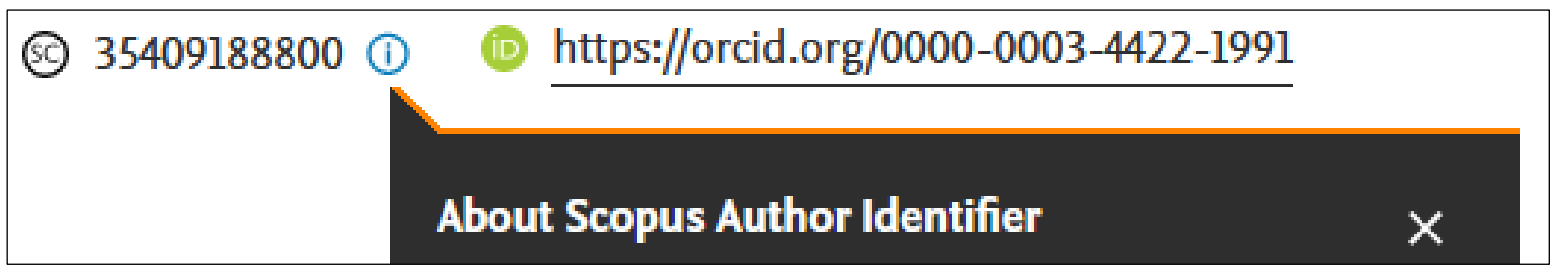

- Prendre en compte des mécanismes supplémentaires de repérage et d'identification des chercheurs francophones à travers les citations d'auteurs et les systèmes de référencement et les réseaux sociaux de recherche internationaux comme HAL, SCOPUS, ORCID, Academia, etc. ;

- Ajouter à « Ma Bneuf » un générateur automatique de CV dynamique à partir des données du profil ;

- Procéder à une restructuration des connecteurs entre les services « Ma Bneuf », « Atlas » et « Réseau social » et une redéfinition des règles d'inscriptions dans chaque service qui prévoient :

    - Une distinction entre chercheur publiant et chercheur non publiant sur la base des contributions à « Ma Bneuf », avec une redirection des non publiants vers le réseau social pour la création de communautés d'intérêt ;

- Une amélioration des mécanismes de recherche en ajoutant des techniques avancées de filtrage comme le tri alphabétique, les moteurs de recherche sur zones, la hiérarchisation fine des contenus, la recherche par expression exacte. Ces critères s'appliquent aussi bien pour les « ressources numériques » de la BNEUF que pour l'« Atlas de l'expertise », « Ma Bneuf » ou « Ma bibliothèque » ;

- Un meilleur contrôle des connecteurs entre l'Atlas et le Réseau social sur le principe de la saisie unique et l'usage d'un Identifiant numérique unique ;

- L'usage généralisé des taxonomies contrôlées chaque fois qu'il est demandé aux utilisateurs de remplir une fiche de métadonnées comme dans l'autoarchivage des ressources (à travers le module « Ma BNF ») ou pendant l'auto-inscription dans l'Atlas de l'expertise ;

- L'usage d'un système de statistiques pour mesurer la notoriété (publications dans BNEUF) des chercheurs et l'impact d'usage de leurs ressource (*Google analytics, Almétrics*, …) ;

Changer le mode d'inscription des experts et du signalement de leurs établissements dans l'Atlas de l'expertise et conditionner leur intégration par des critères sélectifs comme la contribution au contenu de la BNEUF. L'Atlas de l'expertise sera ainsi un vivier d'experts confirmés par leurs productions et non des visiteurs sans traces dans le système BNEUF autres que leurs noms.

## 3.5 Le module « Mon réseau social »

Accessible uniquement aux inscrits, « Mon réseau social » est un module externe à l'environnement de la BNEUF avec laquelle il interagit a priori via des API. Son organisation est standardisée, conçue selon les principes courants des réseaux sociaux conventionnels.

Sa page de connexion offre les rubriques ordinaires pour expliquer les règles de confidentialité, les conditions générales d'utilisation et une présentation du service (lien récursif !).

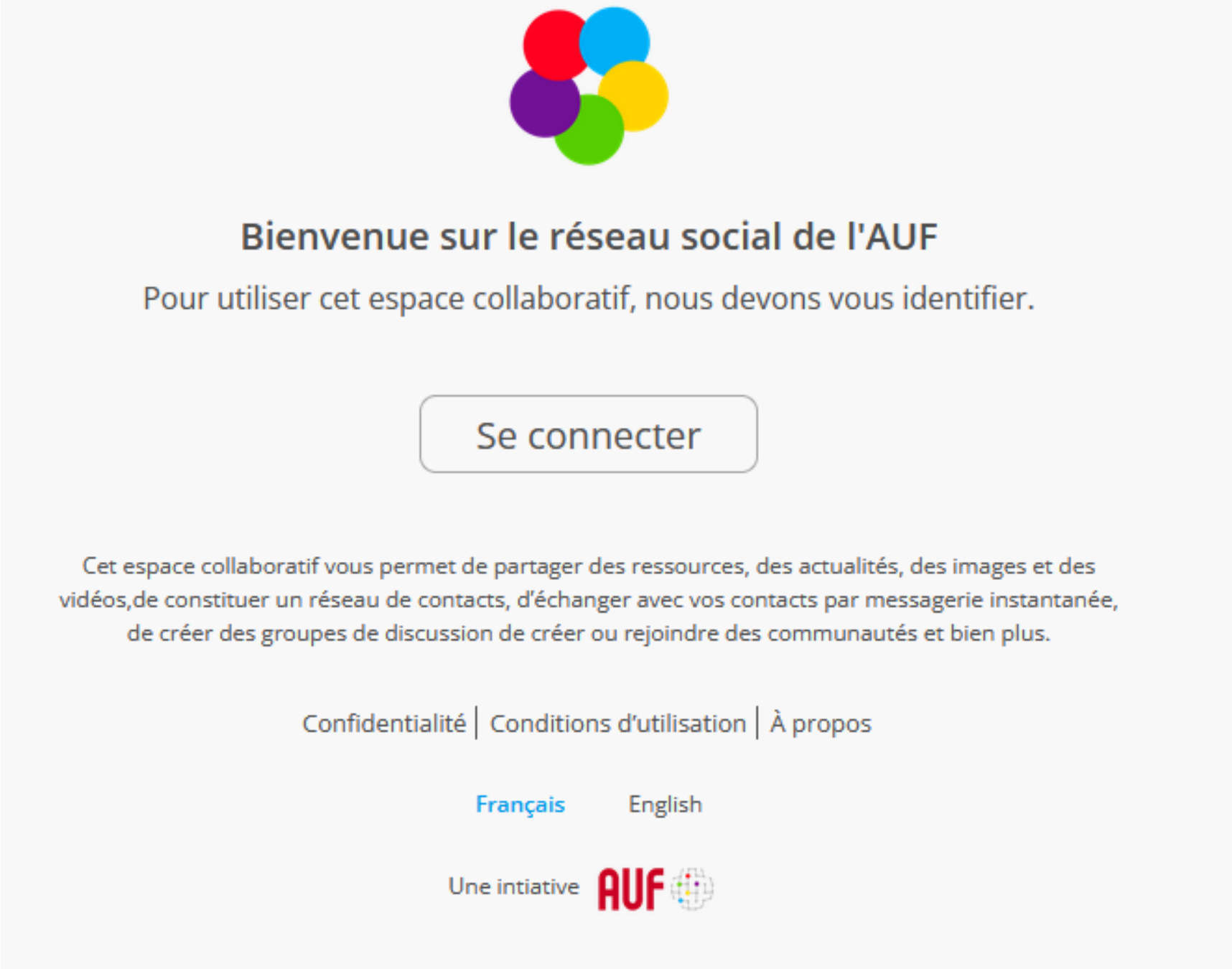

### 3.5.1 *Structure et organisation*

L'interface du réseau social est une copie simplifiée des interfaces plus complexes de réseaux comme Facebook ou LinkedIn.

Elle propose un menu horizontal avec des fonctions basiques d'un moteur de recherche, un lien d'accueil (Home), un bouton de signalement de messages, une cloche de notifications, puis l'image portrait de l'abonnée avec un sous-menu pour le changement de langue et l'affichage des paramètres du compte.

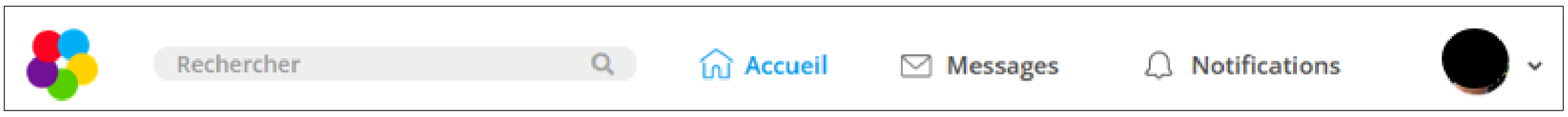

L'interface propose aussi un menu latéral avec une série de fonctionnalités standards : « Fil d'actualité », « Mes publications », « Mon profil », « Connexions », « Notifications », « Communautés ».

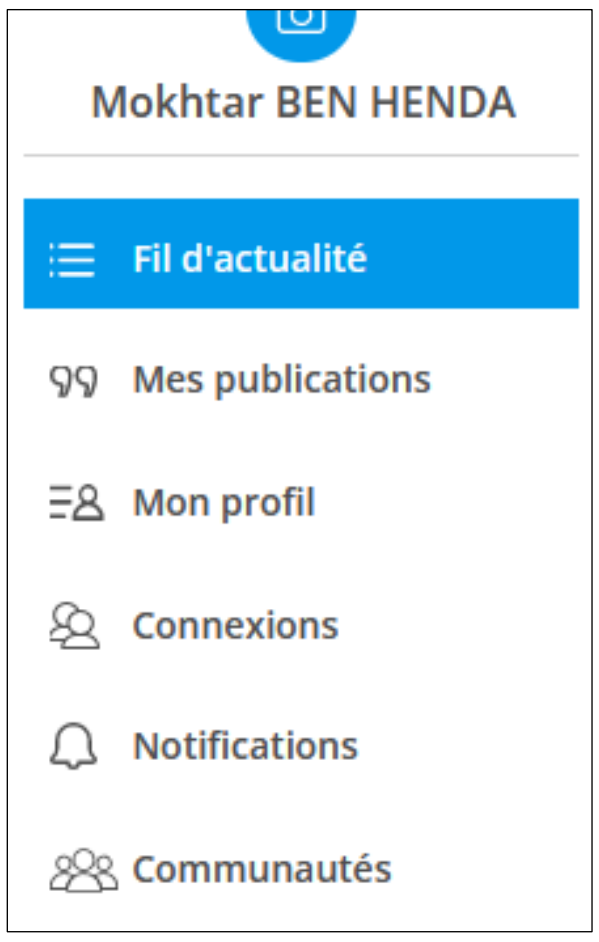

Pour éviter la simple description matérielle de l'interface et de ses services, la rubrique qui mérite un intérêt particulier est celle des « Communautés » :

- La rubrique « Communautés » est censée créer la différence avec les autres réseaux sociaux ;
- Elle est surtout censée créer des connexions avec les autres modules de l'écosystème BNEUF ;
- Elle propose un premier onglet pour gérer les communautés de l'abonnée (« Mes communautés »), celles que l'utilisateur a créé et continue à gérer et celles dont il est membre ;
- La rubrique propose aussi un deuxième onglet qui permet de découvrir les communautés créées par d'autres abonnés et de pouvoir s'y inscrire ;

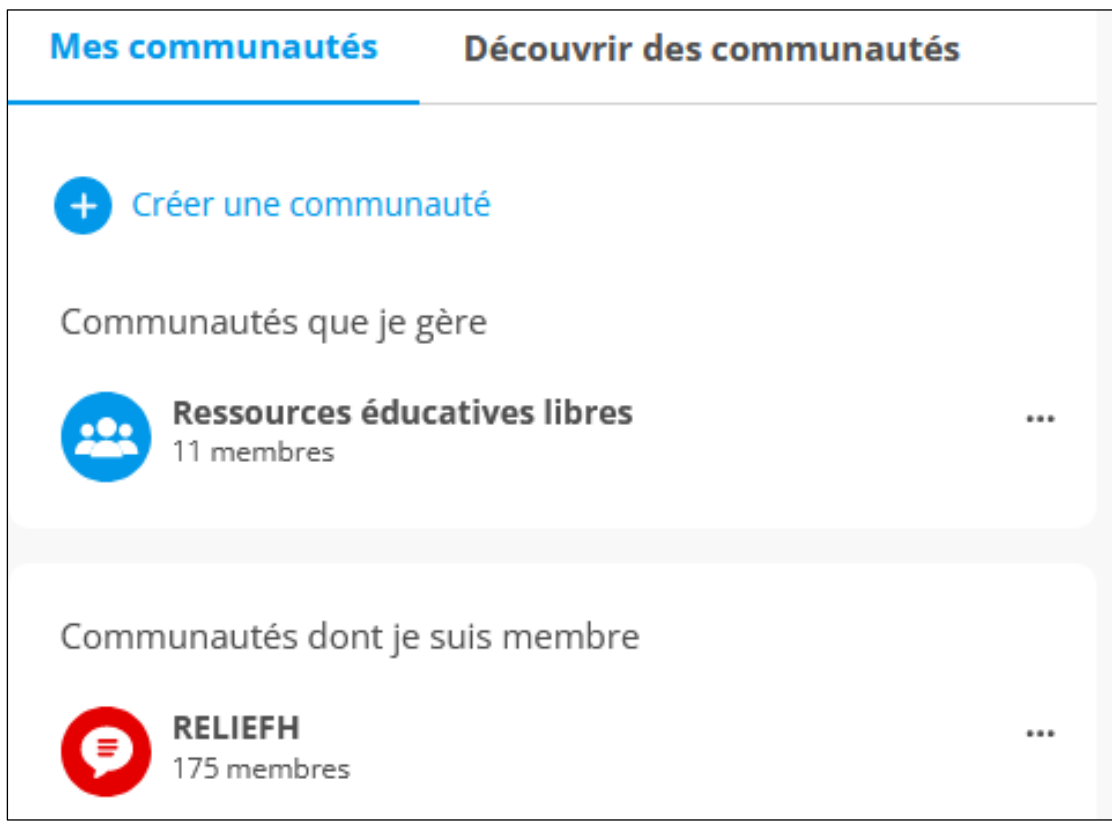

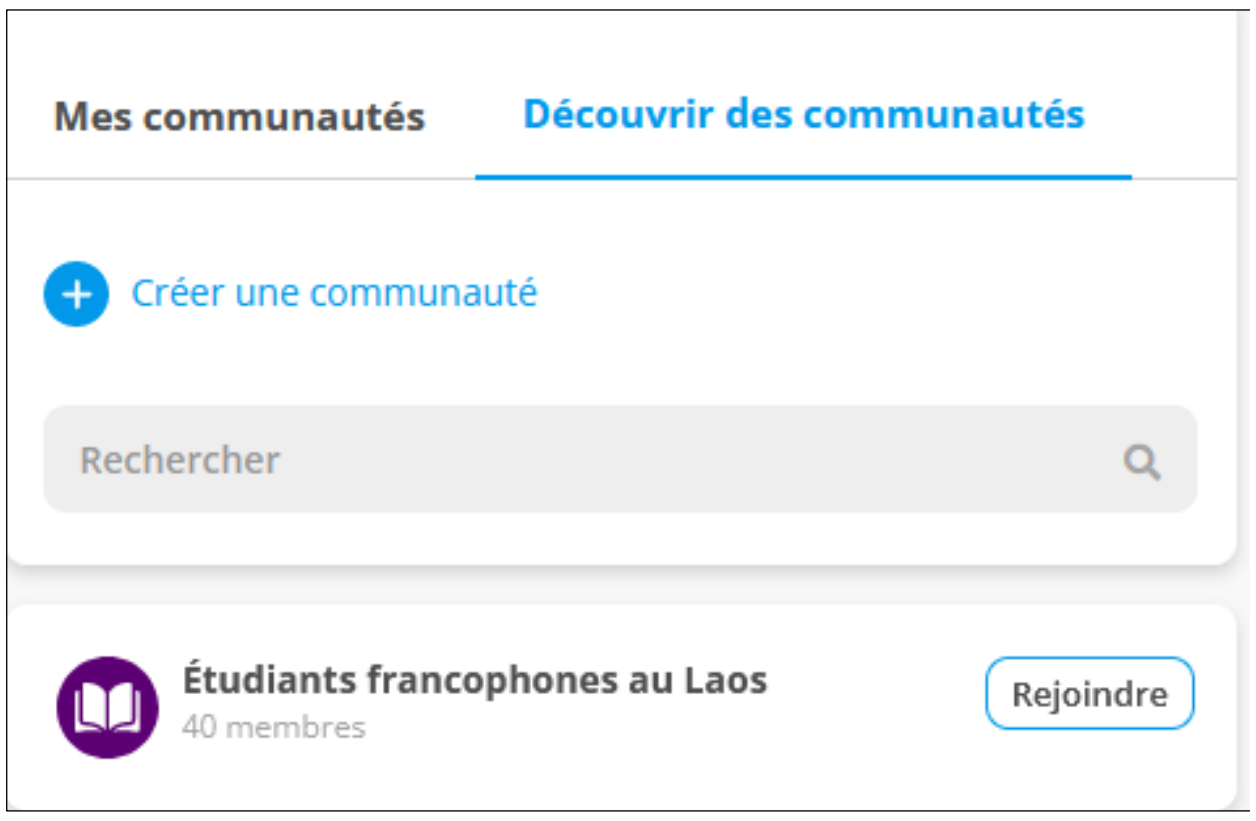

### *3.5.2 Remarques générales*

Les fonctions de tout réseau social spécialisé ne sauraient créer une distinction vis-à-vis des réseaux sociaux publics qu'à condition de créer et offrir une valeur ajoutée spécifique. Or, la valeur ajoutée que pourrait proposer le réseau social francophone de la BNEUF pour des utilisateurs déjà ancrés dans des pratiques durables avec des réseaux sociaux publics comme FB, Twitter, LinkedIn, WhatsApp, Instagram, Tiktok, etc. ne peut venir que de deux solutions : la capitalisation d'une matière issue des autres modules de la BNEUF ou la proposition d'un service original que les autres réseaux sociaux n'offrent pas !

À ce titre, un réseau social francophone autour de la BNEUF doit donc offrir des services spécifiques produisant de la valeur ajoutée concurrentielle. Les mesures suivantes sont proposées à titre indicatif :

- Faire du réseau social francophone un vecteur de promotion des produits et activités scientifiques de ses membres, qu'ils soient inscrits ou non dans l'Atlas de l'expertise ;
- Appuyer le réseau social (et le faire connaître aux membres adhérents) par une stratégie de communication étudiée, par exemple en faire des liens de renvoi à partir des pages Web officielles de l'AUF, de ses directions régionales et de ses partenaires ;
- Ajouter dans le réseau social une rubrique intitulée « Mes publications dans BNEUF » en plus des publications dans le réseau social. Cette fonction filtrera des informations (quand elles existent) concernant l'abonné au réseau social à partir du module « Ma bneuf ». Elle jouera ainsi un rôle stimulant pour inciter plus d'inscrits au réseau social à contribuer à la BNEUF ;
- Ajouter des fonctions de communication synchrone (Chat ou Viso) type « Messenger » pour autoriser les discussions audio-visuelles en temps réel entre les membres ;
- Créer des connecteurs avec l'Atlas de l'expertise pour pouvoir envoyer des invitations à des chercheurs confirmés ;

## 3.6 Le module « Ma bibliothèque »

Le module « Ma bibliothèque » est mis à l'écart des autres modules bien qu'il fasse partie du groupe des modules sous « bneuf.auf.org ». Il aurait pu être dans le bouton damier en haut de la page d'accueil de la BNEUF.

Ce module constitue l'espace privé de tout utilisateur inscrit où sont gardées les traces de ses pratiques de consultation des ressources numériques de la BNEUF. Il complète le module « Ma bneuf » qui garde les données du profil de l'utilisateur inscrit.

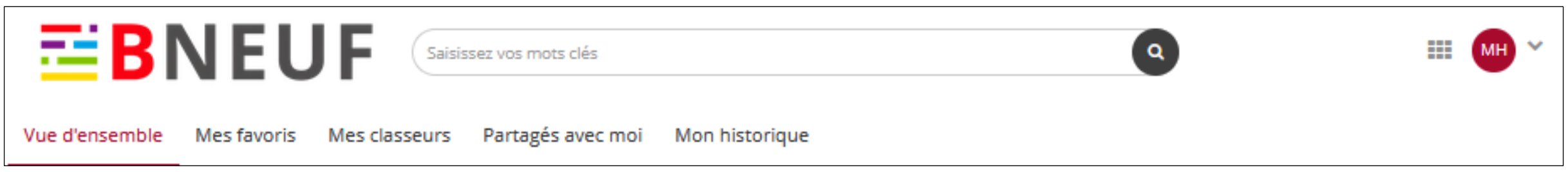

### *3.6.1 Structure et organisation*

Le module est structuré en cinq groupes d'information répartis entre cinq onglets superposés :

- Un onglet « Vue d'ensemble » qui résume le contenu des quatre autres onglets ;
- Un onglet « Mes favoris » qui liste les ressources choisies comme favoris par l'utilisateur pendant sa navigation dans les ressources numériques ;
- Un onglet « Mes classeurs » qui liste les sous-ensembles (rangés dans de classeurs) des ressources choisies par l'utilisateurs pendant sa recherche dans le module documentaire ;
- Un onglet « Partagé avec moi » qui liste l'ensemble des ressources que d'autres utilisateurs ont partagé à travers le bouton « partager » au pied de la vignette ressource dans le module documentaire ;
- Un onglet « mon historique » qui liste un ensemble de ressources (a priori) consultées par l'utilisateur ;

### *3.6.2 Remarques générales*

Les cinq onglets présentent des caractéristiques autour desquelles on peut faire les remarques suivantes :

- Le classeur nécessiterait une fonction d'exportation de son contenu pour un usage personnalisé (i.e. bibliographie). Il est utile, voire fondamental d'ajouter cette fonction d'export dans un/des formats normalisés indiqués plus haut (BibText, Endnote, APA, MLA, etc.) ;

- Une vignette de contenu d'un classeur hérite des paramètres de celle d'une ressource numérique dans BNEUF en ce qui concerne les boutons dynamiques (détails, partage, choix de classeur, favoris) avec un bouton supplémentaire de retrait de la ressource du classeur ainsi que de l'ordre d'affichage (date, pertinence, alphabétique, score, consultation) ;

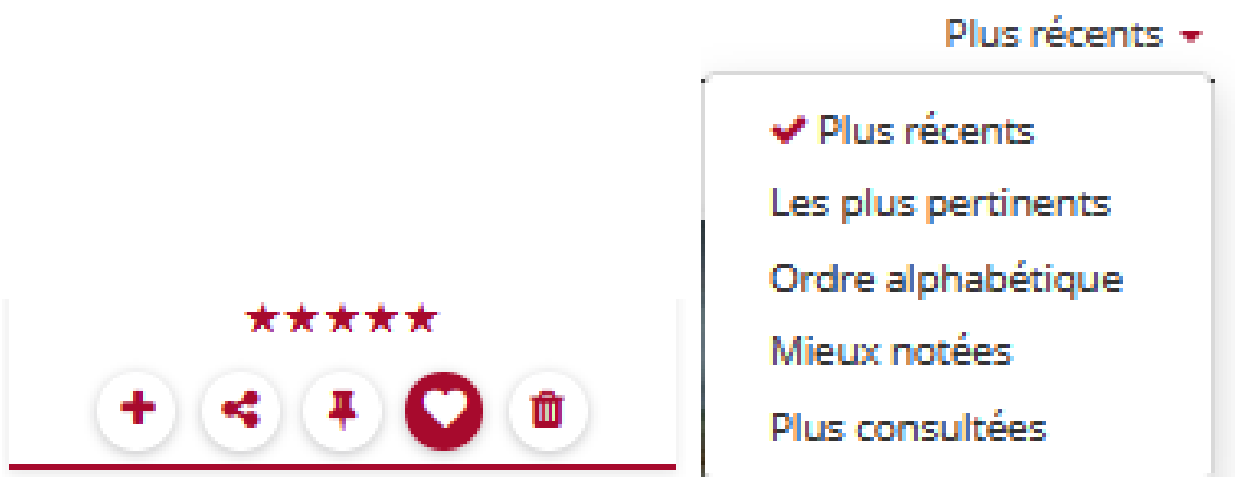

- Toutes les améliorations préconisées dans l'ordre d'affichage pour le module documentaire doivent être aussi appliquées au classeur ;

- La création d'un nouveau classeur est, quant à elle, soumise à une contrainte de nom qui doit contenir au moins 10 caractères. Cette limitation est-elle justifiée ? Sinon il serait utile de la supprimer pour permettre des noms de dossiers plus courts selon le choix de l'utilisateur ;

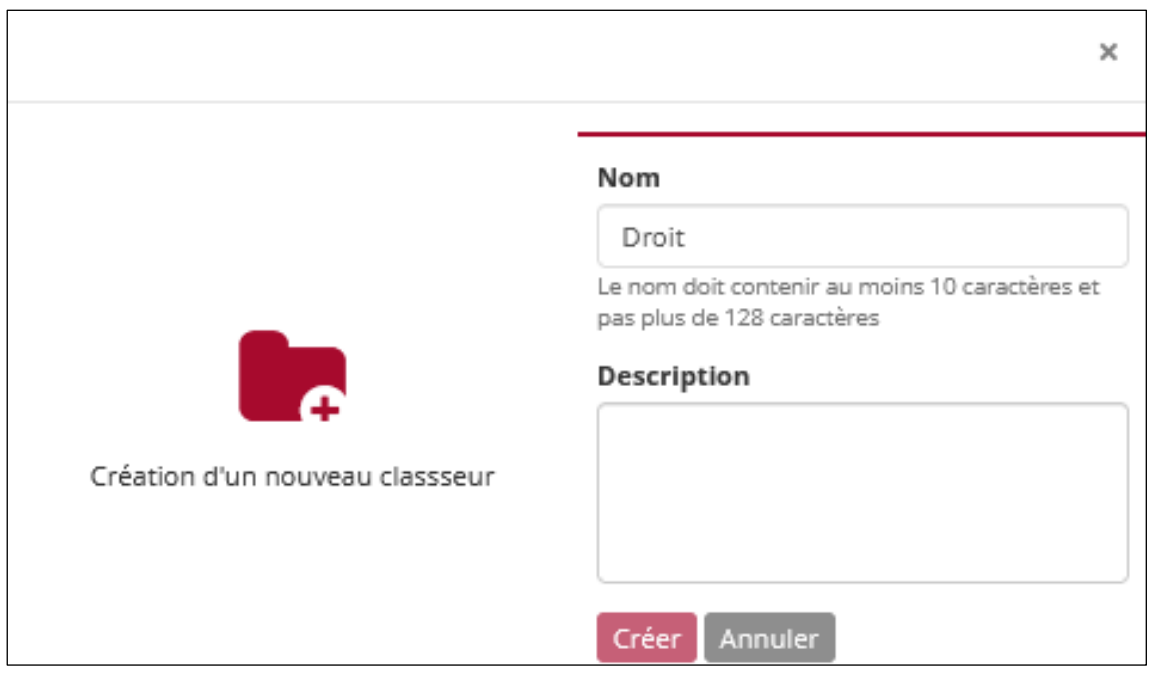

- Les différences entre les deux fonctions de partage : « Invitation » d'un utilisateur et son « Ajout » à un classeur reste floue. Ajouter des bulles d'aide sur les icones de ces fonctions pour expliquer leurs objectifs, aiderait à dissiper l'ambigüité ;

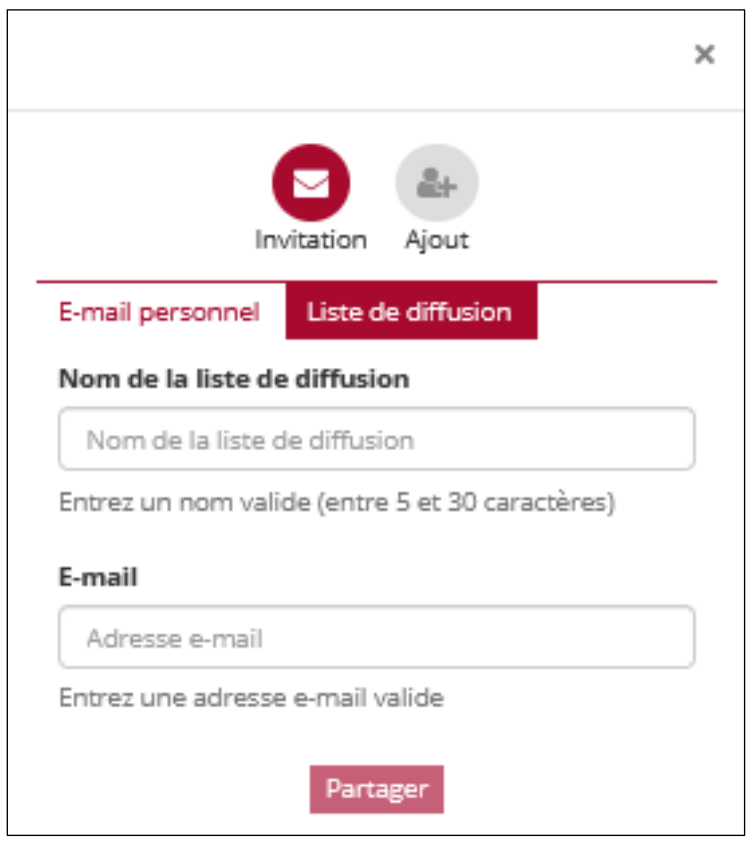

- L'onglet « Mon historique » n'explique pas de quel historique il s'agit : est-ce un historique de navigation, de résultats de recherche, ou d'autre chose ? Une bulle d'aide ou un texte explicatif ou une nomenclature plus précise permettraient de clarifier ces nuances ;

- Les vignettes des ressources dans « Mes classeurs » ou dans « Mon historique » peuvent être traitées par lots pour effectuer des opérations de type suppression, export, partage, envoie par mail, etc. en activant une sélection par case à cocher sous (ou à côté de) chaque ressource ;

# 4 CONCLUSION

Cette partie du rapport a proposé un diagnostic technique du mode opératoire des modules du dispositif BNEUF. Il a pris comme point de départ un diagnostic précédent réalisé par la direction du numérique de l'AUF pour identifier les points forts et les points faible du dispositif.

Ce rapport n'a pas abordé deux aspects particuliers : le socle technologique et l'ingénierie informatique qui participe de la conception du système de la BNEUF et de son fonctionnement. Ces aspects sont du ressort de l'équipe informatique de l'AUF. L'autre aspect est celui de la gouvernance du dispositif qui est reporté à la deuxième partie de e rapport.

Des recommandations générales sont toutefois nécessaires à rappeler avant d'entamer la seconde phase. Il s'agit de quatre modalités essentielles à revoir dans la mise à niveau du dispositif de la BNEUF :

- Les articulations entre les cinq modules « Ressources numériques », « Atlas de l’expertise », « Réseau social », « Ma bibliothèque » et « Ma Bneuf » doivent être renforcées et rendues aussi transparentes et fluides que possible ;

- Les zones d’interfaces où l’utilisateur est supposé faire une saisie de données, doivent fournir le maximum de consigne et d’explication pour réduire les ambigüités et contourner les blocages. Techniquement, le dispositif a les capacités programmatiques pour ajouter des fonctions à valeurs ajoutées comme le tri alphabétique, un moteur de recherche sur une zone de contenu, une bulle d’aide sur une icône, etc. ;

- Les formulaires de métadonnées gagneraient à être consolidés par des listes fermées de termes en langages contrôlés, de plans de classifications thématiques ou disciplinaires, de listes d’autorité d’auteurs et d’établissements, etc. Ce processus permet d’éviter les redondances sémantiques dans la description des ressources, la dispersion dans le classement des ressources, les bruits et les silences dans les résultats de recherche, etc.

- Il est aussi important de rappeler un problème ergonomique transversal : à plusieurs endroits, l’interface ne permet pas de garder les traces des étapes d’une navigation séquentielle ou des sous-ensembles d’une recherche avancée. Le bouton « retour en arrière » du navigateur remet l’utilisateur directement à la page d’accueil de la BNEUF et non à l’étape précédente d’un processus de navigation ou de recherche. Ce dysfonctionnement supprime par exemple toute combinaison de recherche en cours de construction.

# PARTIE II : ANALYSE MANAGÉRIALE

## 1 PRÉAMBULE

Les analyses diagnostiques (interne et externe) qui ont fait l'objet de la première partie de ce rapport ont permis de dégager dans le système BNEUF des services et fonctions qui nécessitent une optimisation d'ordre technique. BNEUF a également besoin de nouvelles mesures opérationnelles relevant d'une nouvelle forme de gouvernance au sein d'un écosystème général défini dans la nouvelle stratégie de l'AUF pour 2021-2025. L'écosystème défini dans la nouvelle stratégie de l'AUF est marqué de deux nouveaux facteurs auxquels la BNEUF devrait définir des points d'ancrage : d'une part la Francophonie scientifique comme orientation stratégique produisant et diffusant de la connaissance francophone et d'autre part une plateforme collaborative mondiale de services intégrés comme socle technologique fédérant l'ensemble des flux opérationnels (workflow) de l'AUF et de ses partenaires régionaux. La BNEUF devrait être l'un des constituants essentiels de l'écosystème de l'AUF en jouant un rôle central dans la production et diffusion de la connaissance dans la Francophonie scientifique.

Dans cette deuxième partie du rapport, il est question de soulever des points de commentaires sur le modèle de gouvernance du dispositif BNEUF à deux niveaux : un niveau technique pour une meilleure qualité de ses ressources et services utilisateurs et un niveau stratégique et managérial pour une optimisation de ses processus et mécanismes opérationnels élargis à son écosystème de gouvernance.

Les aménagements prévus sur le dispositif BNEUF nécessitent en revanche la mobilisation d'une équipe et d'un coordinateur d'équipe dont le profil sera proposé dans une fiche de poste à la fin de cette partie. Un coordinateur d'équipe du genre « gestionnaire de l'information et de la documentation » pourra prendre en charge le processus de rénovation et de contrôle qualité de la BNEUF.

## 2 OPTIMISATION DES PROCESSUS TECHNIQUES DE LA BNEUF

Le rapport de diagnostic interne de la Direction du numérique de l'AUF cité en référence ainsi que le diagnostic détaillé dans la première partie de ce rapport ont mis l'accent sur une série de dysfonctionnements techniques et procédurales (cf. partie I, 2.2) qui feront l'objet de recommandations dans les points suivants.

### 2.1 Épuration : gérer les doublons et les déchets documentaires

BNEUF fonctionne selon les principes technologiques des archives ouvertes qui collectent les métadonnées à partir de différents fournisseurs de données à l'aide du protocole OAI-PMH et fournissent une interface de recherche unifiée. La création de services de recherche fédérée pose toutefois de nombreux défis dont la pertinence des ressources et leurs doublons. Ces problèmes sont très courants dans les bibliothèques et les bases de données bibliographiques. La notion de « déchets documentaires » (analogiques ou numériques) englobe les deux types : d'une part les doublons et d'autre part les ressources inutiles (inadaptées aux besoins) ou mal exploitées (mal indexées). Des études montrent que près de 80 % des données en ligne ne sont pas structurées. Cela signifie que ces données n'ont aucune classification logique et donc ignorées dans les statistiques d'usage.

Il peut s'agir aussi de ressources en double copies provenant de sources différentes sans contrôle de duplication. Dans tous les cas de figure, il s'agit d'un gaspillage couteux de données et un alourdissement fortuit du système pour lesquels des stratégies de remédiation doivent être appliquées. Les chiffres d'usage de la BNEUF font ressortir des ordres de grandeurs qui posent ce problème de façon aigüe. Le rapport de la DN mentionne l'exemple du faible référencement des ressources par des entrées disciplinaires. Plus de 99% des fonds (environs 61 000 ressources sur près de 16 millions) passeraient ainsi sous silence dans un classement thématique.

***Mesures proposées :***

- Pour le fond actif, toutes les données existantes dans BNEUF doivent être examinées d'un point de vue fréquence d'usage pour en déterminer la pertinence. Le résultat déterminera si elles doivent être conservées ou plutôt supprimées définitivement. Les données utiles ou potentiellement utiles peuvent être ensuite inventoriées et classées/cataloguées de manière plus rationnelle ;

- Les doublons doivent être supprimés (ou fusionnés) pour améliorer à la fois l'efficience et l'efficacité du système de recherche d'informations. La détection, fusion ou suppression potentielle des doublons est souhaitable pour un certain nombre de raisons, telles que pour réduire le besoin de stockage et de calcul inutiles, et pour fournir aux utilisateurs des résultats de recherche épurés. Dans BNEUF, en raison des sources variées de moissonnage de métadonnées, des doublons peuvent survenir entre les points de moissonnage. La duplication des références ne peut donc être identifiée qu'à l'aide des métadonnées des ressources ;

- Explorer les méthodes de détection des doublons et évaluer leurs performances quand elles sont appliquées aux données bibliographiques. Il existe pour cela différentes approches de traitement des données et diverses solutions de stockage, de gestion et d'analyse de données (ex. Druide, ClickHouse, Cassandra, Prometheus et Elasticsearch). Ces approches et solutions présentent différents avantages et inconvénients. Il est important de les évaluer méticuleusement pour s'assurer que la solution choisie corresponde aux exigences spécifiques du dispositif technologique de BNEUF ;

- Tester ou créer pour cela des algorithmes adaptés pour épurer efficacement de grandes quantités de données. De nombreux algorithmes de détection des doublons existent, mais sont gourmands en calcul pour les grandes bases de données à sources variées de métadonnées. Une inspiration du modèle Zotero peut constituer une piste. Zotero utilise actuellement les champs titre, DOI et ISBN pour déterminer les doublons. Si ces champs correspondent (ou sont absents), Zotero compare également les années de publication (si elles sont à moins d'un an l'une de l'autre) et les listes d'auteurs/créateurs (si au moins un nom de famille d'auteur et la première initiale correspondent) pour déterminer les doublons. L'algorithme de Zotero résout toujours les éléments en double en les fusionnant, plutôt qu'en supprimant l'un des doublons. Les fusions conserveront toutes les métadonnées des éléments fusionnés. La suppression d'un élément entraînera la perte de ces données ;

- Pour le fond futur, établir une politique de sélection de contenus selon des indicateurs de pertinence thématique, disciplinaire et typologique dans les catalogues de ressources moissonnées pour éviter le fourre-tout documentaire. Les écarts entre les 34 catalogues de ressources moissonnées actuellement ne peut qu'être source de problème : plusieurs catalogues fonctionnent en mode cascade, les uns versants dans les autres. Les ressources des catalogues moins lotis sont fort probablement référencées dans les catalogues plus garnis, ce qui engendre l'existence de doubles copies dans l'absence d'un algorithme de gestion des doublons ;

- Il est conseillé de mettre en place des politiques claires d'organisation et de conservation des données (avec manuels de procédures) : détails concernant les données à stocker, où les stocker, comment classer les données et combien de temps les conserver. Il est également utile d'en faire une politique consistant à ajouter des métadonnées à tous les fichiers stockés pour faciliter la découverte et l'évaluation des données. Avoir une politique claire et complète sur l'organisation et la conservation des données a également l'avantage de faciliter l'automatisation des processus et de se conformer aux réglementations sur les données ;

- Adopter le concept de « source unique de vérité » c.à.d. avoir un référentiel central ou un index de toutes les données d'une collection. Cela garantit que les copies en double ou inutiles sont évitées et facilite également la recherche de données chaque fois que cela est nécessaire pour l'évaluation, la conservation, la fusion ou la suppression.

## 2.2 Définir des interfaces Ux-Design (expérience utilisateur)

Les portails et catalogues de bibliothèques, à l'instar de beaucoup de systèmes et dispositifs numériques, tendent à migrer progressivement vers des solutions participatives qui capitalisent sur les pratiques des utilisateurs : UX-Design. C'est le principe de l'adaptation ou personnalisation de l'interface homme-machine selon deux approches :

1) Des utilisateurs qui peuvent modifier les paramètres des interfaces utilisées à partir des données que le système a collecté sur leurs comportements ;

2) Un système qui répond au mieux aux besoins et aux préférences des utilisateurs grâce à des conceptions centrées sur deux leurs profils et leurs commentaires.

Pour les concepteurs UX, la possibilité d'adapter les interfaces aux profils des utilisateurs rend le processus de consultation des contenus plus attractif et conforme à ce que les utilisateurs s'attendent à voir. La personnalisation aide également les utilisateurs à passer plus de temps sur les contenus. Les services en ligne d'aujourd'hui et plus encore à l'avenir pourront facilement utiliser l'assistance de l'IA pour prédire davantage les étapes d'un utilisateur et personnaliser son expérience d'usage.

Les concepteurs UX peuvent également utiliser les données de rebondissement (feedback) pour mieux comprendre le comportement et les préférences des utilisateurs et prendre des décisions plus éclairées sur la façon de concevoir des interfaces utilisateur personnalisées. Cela permet des changements de conception plus rapides et des améliorations dans les domaines où les utilisateurs ont des difficultés ou dans lesquels la conception ne répond pas à leurs besoins. Cela peut se faire aussi par une « cartographie du parcours utilisateur », une technique utilisée pour identifier les points de contact entre l'interface et l'utilisateur et les interactions que ce dernier a eu avec un produit.

Au vu de ces tendances nouvelles en ergonomie d'interfaces, l'une des formes d'optimisation de BNEUF est dans l'ajout d'une fonction de recommandations ciblées, adressées à chaque utilisateur selon sa recherche en cours ou, pour les inscrits, selon leurs historiques de recherche, leurs profils et leurs intérêts déclarés. Il s'agit en back-office d'ajouter un module supplémentaire à l'architecture ORI-OAI de BNEUF qui contiendrait en plus des informations personnelles des inscrits, deux champs intitulés « intérêts » et « compétences ». Ces champs peuvent contenir une ou plusieurs valeurs numériques ou taxonomiques. En cas d'intérêts multiples, ils doivent être ordonnés selon leur degré d'importance et seront donc représentées par des indices correspondants à la classification utilisée (Dewey ou CDU) dans BNEUF.

La figure suivante présente un exemple de modification (modélisation) de l'architecture ORI-OAI pour ajouter des fonctions permettant de répondre à des besoins d'information personnalisés. Une passerelle spécifique entre les deux modules « Moteur de recherche » et « Profil utilisateur » permet de filtrer, organiser et recommander des ressources pédagogiques personnalisées lors d'opérations de recherche sur l'interface graphique de BNEUF.

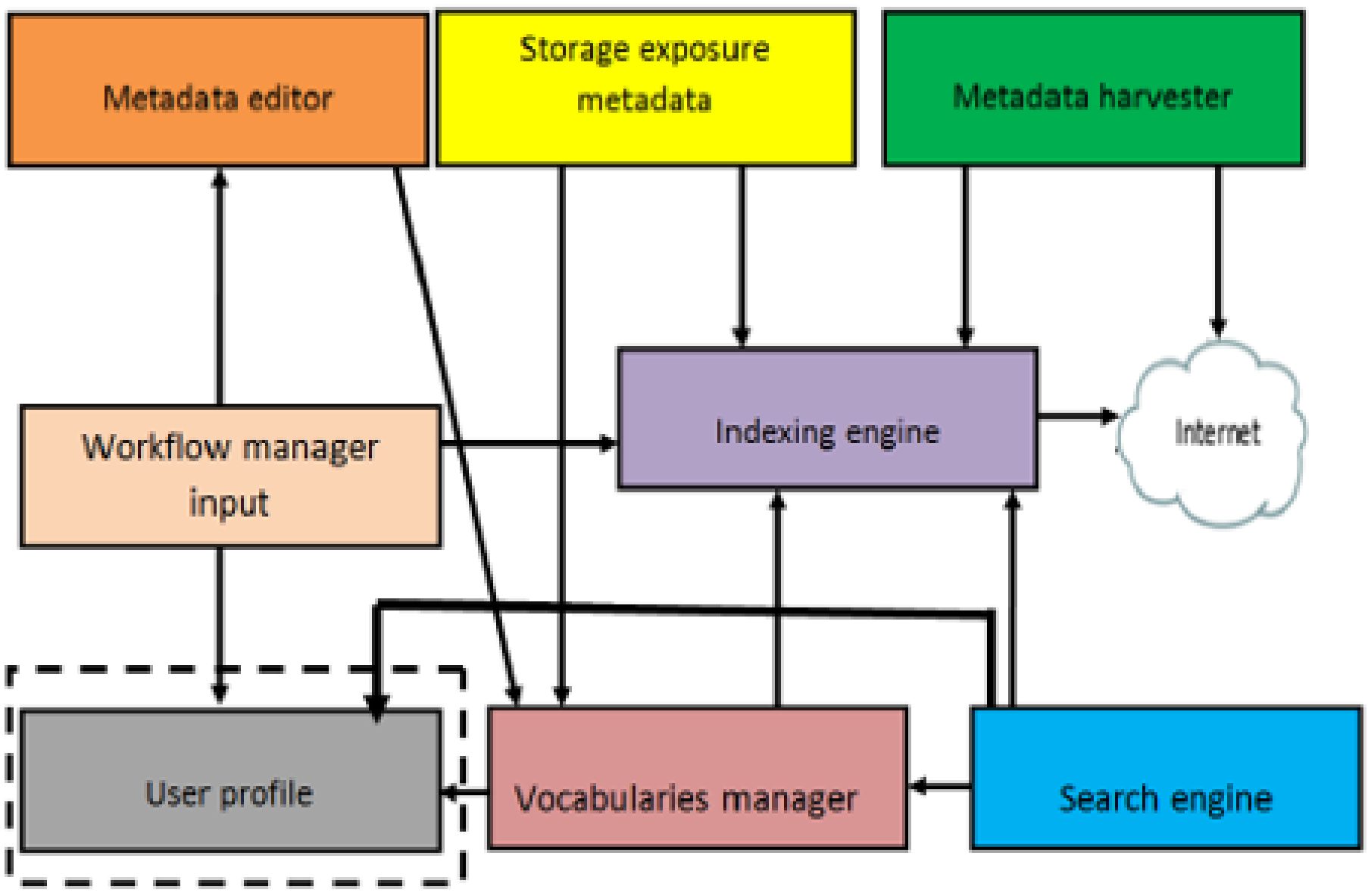

Architecture of ORI-OAI modified model

Source: Slimani et al. "[Personalized search and recommendation in a digital educational resources repository: The case of ORI-OAI]"

***Mesures proposées :***

- Inscrire la BNEUF dans une optique de nouvelle ingénierie de système basée sur la différence entre « Interface utilisateur » (UI) comme interface finie et « Expérience Utilisateur » (UX) comme interface optimisée et adaptable aux expériences utilisateurs prenant en compte l'UI mais aussi l'utilisabilité, la structure, le contenu, la rédaction, la recherche utilisateur, l'analyse, etc. ;

- Prévoir pour cela de constituer une base de données anonymisées des usages, alimentée en continu par l'activité en ligne des utilisateurs, de leurs requêtes, des modes et conditions d'usages des ressources et de leur typologie, afin d'optimiser le partage et d'accroitre la pertinence des réponses proposées aux usagers ;

- Suivre l'approche « flux d'utilisateurs » (User flow) dans la conception UX qui consiste à mémoriser le chemin de l'utilisateur pour lui permettre de terminer, sur plusieurs phases, un objectif spécifique de recherche ou de navigation sur BNEUF ;

- Éviter de submerger les utilisateurs avec trop d'informations grâce au principe du « minimalisme fonctionnel » en supprimant sur écran les éléments inutiles pour l'utilisateur. Cela ne signifie pas que les fonctions de l'interface doivent être limitées, mais qu'elles doivent être utiles et pertinentes ;

- Faire des « tests d'utilisabilité » avec de vrais utilisateurs pour surmonter les biais de faux consensus. Cela permet de produire des interfaces adaptées aux besoins réels des utilisateurs et non ce que le concepteur juge utile ;

- Montrer aux utilisateurs des éléments qu'ils peuvent reconnaître peut améliorer la convivialité. En raison des limitations de la mémoire humaine, il faut s'assurer que les utilisateurs peuvent reconnaître automatiquement comment utiliser certaines fonctionnalités de l'interface au lieu de leur faire rappeler ces informations. Utiliser pour cela les info-bulles pour gérer les ambiguïtés. Le but est de minimiser la charge cognitive et le temps de découverte en rendant les informations et les fonctions d'interface visibles et facilement accessibles.

## 2.3 Définir une ligne éditoriale adaptée et évolutive

Des aspects relatifs à la politique éditoriale de BNEUF sont évoqués sous le point 3.2.4. Certains sont repris ici pour insister sur leur importance :

- *Des métadonnées adaptées* : la BNEUF utilise un schéma de métadonnées sous forme d'un profil d'application utilisé au moment de la soumission en autoarchivage d'une ressource à ajouter dans le catalogue BNEUF. Or, un profil d'application de métadonnées, qui fera la singularité de la BNEUF, devrait faire l'objet d'une étude de pertinence plus poussée (cf. partie II-2.4) ;

- *Des taxonomies contrôlées* : un profil d'application est une structure de forme qui doit être optimisée par une rigueur de contenu, c.à.d. des métadonnées pour harmoniser les différents processus de la chaine éditoriale : catalogage, indexation, recherche, échange et conversion.

- *Des données ouvertes* : la BNEUF doit s'ouvrir aux systèmes et réseaux de référencement scientifique et aux sociétés savantes autour de pratiques éditoriales de découvrabilité et de valorisation. C'est particulièrement utile pour la partie Atlas de l'expertise francophone qui permettrait aux chercheurs inscris une meilleure visibilité. Dans ce contexte, les identifiants numériques et les DOI des publications agissent comme des connecteurs avec les systèmes scientifiques de référencement comme Scopus, Hal, Orcid, arXiv, etc. Une étude comparative avec des portails documentaires ou archives ouvertes comme HAL, Sudoc, Gallica, etc. devrait donner des pistes pour la BNEUF.

- *Des ontologies et des réseaux sémantiques* : Il faut reconnaître qu'en plus des techniques SEO classiques, un travail d'ingénierie doit être fait sur un système d'information pour que ses processus éditoriaux puissent produire de nouvelles formes de représentation plus élaborées. Les standards de description des métadonnées et les syntaxes logiques qui peuvent les mettre en œuvre ne cessent de se perfectionner : Warwick Framework, RDF, etc. Un tel effort de normalisation renforce l'effort sémantique qui s'appuie sur la construction d'ontologies où l'effort de classification encyclopédique est en pleine cohérence avec les fonctions majeures des bibliothèques. D'où la possibilité de construire des ontologies, ou du moins des hiérarchies de catégories permettant de faire en sorte que des machines puissent « comprendre » et non seulement « lire » les contenus des ressources comme de simples données stockées dans des silos. L'ontologie, ici est cruciale puisqu'elle donne du sens à la connaissance véhiculée traditionnellement par l'ensemble des schémas de description et thésaurus. Il s'agit par exemple d'évaluer et de sélectionner des documents afin d'en valoriser au mieux l'usage heuristique ou culturel (recherche d'informations pertinentes, création de collections thématiques, gestion des droits d'accès, etc.). Elle peut être utilisée dans BNEUF pour permettre d'effectuer des raisonnements logiques par/pour les utilisateurs. Or, prévoir cette perspective pour la BNEUF doit être suffisamment formalisée pour que des classes d'objets, des représentations de connaissances communes ou partagées puissent s'exprimer dans une syntaxe utilisable par des machines automatisées.

Dans la continuité des points précédents, l'équipe de la BNEUF peut prévoir des mécanismes évolués du Web des données et du Web sémantique pour relier les ressources entre elles en vue de constituer des réseaux d'informations connectées sous forme de cartes et graphes de points, de barres, de courbes, de graphiques animés, de cadrans et jauges, de cartes thermiques (comme l'arrière-plan du module Atlas) du type *sparkline* ou *fever chart*, etc. Pour la BNEUF, cela permettrait, par exemple, de produire des cartographies de groupes d'institutions universitaires, de groupes et d'unités de recherche, de parcours de formations, de profils d'usagers, de projets de recherche, etc. Cela ouvre aussi la perspective d'améliorer le processus éditorial de la BNEUF grâce aux technologies avancées de l'IA, du Web sémantique, et de la *DataViz* pour communiquer des chiffres ou des informations brutes en les transformant en objets visuels adaptées à chaque situation. Cela a pour but de rendre les données accessibles, comparables, compréhensibles et d'en tirer des enseignements.

En définitive, il faut rappeler que tout projet éditorial à succès documente ses processus, facilitant ainsi la garantie d'un contenu de qualité et le transfert des connaissances existantes aux utilisateurs. Pour que l'édition numérique soit rentable, elle a besoin d'un bon « flux de travail éditorial » qui élimine l'incertitude et réduit le gaspillage. Documenter un flux de travail pour la BNEUF est la première étape pour comprendre si les systèmes et les processus en place fonctionnent réellement. Un « workflow éditorial » BNEUF englobe les processus qui déterminent la façon dont une publication créée, éditée et publiée parvient jusqu'au lecteur. Il s'agit plus concrètement d'un document de référence destiné à formaliser les règles et les procédures auxquelles doivent se soumettre les membres d'un projet éditorial, en l'occurrence la communauté BNEUF.

***Mesures proposées :***

- Définir un processus d'assurance qualité comme méthode proactive visant à prévenir les problèmes de qualité des catalogues à moissonner et des métadonnées à collecter avant leur intégration dans le dispositif BNEUF ;
- Mettre en place un processus de contrôle qualité rétroactif sur le contenu de la BNEUF : le flux de travail éditorial sur l'état actuel de la BNEUF doit trancher entre deux options : soit publier des éléments de contenu moins nombreux mais plus percutants (i.e. l'épuration des déchets documentaires), soit se concentrer sur le moissonnage du plus grand nombre de catalogues de ressources au détriment de la profondeur et la qualité de la couverture ;
- Réduire les inefficacités opérationnelles : travailler sans flux de travail établi signifie que le contenu est produit à l'aide de processus archaïques et mal structurés. Cela produit des inefficacités opérationnelles telles que des redondances, des ambigüités, des doublons, des bruits et silences informationnels, et d'un contrôle de qualité inférieur à la moyenne ;
- Faire l'inventaire de l'existant pour identifier les catégories de ressources présentes dans les catalogues de ressources et chez les partenaires universitaires et déterminer si elles sont adaptées au public cible. Une étude de démarche qualité est à prévoir pour étudier la pertinence des ressources dans les 34 Catalogues moissonnés par BNEUF. Le résultat permettra de définir une stratégie éditoriale convergente en amont du processus de moissonnage et éviterait l'ensemble des redondances soulignées dans la première partie de diagnostic ;

- Focaliser sur la production éditoriale des pays partenaires francophones du Sud. Un nouvel équilibre en faveur des pays francophones du Sud dans l'offre documentaire de la BNEUF doit être considéré sérieusement comme un indicateur qui devrait crée la différence de la BNEUF avec ses « concurrents ». La BNEUF est dans un contexte compétitif mondial qu'il faut traiter avec une logique élémentaire de marché, celle de l'offre et de la demande. Il faut traiter cette dimension en répondant aux deux questions suivantes :
    - Quels sont les indicateurs qui caractérisent la BNEUF comme un service francophone différencié, destiné à servir les EES du Sud comme population cible privilégiée ?
    - Que devrait proposer la BNEUF à son public francophone du Sud pour le fidéliser à ses services et contenus ?
- Faire la promotion des contenus de la BNEUF sur les réseaux sociaux avec des « call-to-action » pour inciter les utilisateurs à partager les ressources, à s'inscrire à une newsletter, etc.
- Mieux cerner le public cible en créant des profils types. La BNEUF ne peut pas être un service « public large » comme une bibliothèque publique ou une bibliothèque nationale. Son caractère scientifique et particulièrement universitaire (enseignement/recherche) réduit son champ opérationnel. Or, le contexte universitaire, bien que stéréotypé dans le profil de ses acteurs, diffère d'un pays à l'autre et peut engendrer des dispersions dans l'identification des publics cibles ;
- Vérifier la « découvrabilité » des ressources via des outils de suivi du positionnement (*SeeYouRank, MyPoseo, IndexWeb*) ;
- Effectuer un suivi et mesurer des actions pour valider des KPIs définis au préalable sur les usages (ressources consultées, taux de rebond, temps passé sur le portail, taux de clics, nombre de partages, etc.).

## 2.4 Définir un profil d'application de métadonnées

On ne se pose plus de questions sur l'importance des métadonnées (et leurs schémas) dans les systèmes d'information et l'exploitation des bases de données. Les schémas de métadonnées prennent tout leur sens dans un contexte d'accès, d'organisation et de gestion des données propices à la diffusion et à la mutualisation de l'information à large échelle comme pour la BNEUF. Or, les schémas de métadonnées sont souvent convertis en « profil d'application », un ensemble d'éléments de métadonnées qui ont diverses origines et qui peuvent être mis en correspondance pour satisfaire les besoins d'une communauté spécifique tout en jouant la carte de l'interopérabilité. Il s'agit par exemple de collecter (moissonner) et combiner de l'information provenant de différentes sources pour en livrer de nouvelles interprétations ou bien de poser différents filtres sur de l'information pour la recontextualiser. C'est exactement le cas de la BNEUF qui, à la fois, moissonne les métadonnées de 34 catalogues différents mais en recueille d'autres à travers un profil d'application personnalisé pour récupérer des métadonnées provenant des dépôts individuels avec un objectif de recontextualiser ces métadonnées sous forme de produits dérivés.

La BNEUF, comme bibliothèque numérique destinée à l'enseignement supérieur, utilise un profil d'application constitué des 13 champs suivants : « Titre », « Auteur », « Résumé », « Mots-clés »,

« Droits d'auteurs », « Langue du document », « Difficultés », « Type », « Catégorie », « Durée d'exécution », « Contenu », « Niveau », « Pays ». On y constate un croisement fort avec les 15 champs de la norme Dublin Core, qui constituent eux-mêmes le noyau dur du LOM (*Learning Object Metadata*), ce qui lui confère une grande marge d'interopérabilité.

Cependant, il n'est pas évident que tous les catalogues moissonnés par BNEUF ont des schémas de métadonnées d'un niveau élevé de convergence avec un format pivot comme Dublin Core ou LOM, d'où l'étude nécessaire des caractéristiques des Catalogues moissonnés pour définir un format d'échange ou un convertisseur capable de faire converger toutes les divergences des métadonnées moissonnées vers un profil d'application unifié propre à BNEUF. Les convertisseurs de métadonnées sont abondants dans le monde des bibliothèques et leur utilisation est très répandue entre catalogues en ligne pour l'importation des métadonnées lorsque le format bibliographique source et le format bibliographique cible sont différents. Cette solution traite la métadonnée source avant de l'importer.

Un profil d'application bien étudié, propre à la BNEUF est dès lors nécessaire pour harmoniser les opérations de moissonnage ORI-OAI (DC & LOM) et autres catalogues sources (autres formats). Les métadonnées hétérogènes provenant de sources variées, passeraient ainsi par un filtre de conversion pour être adaptées à un profil BNEUF. Le principe est d'ouvrir les champs du possible en empruntant divers éléments de métadonnées à différents standards et en les articulant (approche dite de "mix et match") pour en produire une nouvelle organisation d'éléments de métadonnées, particulièrement adaptée à une visée applicative cible. Un tout premier principe, ciblant l'ouverture et l'interopérabilité des profils, est de n'exploiter que des schémas de métadonnées existants, ou à défaut de maintenir de manière ouverte et sur le long terme un nouveau schéma de métadonnées qui vient couvrir les éléments de métadonnées nouvellement introduits. Il s'agit dans ce cas de définir la meilleure articulation possible entre plusieurs schémas de métadonnées et à poser les contraintes sur les éléments de ces schémas à même de prendre en charge la spécificité des domaines couverts.

***Mesures proposées :***

- Prévoir une étude approfondie du profil d'application utilisé pour la proposition de nouvelles ressources dans le module « Ma BNEUF ». L'étude se focalise sur les points suivants :
    - Le choix des champs de descriptions : chaque champ doit faire l'objet d'un argumentaire d'utilité et de pertinence ;
    - Revoir les caractéristiques de chaque champ : occurrence (unique ou répétitif), cardinalité (obligatoire ou facultatif), modalités de sa saisie (libre ou contrôlée), etc. ;
    - Prévoir la suppression de certains champs ou l'ajout d'autres selon des argumentaires étudiés ;
- Définir un nouveau profil d'ensemble de description des ressources qui permet de choisir et de contraindre les éléments de métadonnées retenus avec la production d'un dictionnaire décrivant ces éléments et les contraintes qui vont venir s'y appliquer (cardinalité, multiplication, etc.) ;
- Produire une documentation annexe qui viendrait compléter de manière optionnelle la spécification du profil d'application retenu pour en faciliter l'exploitation. De même, il peut être utile de fournir certaines préconisations concernant les syntaxes adaptées dans le contexte du profil. Cette documentation est à diffuser auprès de la communauté francophone qui chercherait à proposer de nouvelles ressources. Elle fera aussi partie de la stratégie incitative de mobilisation et de partenariat avec les EES (cf. partie II-3.2) ;

- Utiliser le nouveau profil d'application comme filtre central pour la conversion des métadonnées provenant des Catalogues de ressources. Il s'agit au fait de créer une convergence et une interopérabilité entre des métadonnées provenant de sources divergentes avant leur entrée dans l'entrepôt de métadonnées de BNEUF.

## 2.5 Gérer plus rationnellement les licences CC et les droits d'auteurs

Le rapport de la Direction du Numérique qualifie ce point dans ces termes : « *Les modalités de protection de la propriété intellectuelle des ressources mises en ligne ne sont pas clairement définies. Seulement une petite partie concernant principalement les ressources éducatives libres donne des détails sur les licences et les conditions d'accès aux ressources* ».

Ce point a été également soulevé sous le point 2.2 de la première partie comme point sensible qui a été négligé dans la gestion des données de la propriété intellectuelle des ressources. En effet, sur BNEUF, plusieurs exemples démontrent une pratique aléatoire dans le signalement des données de licences et de droits d'auteurs, certes héritées des catalogues des ressources, mais qui ne la désengage pas de la responsabilité juridique de les rediffuser sur son portail. Le fait de donner accès à des ressources numériques doit se faire dans un cadre réglementaire qui rend la BNEUF responsable au regard des auteurs des ressources qu'elle diffuse. Cette responsabilité se détermine tout particulièrement par rapport aux ressources diffusées ouvertement alors qu'elles sont protégées par des droits d'auteurs. Pourtant, la culture des droits d'auteur et des licences est bien connue dans le domaine de l'édition numérique avec des règles bien établies et un jargon bien connu et répandu.

La source du problème est sans doute située à l'origine du système de référencement au niveau des catalogues de ressources moissonnées. Il n'est pas évident que tout les opérateurs et commis de saisie des métadonnées des ressources ont la formation nécessaire pour gérer cet aspect légal, raison pour laquelle on constate des métadonnées de droits d'auteurs du genre « gratuit » ou « Cairn », etc. alors que les mentions de droits d'auteurs courants sont réduits quasiment à trois forme « Protégé par droit d'auteur », « Creative Commons » ou « Domaine public » exprimés dans ces trois formes précises ou dans des formes dérivées (Copyright, CC, C0) mais qui expriment sans ambiguïté la catégorie des droits associés à une ressources. Le mélange entre des ressources libres, des ressources du domaine public et des ressources sans mention de licence peut aussi être à l'origine de cette confusion.

L'harmonisation de ces variantes (comme plusieurs autres dans BNEUF) passe à travers un processus de « mappage » des données qui consiste à mettre en correspondance les variables d'un champ de données vers une forme unifiée de ces valeurs.  C'est la première étape pour faciliter tout traitement de données tant pour la migration, l'intégration, l'extraction, la synchronisation automatique de données. La qualité du mappage conditionne la qualité des données qui seront analysées pour en tirer des informations exploitables.

***Mesures proposées :***

- Commencer par changer sur « Ma Bneuf » les options actuelles du champs « Licence » (Libre de droits/gratuit/payant) qui n'expriment pas la dimension réelle des option des licences et des droits d'auteurs. Avec la prolifération des ressources éducatives libres très ancrées dans la culture des Creative Commons, les six options des CC, en plus du domaine public et des droits d'auteur, sont souvent proposées ;

- Prévoir un module sur BNEUF (Ma Bneuf) pour expliquer les principes des droits d'auteurs et des licences *creative commons* et du domaine public. Ce module aiderait les dépositaires volontaires à s'informer sur la typologie des droits existants et à les engager dans leurs choix de licences pour leurs ressources ;
- Rédiger une charte de droit d'auteurs pour les partenaires créateurs de ressources et producteurs de métadonnées. Cette charte fera partie d'une littérature d'accompagnement pour les partenaires à mobiliser autour de la BNEUF ;

## 2.6 Production de marques blanches

En tant que concept d'origine entrepreneuriale, le principe de la marque blanche se dit d'une solution ou d'un produit cédé ou loué à une entreprise sur lequel cette dernière peut apposer sa propre marque et donc en revendiquer la paternité auprès de ses clients. Le concept de marque blanche a pourtant prouvé ses avantages dans des secteurs d'activité moins marqués par le marketing et le profit. Avec le numérique, de nombreux éditeurs de services ou de solutions Internet ont également bâti des stratégies sur ce principe sous forme de technologies ou contenus distribués à grande échelle sur des sites affiliés. L'un des exemples réussis qui pourrait illustrer cette nouvelle tendance est celui la Bibliothèque nationale de France (BnF) avec son "Gallica Marque blanche"[3], un dispositif lancé en 2013 qui permet à toute institution de créer une bibliothèque numérique en réutilisant l'infrastructure de la bibliothèque numérique de la BNF (Gallica). En parallèle, la BnF enrichit ses collections nationales numérisées de l'apport des fonds des Marques Blanches.

L'expérience de l'AUF avec l'IFEF pour la création de la marque blanche RELIEFH portant sur des ressources dédiées au thème l'égalité femme-homme constitue un projet pilote à reproduire avec d'autres acteurs et dans d'autres domaines. Plusieurs autres sujets porteurs constituent des alternatives possibles à négocier en mode marque blanche avec des développeurs de contenus : les 17 axes des Objectifs de développement durable avec leurs ramifications sur des thèmes génériques comme le réchauffement climatique, les droits de l'Homme, la migration clandestine, l'énergie renouvelable, l'éducation pour tous, etc. ou des thèmes spécifiques pour la Francophonie comme les langues vernaculaires, le français dans le monde, la Francophonie scientifique, la recherche en français, etc.

***Mesures proposées :***

- Définir une liste de thèmes prioritaires à proposer en maque blanche. Cela se détermine en fonction des orientions de l'AUF comme exprimées dans sa stratégie d'avenir ;
- Identifier des acteurs potentiels travaillant dans l'un des domaines prioritaires identifiés ou potentiellement intéressé de partager ou collaborer dans la production de ressources dans un ou plusieurs de ces domaines ;
- Préparer des modèles de termes de références pour négocier des accords-cadres avec des éditeurs commerciaux en marque blanche. L'accord-cadre avec l'IFEF peut constituer un point de repère ;

---

[3] Présentation de Gallica marque blanche (2020) : https://www.bnf.fr/fr/mediatheque/presentation-de-gallica-marque-blanche

- Recenser et étudier les projets en cours de marque blanche dans le domaine de l'édition numérique pour identifier des partenaires potentiels avec lesquels négocier des conventions de partenariat. En voici deux cas concrets :
    - « Gallica Marque blanche » est un modèle très proche de BNEUF qu'il faudrait étudier pour s'inspirer de son mode opératoire[4]. En tant que Catalogue de ressources dominant dans la BNEUF, un terrain de discussion se présente à l'AUF pour engager des discussions avec la BNF sur la voie d'une solution commune de marque blanche ;
    - Le projet de bibliothèque personnelle d'ebooks développé par Numilog et ePagine qui « sous-traite » à des libraires la production en mode marque blanche de ebooks. Ces libraires étaient en attente d'alternative aux plateformes des géants de l'Internet et de la technologie sans faire disparaître leurs identités de libraires. Dans ce projet, les protocoles utilisés sont standards et interopérables. En revanche, cette interopérabilité concerne les commandes et non la lecture, ce qui constitue une bonne piste à la BNEUF pour négocier avec ces libraires une marque blanche de lecture ;
- Envisager une stratégie de sensibilisation aux marques blanches auprès des partenaires universitaires du Sud en focalisant sur les avantages qu'ils pourraient avoir par la validation de leurs collections documentaires et leurs métadonnées, de leur intégration dans la collection et entrepôts numérique de la BNEUF, du développement d'un site spécifique à leurs couleurs pour accueillir leurs documents avec une gestion autonome des activités éditoriales de leurs contenus.

## 2.7 Points sur la conception d'un écosystème intégré

L'enjeu primordial pour la BNEUF est de devenir l'un des agrégats liés à l'écosystème d'information intégré que l'AUF envisage de mettre en place dans le cadre de sa stratégie 2021-2025. Cela consiste à définir les choix technologiques et méthodologiques pour une articulation entre les constituants actuels de la BNEUF d'une part et une plateforme collaborative mondiale de services intégrés d'autre part afin de créer un « carrefour virtuel créant du lien au sein de la communauté scientifique francophone internationale ». Différente d'une solution ERP ou PGI (Progiciel de Gestion Intégré), une plate-forme collaborative mondiale de services intégrés se baserait plutôt sur des API qui créent des connexions entre des systèmes et plates-formes hétérogènes distribués.

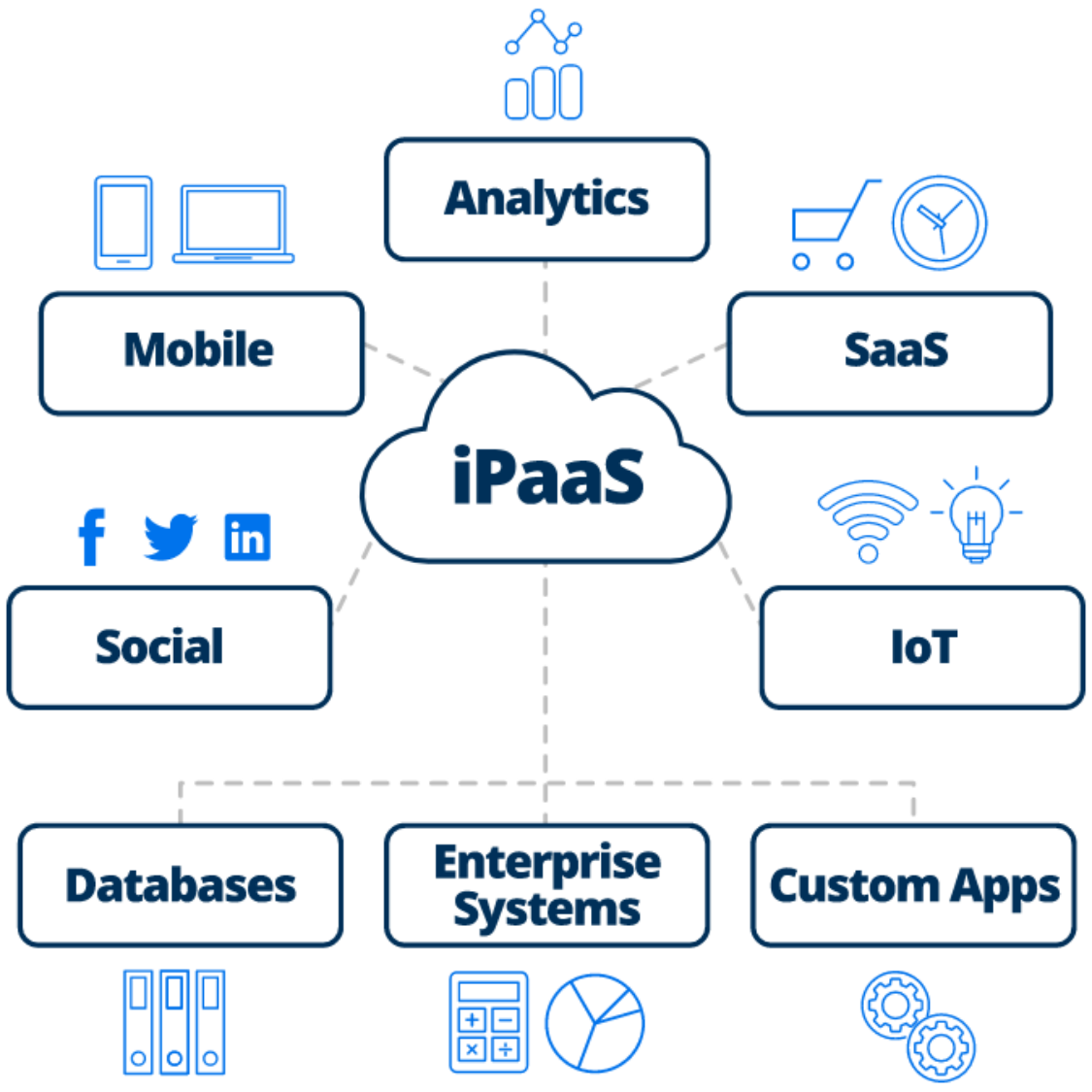

Rappelons qu'un écosystème numérique intégré est un réseau de technologies, de plateformes et de services numériques interconnectés, alimentés par des API et doivent fournir une interopérabilité transparente entre les plateformes, les logiciel et les appareils des utilisateurs finaux. Les écosystèmes numériques n'existent pas isolément. Il s'agit plutôt d'un amalgame d'outils et de services hétérogènes qui se combinent dans une seule expérience intégrée pour offrir facilité, commodité et qualité aux utilisateurs. Techniquement, outre des API, l'intégration basée sur le cloud (iPaaS) se démarque comme une forme d'activité d'intégration de systèmes fournie comme un service de *cloud computing* qui traite les données, les processus, l'architecture orientée services et l'intégration d'applications.

[4] Gallica marque blanche - été 2022 – BnF, https://www.google.com/url?sa=t&rct=j&q=&esrc=s&source=web&cd=&ved=2ahUKEwjPubj8i7P9AhUDXqQEHXfMDAIQFnoECA4QAQ&url=https%3A%2F%2Fwww.bnf.fr%2Fsites%2Fdefault%2Ffiles%2F2020-07%2FGallica_marque_blanche_ete20.pdf&usg=AOvVaw31PGg1AVof3yHM3vVlNIQF

La nouvelle génération d'iPaaS est construite sur des plates-formes plus puissantes et flexibles. Leur architecture leur permet de s'adapter facilement à différentes configurations. Le rapport annuel de Gartner Inc. Pour 2022 cite un certain nombre de fonctionnalités clés du modèle de référence qui rassemblent notamment des outils et des technologies de support destinés à l'exécution des flux d'intégration, à la gestion du cycle de vie et du développement des intégrations, à la gestion et au suivi des flux d'applications, à la gouvernance et aux fonctionnalités essentielles du cloud (multilocation, élasticité, autoapprovisionnement, etc.).

Il y a plusieurs configurations d'écosystèmes intégrés. Les plus avancés sont les écosystèmes intégrés d'information et de plateformes numériques. Ils peuvent impliquer des millions de partenaires et peuvent également intégrer une multitude de solutions numériques. Ce type d'écosystème numérique est fortement axé sur l'approche « data first » qui consiste à exploiter les informations qui circulent sur les différents services connectés afin de concevoir de nouveaux livrables à partir des données générées. Un écosystème intégré réussi doit cultiver un réseau de collaboration qui motive un grand nombre de parties ayant des intérêts similaires (comme les universités) à unir leurs forces et à poursuivre des objectifs similaires sans qu'il y ait de centralisation ni dans les dispositifs ni dans les contenus.

La mise en place d'un écosystème intégré d'information autour d'une plate-forme collaborative mondiale est une lourde tâche qui fait appel à des technologies diverses et des compétences multiples. Elle nécessite une étude approfondie pour mettre en place un cahier des charges qui permettrait de sélectionner un archétype d'écosystème de données avant de se concentrer sur la mise en place de la bonne infrastructure pour soutenir son fonctionnement. L'élaboration d'un tel cahier des charges dépasse les limites de ce rapport. En revanche, il est possible d'émettre quelques hypothèses d'orientation à ce sujet.

Comme condition absolue, un écosystème intégré ne peut pas tenir sa promesse sans garantir l'accès fluide aux données. Cet élément critique repose sur la conception de l'architecture des données. Plusieurs questions primordiales sont à poser à ce sujet lors de la mise en place de l'écosystème de données autour de la plateforme collaborative à services intégrés :

- *Comment échanger des données entre les partenaires de l'écosystème* ? L'expérience montre que les mécanismes standard d'échange de données suivent généralement trois étapes : établissement d'une connexion sécurisée, échange de données via des navigateurs et des clients, et le stockage centralisé des résultats si nécessaire ;
- *Comment gérer l'identité et l'accès* ? Deux stratégies peuvent être appliquées pour sélectionner et mettre en œuvre un système de gestion des identités. L'approche la plus courante consiste à centraliser la gestion des identités via des solutions les plus robustes du marché comme Okta, OpenID ou Ping. Une approche émergente consiste à décentraliser et fédérer la gestion des identités, par exemple en utilisant des mécanismes de registre blockchain ;
- *Comment définir les domaines de données et le stockage* ? Alors que traditionnellement, un gestionnaire d'écosystème centralisait les données dans chaque domaine, les tendances les plus récentes favorisent une architecture ouverte de maillage de données (exposition). Le maillage de données remet en question la centralisation conventionnelle de la propriété des données au sein d'une partie en utilisant les définitions et les actifs de domaine existants au sein de chaque partie en fonction de chaque cas d'utilisation ou produit. Certains cas d'utilisation peuvent encore nécessiter des définitions de domaine centralisées avec un stockage central. En outre, des normes mondiales de gouvernance des données doivent être définies pour garantir l'interopérabilité des actifs de données.

- *Comment gérer l'accès aux données non locaux et comment éventuellement les consolider* ? La plupart des cas d'utilisation peuvent être mis en œuvre avec des chargements de données périodiques via des interfaces de programmation d'application (API). Cette approche se traduit par une majorité de cas d'utilisation ayant un stockage de données décentralisé. Poursuivre cet environnement nécessite deux catalyseurs : un catalogue d'API central qui définit toutes les API disponibles pour assurer la cohérence de l'approche, et une gouvernance de groupe solide pour le partage des données.
- *Comment redimensionner l'écosystème, compte tenu de sa nature hétérogène et faiblement couplée* ? Permettre un accès rapide et décentralisé aux données ou aux sorties de données est la clé pour faire évoluer l'écosystème. Cet objectif peut être atteint en ayant une gouvernance solide pour s'assurer que tous les participants de l'écosystème font ce qui suit :
  - Rendre leurs données détectables, adressables, versionnés et dignes de confiance en termes de précision ;
  - Utiliser une sémantique autodescriptive et des normes ouvertes pour l'échange de données ;
  - Prise en charge des échanges sécurisés tout en permettant un accès à un niveau granulaire.

> En définitive, le succès d'une stratégie d'écosystème de données dépend de la disponibilité des données et de la préparation de l'API pour permettre l'intégration, la confidentialité et la conformité des données et l'accès des utilisateurs dans une configuration distribuée. Cette gamme d'attributs est à transmettre et faire appliquer par les partenaires intéressés à se joindre à la plate-forme collaborative mondiale.

# 3 RENFORCEMENT DU MODÈLE DE GOUVERNANCE DE LA BNEUF

Après avoir exploré des pistes de remédiation d'ordre technique sur le dispositif BNEUF, la section suivante focalisera des aspects d'ordre managérial. Elle contiendra des propositions sur des aspects permettant de gérer l'écosystème général de la BNEUF et son modèle de gouvernance.

## 3.1 Définir un modèle économique hybride

L'extrait suivant du rapport de diagnostic de la DN explique à quel point le modèle économique BNEUF pose problème : « *Sans informations claires en amont sur les conditions d'accès aux ressources, les usagers se voient être régulièrement redirigés vers des ressources qu'ils pensent pourvoir accéder gratuitement alors qu'elles sont payantes. Par ailleurs, vu que le modèle économique envisagé pour assurer la gratuité des ressources payantes n'est pas opérationnel, la pertinence de la présence des ressources payantes sur le portail peut se poser* » (4.2, p.15).

Dans une configuration initiale de l'initiative IDNEUF, il a été souligné que « *L'accès à la plupart des ressources est gratuit. Pour les autres ressources privées et payantes (env. 10%), un modèle économique de tiers payant est en réflexion pour assurer la gratuité d'accès, ou un tarif d'accès modique, à tous les étudiants et enseignants* ». L'hypothèse envisagée était alors de parvenir au bout de 5 à 8 ans à un autofinancement en valorisant des services générateurs de ressources propres. Ceci confirme que la question d'un modèle économique a été posée dès les débuts du projet IDNEUF dans les termes d'une hybridation de modèles entre un accès gratuit et un accès payant. Il restait juste à définir les conditions dans lesquelles cette hybridation peut avoir lieu.

Ce qui est néanmoins certain, c'est que les technologies numériques sont au cœur de l'hybridation des modèles économiques de l'édition et de l'accès à l'information en ligne. L'hybridation est en effet une option qui rencontre un certain succès qu'il faut regarder de près car elle s'avère être un puissant levier de création de valeur dans un environnement économique qui se caractérise par une incertitude croissante. Elle s'ajoute à d'autres modèles qu'il serait opportun d'étudier pour en déterminer la pertinence et l'efficience pour la BNEUF. Il s'agit notamment de :

– *Le modèle économique de Google* : La proposition de valeur de Google est de fournir via Internet des annonces publicitaires extrêmement ciblées au niveau mondial. Grâce à AdWords, les annonceurs ont la possibilité de diffuser des annonces et des liens sponsorisés sur les pages de recherche de Google. Les annonces apparaissent à côté des résultats de recherche lorsque les internautes utilisent le moteur de recherche.

– *Le modèle économique du gratuit* : Dans le modèle économique du gratuit, un segment de client important a la possibilité de bénéficier de manière continue d'une offre gratuite. Différentes modalités rendent l'offre gratuite possible. Le segment qui ne paye pas bénéficie d'une autre composante du modèle qui supporte les coûts ou d'un autre segment de clients. Il s'agit là de fournir certains services de base gratuitement à un premier segment de client et de faire payer d'autres services en s'appuyant notamment sur l'activité générée par le premier segment de client. Par exemple, la publicité utilise le modèle de plates-formes multiface.

– *Le modèle freemium* : Dans ce modèle une base importante du client bénéficie d'une offre gratuite et sans engagement. Une petite proportion de clients payants qui souscrit au service premium, subventionne la masse des utilisateurs gratuits. Ceci est possible grâce au faible coût marginal de service d'utilisateur gratuit supplémentaire. Dans le modèle freemium les indicateurs clés sont le coût moyen de service d'utilisateur gratuit et le taux auquel des utilisateurs gratuits sont convertis au client premium.

***Mesures proposées :***

- Définir les critères d'un modèle économique pour la BNEUF qui tiendrait compte des orientations stratégiques de l'AUF en rapport à la fois avec l'éducation ouverte, le logiciel libre et les ressources éducatives libres qu'avec les éditeurs commerciaux qui ont été impliquée dans la stratégie de 2017-2021. Dans le « Livre blanc de la Francophonie scientifique » et la « Stratégie AUF 2021-2025 », ces orientations économiques ne sont pas clairement définies ;

- Si la perspective de réimpliquer les éditeurs commerciaux est encore valable, définir un cadre de négociation pour des prix préférentiels en faveur des établissements membres de la Francophonie comme moyen de sensibilisation et motivation pour adhérer à l'usage et la contribution aux contenus de la BNEUF ;

## 3.2 Définir une politique d'accompagnement incitatif avec les EES :

Mobiliser les acteurs autour d'un dispositif ou d'un projet est une étape essentielle de la mise en place d'un processus participatif. Il s'agit de rassembler des acteurs (personnes physiques et entités morales) en mesure d'apporter un plus au projet. La mobilisation de personnes issues d'horizons culturels, sociaux, et professionnels divers, relève aussi du caractère durable du projet.

La BNEUF a besoin d'une appropriation par les partenaires de l'AUF. Une stratégie de sensibilisation, de mobilisation, de formation et d'accompagnement est à prévoir en direction des partenaires du Sud pour ancrer l'identité et la culture scientifique du dispositif BNEUF dans les implantations francophones mondiales et chez les établissements partenaires.

Or, le rapport de la DN souligne les contraintes des partenaires et utilisateurs « *à contribuer au portail en soumettant leurs ressources éducatives ; il n'est pas possible pour les établissements et professeurs de créer un compte et de déposer directement des ressources. Par ailleurs, aucune procédure formelle de validation des contributeurs n'ait été défini. En plus, le manque d'informations relatives aux fréquentations du portail, et à sa visibilité en général, ne favorise pas la participation des établissements qui ont déjà développé leur propre système* ». Ceci montre la défaillance du modèle de communication actuel autour de la BNEUF et la nécessité de le revoir pour un meilleur rendement.

***Mesures proposées :***

- Définir un plan de communication en direction des communautés francophones du Sud : faire la promotion du service BNEUF sur les sites des Directions régionales de l'AUF et négocier des échanges de logos avec les sites partenaires ;
- Définir une campagne de sensibilisation sur la base des services à valeurs ajoutées de la BNEUF : communiquer aux enseignants-chercheurs francophones les nouveaux services et fonctions à valeurs ajoutées, (une fois créés), comme les Identifiants numériques, les DOI et la connexion avec les réseaux de référencement SCOPUS ou ORCID pour éveiller leur intérêt pour la eRéputation ;
- Définir une stratégie de motivation sous forme de compensations matérielles : l'accès gratuit aux services de la BNEUF ne serait pas suffisant s'il n'est pas doublé d'une compensation matérielle pour l'autoarchivage. Les enseignants-chercheurs qui déposeraient leurs productions (cours, recherche), selon des critères de qualité prédéfinis, pourraient éventuellement bénéficier de crédits ou de bonus à faire valoir pour l'accès à des ressources payantes ;
- Proposer comme services à valeur ajouté sur BNEUF des systèmes de veille intelligente personnalisés de type Almetrics, Google Scholar ou Google Analytics ;
- Passer des contrats-cadres avec des EES francophones du Sud pour optimiser leurs catalogues de bibliothèques en créant des API adaptées pour les relier à la BNEUF. La solution des marques blanches peut être envisagée sous cet angle ;
- Prévoir une formule de « tarif d'accès modique » pour les utilisateurs individuels ou les universités « tiers payantes » pour accéder aux ressources privées dans la BNEUF. Conduire une enquête d'opinion à ce sujet auprès des acteurs concernés ;
- Développer la culture de l'accès ouvert auprès des partenaires du Sud à travers des manifestations scientifiques ou des appels à projets autour des archives ouvertes, de l'éducation ouverte et la science ouverte avec une focalisation de production de ressources téléchargeables sur la BNEUF ;

## 3.3 Créer un comité de pilotage BNEUF

Pour atteindre au mieux ses objectifs, le projet de la restructuration de la BNEUF piloté par la Direction du numérique a besoin d'un cadre organisationnel polyvalent élargi à différents acteurs internes et externes à l'AUF. D'autres directions des services centraux, des institutions comme l'IFIC et les différentes directions régionales sont toutes concernée directement ou indirectement par les services de la BNEUF. Des acteurs externes sont de même à solliciter pour des services complémentaires d'ingénierie, d'expertise et de conseil, de formation et d'expérimentation.

***Mesures proposées :***

- Créer un comité de pilotage pour la rénovation de BNEUF et un comité de suivi des phases d'entretien du dispositif ;
- Le comité de pilotage serait composé de membres internes de la Direction du numérique avec la contribution d'acteurs externes, quand nécessaires, sur des questions documentaires ;
- Un comité de suivi serait composé de représentants de la direction du numérique avec des vis-à-vis régionaux qui assurent la coordination des antennes délocalisées ;
- Le comité de suivi est coordonné par un animateur de dispositif recruté selon les critères de la fiche de poste suivante.

# 4 COMPÉTENCES POUR UNE FICHE DE POSTE (ANIMATEUR DE DISPOSITIF)

La mise en œuvre d'une nouvelle conception des services numériques de l'AUF autour de l'agrégat BNEUF (Bibliothèque, Atlas et réseau social) nécessite des compétences particulières déployées en permanence pour la gestion, maintenance et animation du dispositif. Une fiche de poste est proposée pour définir un profil métier d'un animateur de dispositif qui prendrait l'intitulé de « Gestionnaire de l'information et de la documentation », « Administrateur ou Coordinateur de systèmes d'information », « Technicien.ne à la gestion des documents numériques », etc.

**DESCRIPTION DES TÂCHES**

Gérer un système d'information documentaire par des compétences polyvalentes d'administration de bibliothèques numériques en ligne et d'animation de communautés d'intérêt interdisciplinaires, en coordonner les processus de traitement des ressources et de prestation des services et en faire la promotion auprès des partenaires scientifiques et économiques de la Francophonie.

**PROFIL TYPE**

Le/la cadidat.e. au poste doit avoir une solide connaissance des procédures de gestion numérique des données documentaires. Il/elle doit avoir des compétences spécifiques au domaine des bibliothèques numériques, de l'informatique documentaire et du Web des données :

***Fonctions principales :***

- Coordonner une politique de développement des fonds et des collections documentaires d'une bibliothèque numérique francophone en ligne ;
- Animer un réseau social et des groupes d'intérêt autour de la bibliothèque numérique francophone ;

- Assister et former sur les usages des supports et des outils documentaires du dispositif de la bibliothèque numérique francophone ;
- Définir et réaliser une veille documentaire interdisciplinaire pour le compte de la bibliothèque numérique francophone ;
- Négocier des partenariats avec des universités, centres de recherche, éditeurs, laboratoires, sociétés savantes, portails de ressources ;

***Compétences générales (soft skills) :***

- Faire preuve d'une pédagogie rigoureuse dans la réalisation des plans d'action du projet de la bibliothèque numérique francophone ;
- Faire preuve de clarté et d'un grand esprit de synthèse et d'une capacité d'analyse, d'interprétation et d'organisation de grandes quantités de données liées à et issues de la bibliothèque numérique francophone ;
- Avoir d'excellentes compétences en communication orale et écrite pour traduire des processus complexes en utilisant des termes non techniques ;
- Avoir des bonnes aptitudes de marketing pour promouvoir les produits de la bibliothèque numérique francophone et lui créer des communautés de pratiques ;
- Avoir une bonne capacité d'écoute et de bonne aptitude au dialogue pour exprimer les besoins du projet et négocier des partenariats ;
- Avoir une bonne culture juridique pour savoir traiter des questions relatives aux licences, droits d'auteurs et protection des données personnelles ;

***Compétences métier (savoir-faire) :***

- Pouvoir intervenir dans les processus techniques d'une chaine de gestion documentaire : acquisition, classement et indexation, conservation et diffusion des ressources ;
- Savoir utiliser et maintenir les outils techniques et méthodologiques du traitement de l'information documentaire (thesaurus, plan de classement, taxonomies, langages contrôlées, …)
- Pouvoir structurer l'information : gérer des bases de données, identifier les métadonnées adaptées, maintenir les référentiels nécessaires à la recherche de l'information ;
- Pouvoir mettre en œuvre les solutions techniques de gestion de l'information en prenant en compte les attentes et les pratiques des usagers, les contraintes techniques et économiques de l'institution dans le respect des normes, de la sécurité informatique et du droit de l'information ;
- Pouvoir mettre en place une stratégie de veille documentaire pour alimenter le contenu de la bibliothèque numérique francophone ;
- Avoir une bonne maîtrise des outils de travail collaboratifs (planning partagé, web conférence, réseaux sociaux, ...) ;
- Savoir élaborer des rapports sur les activités et service de la bibliothèque numérique francophone.

…/…